\documentclass[usenatbib]{mn2e}

\usepackage{graphicx}
\usepackage[usenames,dvipsnames]{color}
\usepackage{amssymb}

\newlength{\colwidth}
\newcommand{\NGB}{N_{\rm ngb}}
\newcommand{\FTH}{f_{\rm th}}

\newcommand{\Msolyrkpcsq}{\mbox{M}_{\sun}\,\mbox{yr}^{-1}\,\mbox{kpc}^{-2}}
\newcommand{\Msolpcsq}{\mbox{M}_{\sun}\,\mbox{pc}^{-2}}
\newcommand{\Msolh}{h^{-1}~\mbox{M}_{\odot}}
\newcommand{\Msol}{\mbox{M}_{\odot}}
\newcommand{\Myr}{\mbox{Myr}}

\newcommand{\kpch}{h^{-1}~\mbox{kpc}}
\newcommand{\pch}{h^{-1}~\mbox{pc}}
\newcommand{\cmmt}{\mbox{cm}^{-3}}

\newcommand{\kpc}{\mbox{kpc}}

\newcommand{\ergg}{\mbox{erg}~\mbox{g}^{-1}}

\newcommand{\ergcmmts}{\mbox{erg}~\mbox{cm}^{-3}~\mbox{s}^{-1}}

\newcommand{\K}{\mbox{K}}

\newcommand{\dex}{\mbox{dex}}

\newcommand{\cm}{{\rm cm}}

\newcommand{\kms}{{\rm km}\,{\rm s}^{-1}}
\newcommand{\pc}{{\rm pc}}

\newcommand{\erg}{{\rm erg}}
\newcommand{\yr}{{\rm yr}}

\begin{document}

\title[Galactic outflows with thermal SN feedback]{Simulating galactic
  outflows with thermal supernova feedback}

\author[C. Dalla Vecchia \& J. Schaye]
{Claudio Dalla Vecchia$^{1,2}$\thanks{E-mail: caius@mpe.mpg.de} and
Joop Schaye$^2$\thanks{E-mail: schaye@strw.leidenuniv.nl}\\
$^1$Max Planck Institute for Extraterrestrial Physics, Gissenbachstra\ss{}e 1, 85748 Garching, Germany\\
$^2$Leiden Observatory, Leiden University, P.O. Box 9513, 2300 RA Leiden,
the Netherlands}

\maketitle

\abstract
Cosmological simulations make use of sub-grid recipes for the implementation of galactic winds driven by massive stars because direct injection of supernova energy in thermal form leads to strong radiative losses, rendering the feedback inefficient. We argue that the main cause of the catastrophic cooling is a mismatch between the mass of the gas in which the energy is injected and the mass of the parent stellar population. Because too much mass is heated, the temperatures are too low and the cooling times too short. We use analytic arguments to estimate, as a function of the gas density and the numerical resolution, the minimum heating temperature that is required for the injected thermal energy to be efficiently converted into kinetic energy. We then propose and test a stochastic implementation of thermal feedback that uses this minimum temperature increase as an input parameter and that can be employed in both particle- and grid-based codes. We use smoothed particle hydrodynamics simulations to test the method on models of isolated disc galaxies in dark matter haloes with total mass $10^{10}$ and $10^{12}\Msolh$. The thermal feedback strongly suppresses the star formation rate and can drive massive, large-scale outflows without the need to turn off radiative cooling temporarily. In accord with expectations derived from analytic arguments, for sufficiently high resolution the results become insensitive to the imposed temperature jump and also agree with high-resolution simulations employing kinetic feedback. 
\endabstract



\section{Introduction}

It is widely accepted that star formation (SF) feeds back energy into the
interstellar medium (ISM). The energy released by massive stars, both through stellar winds and core-collapse supernova (SN) explosions, can efficiently suppress SF by evaporating dense, star-forming clouds, by
generating supersonic turbulence and, eventually, by generating
powerful, large-scale outflows that eject gas from galaxies and
enrich the intergalactic medium.

Modern cosmological simulations that follow the formation and
evolution of galaxies still lack both the resolution and the physics that is required to model the multi-phase ISM and individual massive stars or SN explosions. The same is true for simulations of individual galaxies, although the resolution of such models is now sufficient to begin to crudely disentangle the relative roles of the different mechanisms through which massive stars inject energy and momentum \citep[e.g.\ radiation pressure vs.\ SN explosions,][]{Hopkins2012}.
Thus, in
naive implementations of stellar feedback, ``star'' particles representing
simple stellar populations (SSPs) distribute ``SN energy'' over neighbouring
resolution elements at each time step. This procedure is well-known to
be inefficient, in that most of the thermal energy is radiated away
before it can be converted to kinetic energy \citep[e.g.][]{Katz1996}.

Three types of sub-grid recipes are commonly used to solve the
over-cooling problem: injecting the energy in kinetic form
\citep[e.g.][]
{Navarro1993,MihosHernquist1994,Kawata2001,Kay2002,Springel2003,Oppenheimer2006,DallaVecchia2008,Dubois2008,Hopkins2012},
suppressing radiative cooling by hand
\citep[e.g.][]{Gerritsen1997,Mori1997,Thacker2000,Kay2002,Sommer-Larsen2003,Brook2004,Stinson2006,Piontek2011},
and decoupling the different thermal phases by hand
\citep[e.g.][]{Marri&White2003,Scannapieco2006,Murante2010}. Each
solution has its own pros and cons and all require the specification
of sub-grid parameters. The different approaches should converge when
the resolution is increased, although it is not obvious that this will
in fact happen.

The inefficiency of thermal feedback is usually attributed to a lack
of resolution: the energy is deposited in gas that is too dense,
because the hot, low-density, bubbles that fill much of the volume of
the multiphase ISM are missing. However, as we pointed out in
\citet[][hereafter DS08]{DallaVecchia2008}, a more fundamental problem
is the fact that the SN energy is distributed over too much mass,
which implies that the temperature of the SN heated gas is too low and
hence the cooling time too short. Indeed, in reality one SNII is produced for every $\sim 10^2~\Msol$ of stars, and the energy released in the explosion is initially carried by $\ll 10^2~\Msol$ of ejecta. Hence, the ratio of the mass of the ejecta and the mass of the stellar population that released the energy is small $\ll 1$. In contrast, in simulations this ratio is $\gg 1$, unless the mass of star particles is very large compared with that of the surrounding gas resolution elements, which is typically not the case. Increasing the resolution does not
change the ratio between the mass of a star particle and the mass of
the neighbouring resolution elements (although it may in the case of
grid simulations) and hence is unlikely to solve the over-cooling
problem by itself.

The temperature jump of the gas receiving feedback energy can be
increased by storing the energy until it suffices to heat the gas by a
desired amount. Indeed, this strategy is for example used in some
sub-grid recipes for feedback from active galactic nuclei (AGN), which
store the AGN energy in the black hole until the neighbouring gas can
be heated to a desired temperature \citep{Booth2009}. This approach
is, however, not suitable for SN feedback. Storing the energy in a
star particle would not help because standard implementations of
thermal SN feedback are inefficient even if all the SN energy of an SSP is released at once. Storing energy in
gas particles is undesirable because it would make the feedback
non-local, which would for example mean that heavy elements released
by the star particle are less likely to be carried by outflows.

An alternative way to increase the temperature jump is to reduce the
ratio of the heated mass to the mass of a star particle. This can be
done by reducing the number of neighbouring resolution elements that
are heated, but that does not help if even a single resolution element
contains too much mass. Moreover, this approach would result in a
range of temperatures if not all resolution elements have the same
mass. To guarantee the efficiency of the feedback, one would like to
be able to specify the temperature jump of the gas that receives
energy. This can be accomplished by making the thermal feedback
stochastic: the probability that a neighbouring resolution element is
heated will then depend on the desired temperature jump and on the
ratio of the mass of the star particle to that of the neighbouring
gas resolution element.

A stochastic approach to thermal feedback was tried, and found to be
effective, by \citet{Kay2003} in smoothed particle hydrodynamics (SPH)
simulations of groups of galaxies. For each simulation time-step and
each star particle, they integrated the energy released by SNe and
distributed it stochastically to its nearest gas neighbour. They only
considered the limiting case in which the temperature jump was
sufficiently high that the number of particles that could be heated
per time step was less than one.

In this paper we generalise the method of \citet{Kay2003} to work also
for temperature jumps sufficiently low for multiple neighbours to be
heated. Using SPH simulations of isolated disc galaxies embedded in
dark haloes with total mass $10^{10}$ and $10^{12}\Msolh$, we show
that our implementation of thermal feedback is able to strongly
suppress SF, to alter the morphology of the galaxy and to
generate galactic winds. Reassuringly, for our high-resolution
simulations we reproduce the results of DS08, where we simulated the
same disc galaxies with kinetic feedback.

This paper is organised as follows. We compute the energy provided by
core collapse SNe (SNII) in Section~\ref{sec:ener}, where we show that for a standard
initial mass function (IMF), a single star particle produces enough SN
energy to heat a gaseous mass equal to the mass of the star particle
by a few keV, a temperature that is sufficiently high for radiative
cooling to be relatively inefficient. We present our numerical implementation for
thermal SN feedback in SPH in Section~\ref{sec:recipe}, where we also
compute some useful quantities as a function of the free parameters of
the method: the amount of SN energy injected per unit stellar mass and
the desired temperature jump. We also explain how the method could be
adapted for grid simulations. We dedicate Section~\ref{sec:coolexp} to
the derivation of the resolution constraints of our feedback recipe,
which follow from the requirement that the radiative cooling time
exceeds the sound crossing time across the heated resolution
element so that the thermal energy is efficiently converted into kinetic form. After describing our simulations of isolated discs galaxies
in Section~\ref{sec:sims}, we present our results in
Section~\ref{sec:res}. Finally, we summarise and discuss our
conclusions in Section~\ref{sec:disc}.

Videos and high-resolution images can be found at: \texttt{http://www.strw.leidenuniv.nl/DS12/}


\section{Energy provided by SNII}
\label{sec:ener}

Each star particle is treated as a simple (or single) stellar
population. Thus, its stellar content is simply described by an
IMF, $\Phi(M)$. The number of stars per unit stellar mass ending their
life as SNII, $n_{\rm SNII}$, is then the integral of the IMF over the
mass range $[M_0,M_1]$,
\begin{equation}
n_{\rm SNII}=\int_{M_0}^{M_1} \Phi(M)\,{\rm d}M,
\end{equation}
where $M_0$ and $M_1$ are the minimum and maximum initial mass of stars
that will explode as core collapse SNe.  Throughout the paper we will
use a Chabrier IMF and the mass interval $[M_0,M_1]=[6,100]~\Msol$,
although we also report calculations for a Salpeter IMF and for the
widely used mass range of $[8,100]~\Msol$ ($6$-$8~\Msol$ stars explode
as electron capture SNe in models with convective overshoot; e.g.
\citealt{Chiosi1992}). For the Chabrier (Salpeter) IMF we obtain, for
the mass range $[M_0,M_1]=[6,100]~\Msol$, $n_{\rm SNII}=1.736\times
10^{-2}~\Msol^{-1}$ ($1.107\times 10^{-2}~\Msol^{-1}$). For the mass
range $[M_0,M_1]=[8,100]~\Msol$, we have $n_{\rm SNII}= 1.180\times
10^{-2}~\Msol^{-1}$ ($0.742\times 10^{-2}~\Msol^{-1}$).

The total available energy per unit stellar mass provided by SNII,
$\epsilon_{\rm SNII}=n_{\rm SNII} E_{\rm SNII}$, is given by
\begin{equation}
\epsilon_{\rm SNII}=8.73\times10^{15}~\ergg
\left(\frac{n_{\rm SNII}}{1.736\times 10^{-2}~\Msol^{-1}}\right) E_{51},
\end{equation}
where $E_{\rm SNII}\equiv E_{51}\times 10^{51}~\erg$ is the available
energy from a single SNII event and we will assume $E_{51}=1$. The amount of energy from SNII
available in a SSP particle is then $m_{\ast}\epsilon_{\rm SNII}$,
where $m_{\ast}$ is the initial mass of the star particle.

If the energy is used to heat a gas mass $m_{\rm g, heat}$, then the
corresponding temperature increase is given by
\begin{eqnarray}
\Delta T &=& (\gamma-1)\frac{\mu m_{\rm H}}{k_{\rm B}}\epsilon_{\rm SNII}
\frac{m_{\ast}}{m_{\rm g, heat}} \nonumber\\
&=& 4.23\times10^7~\K \left(\frac{n_{\rm SNII}}{1.736\times 10^{-2}~\Msol^{-1}}\right)
\left(\frac{\mu}{0.6}\right)\times \nonumber\\
&& E_{51} \frac{m_{\ast}}{m_{\rm g, heat}},
\label{eq:deltaT}
\end{eqnarray}
where $\gamma=5/3$ is the ratio of specific heats for an ideal
monatomic gas, $k_{\rm B}$ is the Boltzmann constant, $m_{\rm H}$ is
the mass of the proton, and $\mu m_{\rm H}$ is the mean particle
mass. We have assumed the gas to be monatomic and neglected the energy
used to increase the degree of ionisation of the plasma.

The standard SPH approach is to distribute the SN energy over all
neighbours of a star particle.\footnote{The standard approach is to
weigh the contribution to each receiving gas particle by the SPH
kernel. We ignore this weighting here for simplicity and because
differences between SPH neighbours are by definition poorly resolved.}
The heated mass is then $m_{\rm g, heat}=\NGB m_{\rm g}$, where $\NGB$
is the number of neighbouring particles (typically 32--64, we use 48
in our simulations) and $m_{\rm g}$ is the mass of a single gas
particle. Assuming $m_{\ast}=m_{\rm g}$, we can see from
equation~(\ref{eq:deltaT}), that the average temperature increase for the
heated gas particles is $\sim 10^6~\K$, which falls in the temperature
regime for which the cooling time is relatively short
\citep[e.g.][]{Wiersma2009a}. Note that this procedure leads to even
lower temperature increases if $m_{\ast}<m_{\rm g}$, which happens if
multiple star particles are produced by each star-forming gas
particle. Heating only a single gas particle would give a temperature
increase of $\sim 10^{7.5}~\K$ and, as we will show, a much longer
cooling time.  As we will describe in the next section, we will make
the temperature increase $\Delta T$ a free parameter and
stochastically heat neighbouring gas particles.


\section{Thermal feedback implementation}
\label{sec:recipe}

For simplicity and for consistency with DS08, we assume that all the
SN energy produced within a star particle becomes available in a
single time step. Once a stellar particle reaches an age $t_{\rm SN} =
3\times 10^7~\yr$, corresponding to the maximum lifetime of stars that
end their lives as core collapse SNe, it stochastically injects
thermal energy into its surroundings. Another reason why we prefer to
impose this small time delay, is that it prevents the injection of
energy before the release of heavy elements by (the progenitors of)
SNe, a process that happens continuously in our stellar evolution
prescription \citep{Wiersma2009b}.

We note that it is
straightforward to modify our implementation of thermal feedback to
the case where SN energy is released stochastically over multiple time
steps. All one needs to do, is to replace $\epsilon_{\rm SNII}$ in the
equations below by the energy per unit mass that is released by the
star particle over the current time step. We tested this approach and did not find any significant difference.

Before providing a detailed description of our implementation of
thermal SN feedback, we first describe the two free parameters used in
our model.

\subsection{Free parameters}

Our recipe uses two free parameters.  The first parameter, $\FTH$, is the
fraction of the available SN energy that is actually used in
performing the feedback.\footnote{We do not treat the energy per SN,
$E_{51}$, as a free parameter because it is relatively well
constrained by observations and because changes in $E_{51}$ are in any
case degenerate with changes in $\FTH$.}  The value of $\FTH$ is
between zero and unity, and can be used to control the efficiency of
the feedback. Note that this parameter is not particular to our
method, it is common to all implementations of SNII feedback.

We will present simulations that use either $\FTH=0.4$ or
$\FTH=1$. The former value is used for comparison with the kinetic
feedback of DS08, where we used the same value.  It was also used by
\citet{Schaye2010}, who already showed some results of cosmological
simulations that used the prescription for thermal feedback presented
here. For most of our runs we will use $\FTH=1$ because it is an interesting limiting
case and because high values can be justified for thermal feedback
since we are in fact simulating radiative losses (we do not at any
point turn off radiative cooling).

As discussed in the introduction, we wish to control the temperature
increase of the heated gas particles in order to avoid the regime of
short cooling times. We accomplish this by making the temperature
increase $\Delta T$, or rather the increase of the thermal energy per
unit mass $\Delta\epsilon$, the second free parameter. Although
$\Delta\epsilon$ is the parameter we actually use, we will often refer
to the more intuitive corresponding value of $\Delta T$, computed
assuming the gas to be fully ionised ($\mu=0.6$). Our fiducial
temperature increase is $\Delta T = 10^{7.5}~\K$, but we will explore
a range of values.

Simulations employing the recipe described here should use values of $\Delta T$ sufficiently high to avoid catastrophic cooling. As shown analytically in section~\ref{sec:coolexp}, the required minimum value will depend on the resolution. The remaining freedom, particularly the value of $f_{\rm th}$, including possible dependencies on local physical conditions, can be used to calibrate the method by comparing the predictions for the quantities of interest to observations. Naturally, the best-fit parameter values may depend on other numerical and physical parameters.

\subsection{Distributing the energy}

In this section we derive the stochastic formulation for the energy
injection.  The energy released by a single star particle is shared
among a fraction of the $\NGB$ neighbouring resolution elements. We
will hereafter refer to resolution elements as ``particles'' but note
that the same method will also work for grid simulations (replace
``particle'' by ``cell''). We give each gas particle the same
probability $p$ of receiving energy from the star, irrespective of its
mass (and, for the case of SPH, irrespective of its kernel weight). We draw a
random number $0\le r\le 1$ for each star-gas particle pair, and
increase the internal energy of the gas particle by $\Delta\epsilon$
if $r\le p$. We will now derive the value of the probability $p$.

The expectation value for the total amount of energy from SNII
injected by a single star particle in the surrounding medium is
\begin{equation}
\label{eq:toten}
p\sum_{i=1}^{\NGB} E_i=p\Delta\epsilon \sum_{i=1}^{\NGB}m_i,
\end{equation}
where $E_i$ and $m_i$ are, respectively, the total energy given to and
the mass of gas particle $i$. We require the mean injected energy to
equal the energy contributed by the star particle, $\FTH m_\ast
\epsilon_{\rm SNII}$, from which the probability $p$ follows:
\begin{equation}
\label{eq:prob}
p=
\FTH\frac{\epsilon_{\rm SNII}}{\Delta\epsilon}\frac{m_{\ast}}{\sum_{i=1}^{\NGB}m_i}.
\end{equation}
Thus, the probability that a gas particle is heated is proportional to
$\FTH$, the fraction of total available SNII energy that the star
particle shares with the surrounding gas, and inversely proportional
to $\Delta\epsilon$, the amount of thermal energy per unit mass that is given to
each heated gas particle.

The stochastic treatment breaks down if the probability is larger than
one, because in that case the average amount of injected energy is
lower than the required value. Imposing $p\le 1$ puts a constraint on
the value of the parameter $\Delta\epsilon$:
\begin{equation}
\label{eq:econstr}
\Delta\epsilon \ge
\FTH\epsilon_{\rm SNII}\frac{m_{\ast}}{\sum_{i=1}^{\NGB}m_i}\simeq
\frac{\FTH\epsilon_{\rm SNII}}{\NGB},
\end{equation}
where the last equality holds exactly if all particles (gas and stars)
have the same mass, as is usually approximately true for SPH
simulations.

For a Chabrier IMF, assuming $\NGB=48$ and $\FTH=1$, we obtain
$\Delta\epsilon\ge 1.82\times 10^{14}~\ergg$, which corresponds to a
temperature increase $\Delta T \ge 8.8\times 10^5~\K$. For our
fiducial choice for the temperature increase of $10^{7.5}~\K$ we are
well above this lower limit. For the case of grid simulations it is in principle possible that
$\sum_{i=1}^{\NGB} m_i \ll m_\ast$ and hence that $p>1$. In that case
it is necessary to increase $\Delta\epsilon$ above the chosen value in
order to limit the probability to unity.

The sum of the probability over all neighbours gives the expectation
value for the number of heated neighbours:
\begin{equation}
\label{eqn:nheat}
\left<N_{\rm heat}\right>=\FTH\frac{\epsilon_{\rm SNII}}{\Delta\epsilon}\frac{m_{\ast}
\NGB}{\sum_{i=1}^{\NGB}m_i}\simeq\frac{\FTH\epsilon_{\rm SNII}}{\Delta\epsilon},
\end{equation}
where the last equality holds exactly if all particles have the same
mass. The average number of heated neighbours is inversely
proportional to $\Delta\epsilon$ and proportional to $\FTH$. Expressed
in terms of the temperature increase $\Delta T$, the mean number of
heated neighbours is
\begin{eqnarray}
\left<N_{\rm heat}\right> &=& 1.34 ~E_{51}
\left(\frac{n_{\rm SNII}}{1.736\times 10^{-2}~\Msol^{-1}}\right)
\left(\frac{\mu}{0.6}\right) \times \nonumber \\
&& \FTH \left(\frac{\Delta T}{10^{7.5}~\K}\right)^{-1}.
\label{eq:nheat}
\end{eqnarray}

Ideally, the energy should on average be shared with at least one gas
neighbour to make the feedback local to the star particle and to
ensure that the metals released by massive stars can be driven
outwards. By injecting all the SNII energy from a SSP at once, we can satisfy this
constraint for temperature increases that are sufficiently large to
make radiative cooling (initially) inefficient.


\section{Ensuring effective feedback: resolution requirements}
\label{sec:coolexp}

The thermal feedback can only be effective if the heated gas responds
hydrodynamically to the temperature increase before the thermal energy
is radiated away. This implies that the sound-crossing time scale
across a heated resolution element, $t_{\rm s}$, must be short
compared with the radiative cooling time scale in the heated gas,
$t_{\rm c}$. If this condition is satisfied, then the gas will start
to expand adiabatically, doing work on its surroundings and converting
thermal energy into kinetic energy. The ratio $t_{\rm s}/t_{\rm c}$
can be decreased either by increasing $t_{\rm c}$, which can usually
be accomplished by increasing the temperature, or by decreasing
$t_{\rm s}$, which means increasing the temperature and/or the spatial
resolution.

The sound crossing time across a resolution element of length $h$ is 
\begin{eqnarray}
\label{eq:tsound}
t_{\rm s} &=& \frac{h}{c_{\rm s}} = \left(\frac{\mu m_{\rm H}}{\gamma
    k_{\rm B}}\right )^{1/2} \frac{h}{T^{1/2}}\nonumber\\
&=& 1.15\times 10^5~\yr 
\left(\frac{\mu}{0.6}\right)^{1/2}
\left(\frac{T}{10^{7.5}~\K}\right)^{-1/2}
\left(\frac{h}{100~\pc}\right),
\end{eqnarray}
where $c_{\rm s}$ is the local sound speed. This time scale is related to the classical definition of the Courant time step, $\Delta t=C h/c_{\rm s}=C t_{\rm s}$, where $C$ ($<1$) is the Courant factor.

\begin{figure}
\includegraphics[width=0.49\textwidth]{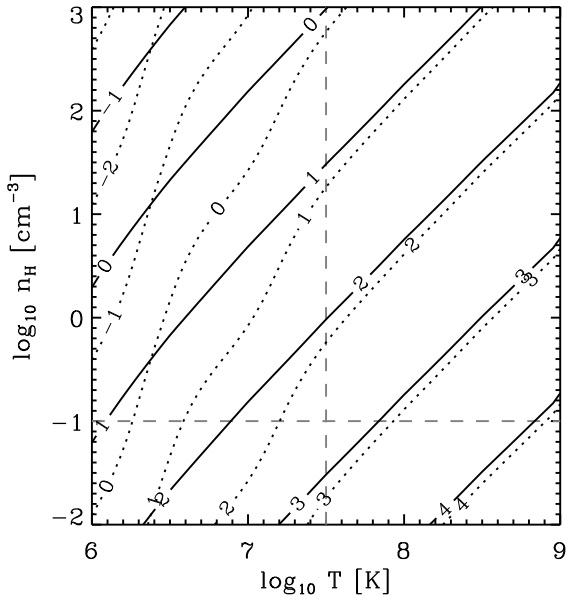}
\caption{Contour plot of the logarithm of the ratio between the
cooling time and the sound-crossing time over the SPH kernel as a
function of the density and the temperature of the gas. The cooling
time scale is computed using tabulated cooling rates for primordial
(solid contours) and solar (dotted contours) element abundances and
assuming collisional ionisation equilibrium. We show the ratio for the
particle mass used in our simulations of the $10^{12}~\Msolh$ halo. The
vertical dashed line marks the fiducial heating temperature of
$10^{7.5}~\K$ and the horizontal dashed line the star formation
density threshold in our simulations.}
\label{fig:tscale}
\end{figure}

We define the radiative cooling time as
\begin{equation}
\label{eq:tcool}
t_{\rm c} = \frac{u}{\Lambda} = \frac{\rho\epsilon}{\Lambda},
\end{equation}
where $u$ and $\Lambda$ are the internal energy and radiative cooling
rate per unit volume, respectively, and $\rho$ is the gas density.
The internal energy per unit volume can be written as
\begin{equation}
\label{eq:inten}
u = \frac{1}{\gamma-1}\frac{k_{\rm B}T}{\mu m_{\rm H}}\rho =
\frac{1}{\gamma-1}\frac{k_{\rm B}T}{\mu X_{\rm H}}n_{\rm H},
\end{equation}
where $n_{\rm H}$ and $X_{\rm H}$ are the hydrogen number density and
mass fraction, respectively. Before showing results for general
cooling functions, we will consider the case of pure Brehmsstrahlung,
which dominates the cooling rate for $T\ga 10^7~\K$ and for which the
cooling rate is given by \citep{Osterbrock1989}:
\begin{eqnarray}
\label{eq:coolrate}
\Lambda &\simeq& 7.99\times 10^{-24}~\ergcmmts
\left(\frac{n_{\rm H}}{1~\cmmt}\right)^2
\left(\frac{T}{10^{7.5}~\K}\right)^{1/2}
\times  \nonumber\\
&& g_{\rm f} \eta_{\rm e}(\eta_{\rm HII}+\eta_{\rm HeII}+\eta_{\rm HeIII}) \nonumber\\
&\simeq& 1.12\times 10^{-23}~\ergcmmts 
\left(\frac{n_{\rm H}}{1~\cmmt}\right)^2
\left(\frac{T}{10^{7.5}~\K}\right)^{1/2} \nonumber\\
&& \frac{(1+X_{\rm H})(1+3X_{\rm H})}{8X_{\rm H}^2},
\end{eqnarray}
where $\eta_{\rm i}=n_{\rm i}/n_{\rm H}$ is the number density of
species $i$ relative to hydrogen and we assumed the plasma to be fully
ionised in the last step. We assumed for simplicity that the Gaunt
factor $g_{\rm f}\simeq1.4$ (consistent with the approximated equation
given by \citet{Theuns1998} for our fiducial temperature increase) and
a primordial composition. Hence, we obtain a radiative cooling time of
\begin{eqnarray}
t_{\rm c} &\simeq& 3.26\times 10^7~{\rm yr} 
\left(\frac{n_{\rm H}}{1~\cmmt}\right)^{-1}
\left(\frac{T}{10^{7.5}~\K}\right)^{1/2}
\times  \nonumber\\
&& \left(\frac{\mu}{0.6}\right)^{1/2}
\left(\frac{f(X_{\rm H})}{0.13}\right),
\end{eqnarray}
where
\begin{equation}
f(X_{\rm H})=X_{\rm H}(1+X_{\rm H})^{-1}(1+3X_{\rm H})^{-1}
\end{equation}
and $f(X_{\rm H}=0.752)\simeq 0.13$. 

Thus, we can write for the ratio between the two time scales
\begin{eqnarray}
\frac{t_{\rm c}}{t_{\rm s}}
&=& 2.8\times 10^{2} 
\left(\frac{n_{\rm H}}{1~\cmmt}\right)^{-1}
\left(\frac{T}{10^{7.5}~\K}\right)
\left(\frac{h}{100~\pc}\right)^{-1}
\times \nonumber \\
&&
\left(\frac{\mu}{0.6}\right)^{-3/2}
\left(\frac{f(X_{\rm H})}{0.13}\right) ,
\label{eq:tratio_h}
\end{eqnarray}

For the case of SPH simulations, the spatial resolution is given by
the gas particle's smoothing kernel which can be approximated as
\begin{equation}
\label{eq:h}
h\simeq\left(\frac{3}{4\pi}\frac{\sum_{i=1}^{\NGB}m_i}{\rho}\right)^{1/3}
\simeq\left(\frac{3}{4\pi}\frac{\NGB\left<m\right>}{m_{\rm H} n_{\rm H}}X_{\rm
  H}\right)^{1/3},
\end{equation}
where $\left<m\right>$ is the average mass of the gas particles.
Substituting the above approximation into equation~(\ref{eq:tratio_h})
we obtain the ratio of time scales
\begin{eqnarray}
\label{eq:tratio}
\frac{t_{\rm c}}{t_{\rm s}}
&\simeq& 98 
\left(\frac{n_{\rm H}}{1~\cmmt}\right)^{-2/3}
\left(\frac{T}{10^{7.5}~\K}\right)
\left(\frac{\left<m\right>}{7\times10^4~\Msol}\right)^{-1/3}
\times \nonumber\\
&& 
\left(\frac{\NGB}{48}\right)^{-1/3}
\left(\frac{\mu}{0.6}\right)^{-3/2}
\left(\frac{g(X_{\rm H})}{0.14}\right),
\end{eqnarray}
where $g(X_{\rm H})=X_{\rm H}^{-1/3}f(X_{\rm H})$ and we used the particle mass appropriate for our simulations of the $10^{12}~\Msolh$ halo. 

We expect cooling losses to be important for $t_{\rm c} \lesssim t_{\rm s}$. The exact value of the ratio $f_{\rm t} \equiv t_{\rm c}/t_{\rm s}$ required to ensure efficient feedback can only be determined using simulations, but we expect it to be similar to 10 and will therefore use this as our fiducial value. This agrees well with the recent, independent work of \citet{Creasey2011}, who derived a resolution criterion for shock capturing in SPH and adaptive mesh refinement (AMR) simulations. Their criterion is based on comparing the rates of
shock heating and radiative cooling in a shock front, and ensures that shock
heating overwhelms cooling in order to avoid numerical over-cooling.
Following the suggestion of DS08, they applied their results to the
injection of thermal energy, and obtained a criterion that basically
translates into $t_{\rm c}/t_{\rm s}>8$.\footnote{Note that
\protect\citet{Creasey2011} derived their criterion by estimating the
velocity of a blast wave at the blast radius of half the mean
inter-particle distance.} 

From the relation between the sound crossing time and the Courant time step, one can estimate the number of simulation time steps it would take the gas to radiate its thermal energy if it cooled isochorically: $n_{\rm step}\sim t_{\rm c}/\Delta t=f_{\rm t}t_{\rm s}/\Delta t=f_{\rm t}/C$ ($\sim30$ for the fiducial value of $f_{\rm t}$ and $C=0.3$). In most implementations of SPH the sound speed in the Courant criterion is replaced by a ``signal velocity'' \citep[e.g.][]{Monaghan1997}, $v_{\rm sig}\ga 2c_{\rm s}$, which would more than double the number of time steps.

Inverting equation~(\ref{eq:tratio}), we find the following maximum density for which the feedback is expected to be effective,
\begin{eqnarray}
n_{{\rm H},t_{\rm c}= f_{\rm t}t_{\rm s}} &=& 31~\cmmt  
\left(\frac{T}{10^{7.5}~\K}\right)^{3/2}
\left(\frac{f_t}{10}\right)^{-3/2}
\times \nonumber\\
&&
\left(\frac{\left<m\right>}{7\times10^4~\Msol}\right)^{-1/2}
\left(\frac{\NGB}{48}\right)^{-1/2}
\times \nonumber\\
&& 
\left(\frac{\mu}{0.6}\right)^{-9/4}
\left(\frac{g(X_{\rm H})}{0.14}\right)^{3/2}.
\label{eq:nhcrit}
\end{eqnarray}
The critical density is thus proportional to $T^{3/2} \left<m\right>^{-1/2}$. 

For the case of AMR simulations, the spatial resolution is given by
the linear size of the grid cell, which is commonly decreased until it is some factor, $f_{\rm J}$, smaller than the local
Jeans length: $h\leq L_{\rm J}/f_{\rm J}$.
Expressing the Jeans length as a function of temperature and density, $L_{\rm J} \equiv c_{\rm s} \sqrt{\pi/(G\rho)}$, and substituting this into equation~(\ref{eq:tratio_h}), we obtain
\begin{eqnarray}
\label{eq:tratioamr}
\frac{t_{\rm c}}{t_{\rm s}}
&\geq& 50 
\left(\frac{n_{\rm H}}{1~\cmmt}\right)^{-1/2}
\left(\frac{T}{10^{7.5}~\K}\right)
\left(\frac{T_0}{10^{4}~\K}\right)^{-1/2}
\left(\frac{f_{\rm J}}{4}\right)
\times \nonumber\\
&& 
\left(\frac{\mu}{0.6}\right)^{-1}
\left(\frac{g\prime(X_{\rm H})}{0.15}\right),
\end{eqnarray}
where $T_0$ is the initial gas temperature (from which the Jeans mass,
hence the spatial resolution, is derived), $g\prime(X_{\rm
  H})=X_{\rm H}^{-1/2}f(X_{\rm H})$, and we assumed that the Jeans
length is resolved with four resolution elements. The equality holds for $h=L_{\rm J}/f_{\rm J}$. Note that the same relation applies to SPH simulations with particle mass at least $f_{\rm J}^3$ times smaller than the Jeans mass in gas with density $n_{\rm H}$ and temperature $T_0$.

Inverting equation~(\ref{eq:tratioamr}), we obtain
\begin{eqnarray}
\label{eq:nhcritamr}
n_{{\rm H},t_{\rm c}= f_{\rm t}t_{\rm s}} &\geq& 25~\cmmt 
\left(\frac{T}{10^{7.5}~\K}\right)^{2}
\left(\frac{T_0}{10^{4}~\K}\right)^{-1}
\left(\frac{f_{\rm J}}{4}\right)^2
\times \nonumber\\
&& 
\left(\frac{\mu}{0.6}\right)^{-2}
\left(\frac{g\prime(X_{\rm H})}{0.15}\right)^{2},
\end{eqnarray}
which is the analogue of equation~(\ref{eq:nhcrit}) for AMR (and for SPH simulations with particle mass at least $f_{\rm J}^3$ times smaller than the Jeans mass in gas with density $n_{\rm H}$ and temperature $T_0$).

We show in Fig.~\ref{fig:tscale} a contour plot of the time-scale
ratio $t_{\rm c}/t_{\rm s}$ in the temperature-density plane for the
fiducial gas particle mass $m_{\rm g}=5.1\times 10^4\Msolh$. We
computed $t_{\rm c}$ using the tabulated values of the collisional
cooling rates of \citep{Wiersma2009a}, which are also used in the SPH
simulations described below. We combine in the same plot the two cases
of primordial (solid contours) and solar (dotted contours) chemical
compositions. The vertical dashed line marks the fiducial heating
temperature of $10^{7.5}~\K$, while the horizontal dashed line
indicates the density threshold for SF in our simulations.
For $T\ge 10^{7.5}~\K$ the difference between primordial and solar
metallicity is very small, but at lower temperatures metals reduce the
ratio $t_{\rm c}/t_{\rm s}$. Once the temperature has dropped
to $10^7$~K for the case of solar abundances (or to values smaller
than $10^6$~K for primordial abundances), collisional excitation processes cause the cooling rate to increase as the temperature drops \citep[e.g.][]{Wiersma2009a}. For such temperatures the equations above, which assumed the cooling rate to be dominated by Brehmsstrahlung, will underestimate the radiative losses and will therefore overestimate the minimum density for which the feedback is expected to be efficient.

Interestingly, for purely adiabatic expansion, i.e.\ $\rho\propto T^{3/2}$, the ratio $t_{\rm c}/t_{\rm s}$ remains constant for the case of SPH (eq.~[\ref{eq:tratio}]). In that case the gas will follow a track parallel to the high temperature part of the contours in Fig.~\ref{fig:tscale} and $t_{\rm c}\propto \rho^{-2/3}$. Hence, if radiative losses were unimportant initially, so that the hot bubble will start to expand adiabatically, then radiative losses will remain unimportant as long as the cooling rate is dominated by Brehmsstrahlung. For AMR, on the other hand, the ratio $t_{\rm c}/t_{\rm s} \propto T^{1/4}$ during the adiabatic phase, which implies that radiative losses may already become important while Brehmsstrahlung dominates.


\section{Simulations}
\label{sec:sims}

\begin{table*}
\begin{center}
\caption{Simulation parameters:
  total mass, $M_{\rm halo}$;
  fraction of SN energy injected, $f_{\rm th}$;
  temperature jump, $\log_{10} \Delta T$;
  total number of particles, $N_{\rm tot}$;
  total number of gas particles in the disc, $N_{\rm disc}$;
  mass of baryonic particles, $m_{\rm b}$;
  mass of dark matter particles, $m_{\rm DM}$;
  gravitational softening of baryonic particles, $\epsilon_{\rm b}$;
  gravitational softening of dark matter particles, $\epsilon_{\rm DM}$;
  thermal feedback included, (Feedback).
  Values different from the fiducial ones are shown in
  bold. \label{tbl:params}}
\scriptsize
\begin{tabular}{ccccccccccc}
\hline
  Simulation & $M_{\rm halo}$ & $f_{\rm th}$ & $\log_{10} \Delta T$ & $N_{\rm tot}$ & 
$N_{\rm disc}$ & $m_{\rm b}$ & $m_{\rm DM}$ & $\epsilon_b$ & $\epsilon_{\rm DM}$ & 
Feedback\\
             & $(\Msolh)$    &           & $(\K)$            &              &              
& $(\Msolh)$  & $(\Msolh)$  & $(\pch)$    & $(\pch)$            & \\
\hline
\hline
  \textit{G10-NOFB}  & $10^{10}$ & --- & --- & 5$\,$000$\,$494 & 235$\,$294 & 
$5.1\times 10^2$ & $2.4\times 10^3$ & 10 & 17 & \textbf{N} \\
\hline
  \textit{G10-040-70}  & $10^{10}$ & $\mathbf{0.4}$ & $\mathbf{7.0}$ & 5$\,$000$\,$494 & 235$\,
$294 & $5.1\times 10^2$ & $2.4\times 10^3$ & 10 & 17 & Y \\
\hline
  \textit{G10-100-65}  & $10^{10}$ & $1.0$ & $\mathbf{6.5}$ & 5$\,$000$\,$494 & 235$\,
$294 & $5.1\times 10^2$ & $2.4\times 10^3$ & 10 & 17 & Y \\
  \textit{G10-100-70}  & $10^{10}$ & $1.0$ & $\mathbf{7.0}$ & 5$\,$000$\,$494 & 235$\,
$294 & $5.1\times 10^2$ & $2.4\times 10^3$ & 10 & 17 & Y \\
  \textit{G10-100-75}  & $10^{10}$ & $1.0$ & $7.5$ & 5$\,$000$\,$494 & 235$\,$294 & 
$5.1\times 10^2$ & $2.4\times 10^3$ & 10 & 17 & Y \\
  \textit{G10-100-80}  & $10^{10}$ & $1.0$ & $\mathbf{8.0}$ & 5$\,$000$\,$494 & 235$\,
$294 & $5.1\times 10^2$ & $2.4\times 10^3$ & 10 & 17 & Y \\
  \textit{G10-100-85}  & $10^{10}$ & $1.0$ & $\mathbf{8.5}$ & 5$\,$000$\,$494 & 235$\,
$294 & $5.1\times 10^2$ & $2.4\times 10^3$ & 10 & 17 & Y \\
\hline
  \textit{G10-100-75-LR08}  & $10^{10}$ & $1.0$ & $7.5$ & \textbf{625$\,$061} & 
\textbf{29$\,$411} & $\mathbf{4.1\times 10^3}$ & $\mathbf{1.9\times 10^4}$ & 
\textbf{20} & \textbf{34} & Y \\
  \textit{G10-100-75-LR64}  & $10^{10}$ & $1.0$ & $7.5$ &  \textbf{78$\,$132} &  
\textbf{3$\,$676} & $\mathbf{3.3\times 10^4}$ & $\mathbf{1.5\times 10^5}$ & \textbf{40} 
& \textbf{68} & Y \\
\hline
\hline
  \textit{G12-NOFB}  & $10^{12}$ & --- & --- & 5$\,$000$\,$494 & 235$\,$294 & 
$5.1\times 10^4$ & $2.4\times 10^5$ & 46 & 79 & \textbf{N} \\
\hline
  \textit{G12-040-70}  & $10^{12}$ & $\mathbf{0.4}$ & $\mathbf{7.0}$ & 5$\,$000$\,$494 & 235$\,
$294 & $5.1\times 10^4$ & $2.4\times 10^5$ & 46 & 79 & Y \\
\hline
  \textit{G12-100-65}  & $10^{12}$ & $1.0$ & $\mathbf{6.5}$ & 5$\,$000$\,$494 & 235$\,
$294 & $5.1\times 10^4$ & $2.4\times 10^5$ & 46 & 79 & Y \\
  \textit{G12-100-70}  & $10^{12}$ & $1.0$ & $\mathbf{7.0}$ & 5$\,$000$\,$494 & 235$\,
$294 & $5.1\times 10^4$ & $2.4\times 10^5$ & 46 & 79 & Y \\
  \textit{G12-100-75}  & $10^{12}$ & $1.0$ & $7.5$ & 5$\,$000$\,$494 & 235$\,$294 & 
$5.1\times 10^4$ & $2.4\times 10^5$ & 46 & 79 & Y \\
  \textit{G12-100-80}  & $10^{12}$ & $1.0$ & $\mathbf{8.0}$ & 5$\,$000$\,$494 & 235$\,
$294 & $5.1\times 10^4$ & $2.4\times 10^5$ & 46 & 79 & Y \\
  \textit{G12-100-85}  & $10^{12}$ & $1.0$ & $\mathbf{8.5}$ & 5$\,$000$\,$494 & 235$\,
$294 & $5.1\times 10^4$ & $2.4\times 10^5$ & 46 & 79 & Y \\
\hline
  \textit{G12-100-75-LR08}  & $10^{12}$ & $1.0$ & $7.5$ & \textbf{625$\,$061} & 
\textbf{29$\,$411} & $\mathbf{4.1\times 10^5}$ & $\mathbf{1.9\times 10^6}$ &  
\textbf{92} & \textbf{158} & Y \\
  \textit{G12-100-75-LR64}  & $10^{12}$ & $1.0$ & $7.5$ &  \textbf{78$\,$132} &  
\textbf{3$\,$676} & $\mathbf{3.3\times 10^6}$ & $\mathbf{1.5\times 10^7}$ & 
\textbf{184} & \textbf{316} & Y \\
\hline
\end{tabular}
\end{center}
\end{table*}

We ran simulations of isolated disc galaxies embedded in dark matter
haloes with total masses of $10^{10}$ and $10^{12}\Msolh$, where
$h=0.73$. The initial conditions are as in \citet{Schaye2008} and
DS08, thus the models do not include gaseous haloes
and all the gas is initially in the discs. We adopted a larger
gravitational softening length than in the previous works for the
massive galaxy. We also modified the
original code as described below.

\subsection{Code and initial conditions}
\label{sec:code}

We use a modified version of the TreePM/SPH code \textsc{gadget}
\citep{Springel2005} for all the simulations presented in
this paper.

We employ the SF recipe of \citet{Schaye2008}, which we
briefly describe here. Gas denser than the critical density for the
onset of the thermo-gravitational instability ($n_{\rm H} \sim 10^{-2}
- 10^{-1}~\cm^{-3}$) is expected to be multiphase and star-forming
\citep{Schaye2004}. We model such gas by imposing a minimum
temperature floor given by an effective equation of state with
pressure $P\propto \rho_g^{\gamma_{\rm eff}}$ for densities exceeding
$n_{\rm H} = 0.1~\cm^{-3}$, normalised to $P/k = 10^3~\cm^{-3}~\K$ at
the threshold. We use $\gamma_{\rm eff} = 4/3$ for which both the
Jeans mass and the ratio of the Jeans length and the SPH kernel are
independent of the density, thus preventing spurious fragmentation due
to a lack of numerical resolution.

We introduce a different definition of star-forming particle. In the
previous works a gas particle was flagged as star-forming if it
crossed the density threshold $n_{\rm H}^{\ast} = 0.1~\cm^{-3}$ while
its temperature was below $T=10^5~\K$. The particle then remained
star-forming until its density fell below the threshold density or the
particle was promoted to a wind particle.

In the present work we proceed as follows. Each particle is free to
cool to lower temperatures, but not below the temperature $T_{\rm
  EoS}$ imposed by the effective equation of state. A gas particle is
star-forming if
\begin{equation}
\left\{\begin{array}{l}
n_{\rm H}\ge n_{\rm H}^{\ast} \\
\log_{10} T < \log_{10} T_{\rm EoS} + \Delta\log_{10} T_{\rm EoS}
\end{array}\right.\,,
\end{equation} where $\Delta\log_{10} T_{\rm EoS}$ is a free
parameter which we set to $\Delta\log_{10} T_{\rm EoS}=0.5~\dex$.

\begin{figure*}
\setlength{\tabcolsep}{0pt}
\begin{tabular*}{0.9\textwidth}{p{0.3\textwidth}p{0.3\textwidth}p{0.3\textwidth}}
\centering$\Delta T=10^{6.5}~\K$ & \centering$\Delta T=10^{7.5}~\K$ & \centering$\Delta T=10^{8.5}~\K$
\end{tabular*}
\includegraphics[width=0.3\textwidth]{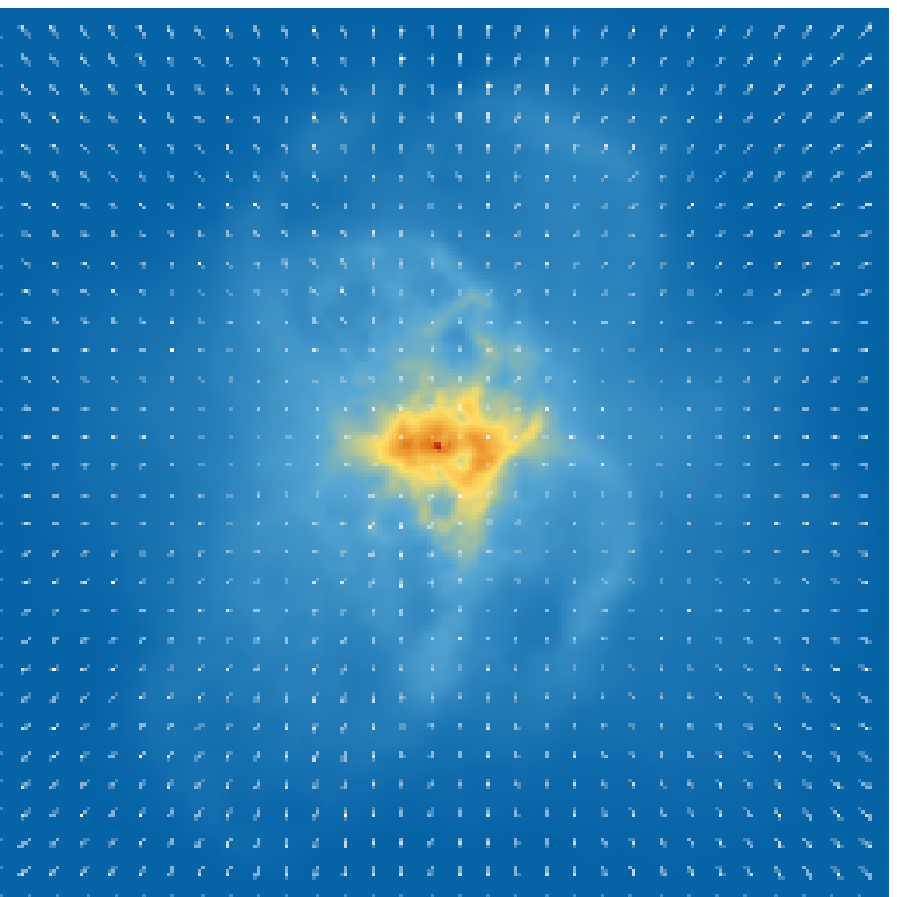}%
\includegraphics[width=0.3\textwidth]{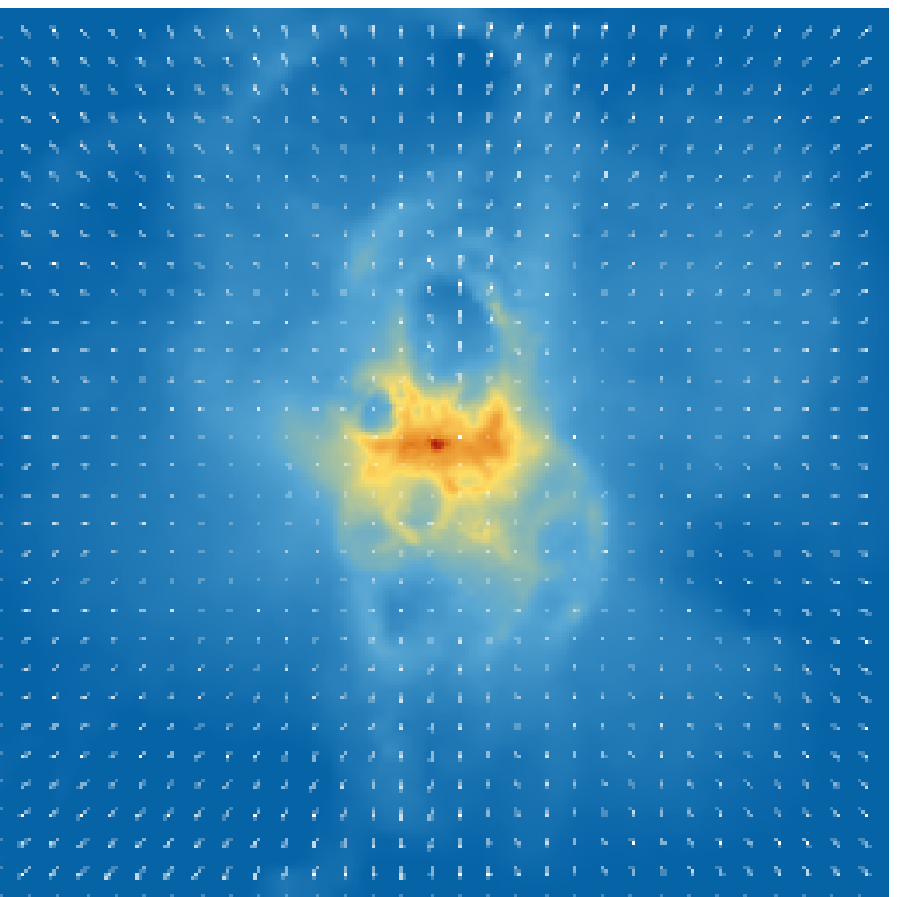}%
\includegraphics[width=0.3\textwidth]{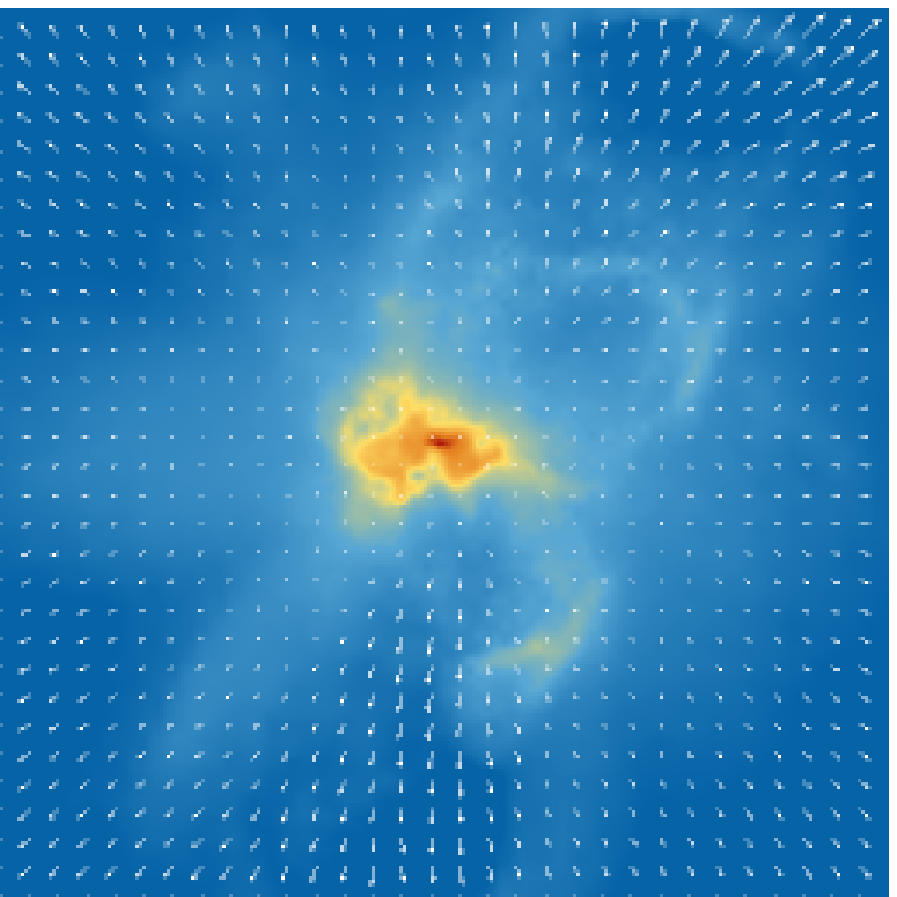}\\
\includegraphics[width=0.3\textwidth]{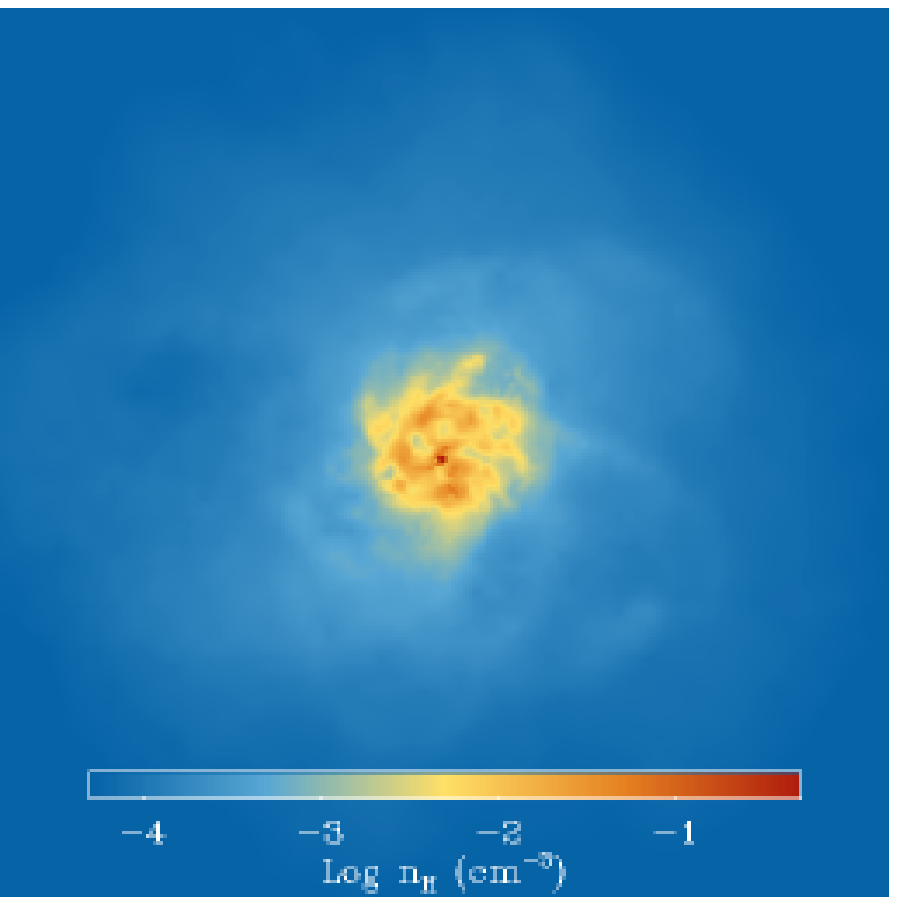}%
\includegraphics[width=0.3\textwidth]{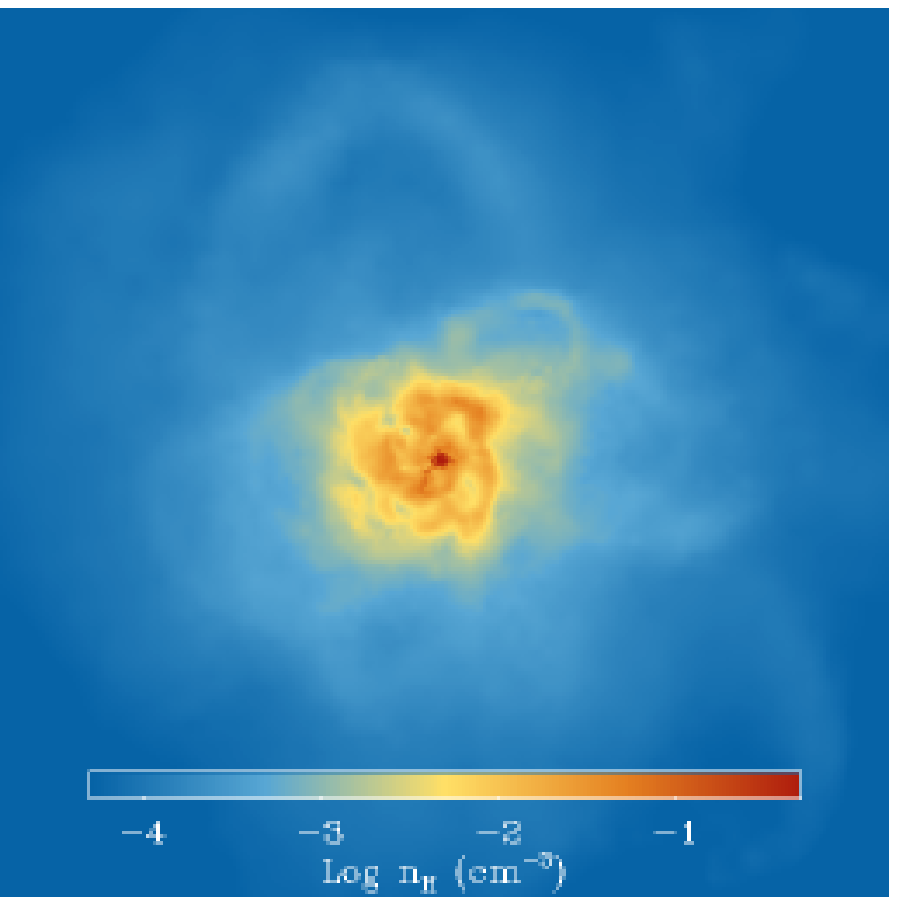}%
\includegraphics[width=0.3\textwidth]{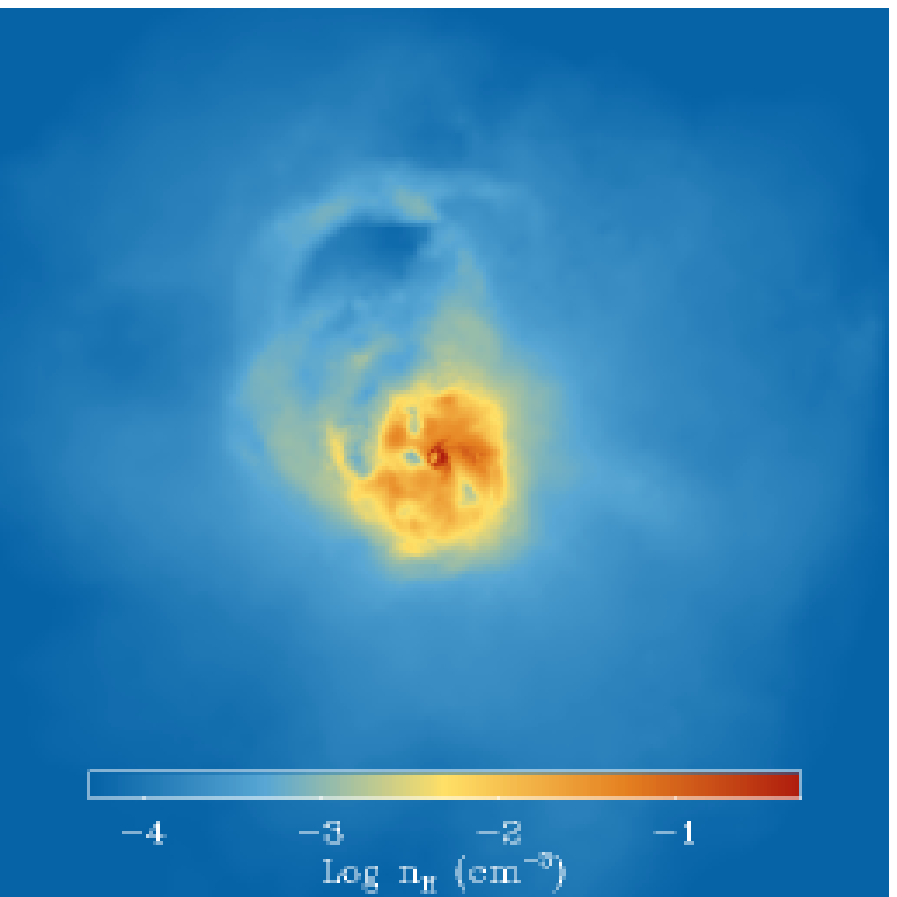}\\
\includegraphics[width=0.3\textwidth]{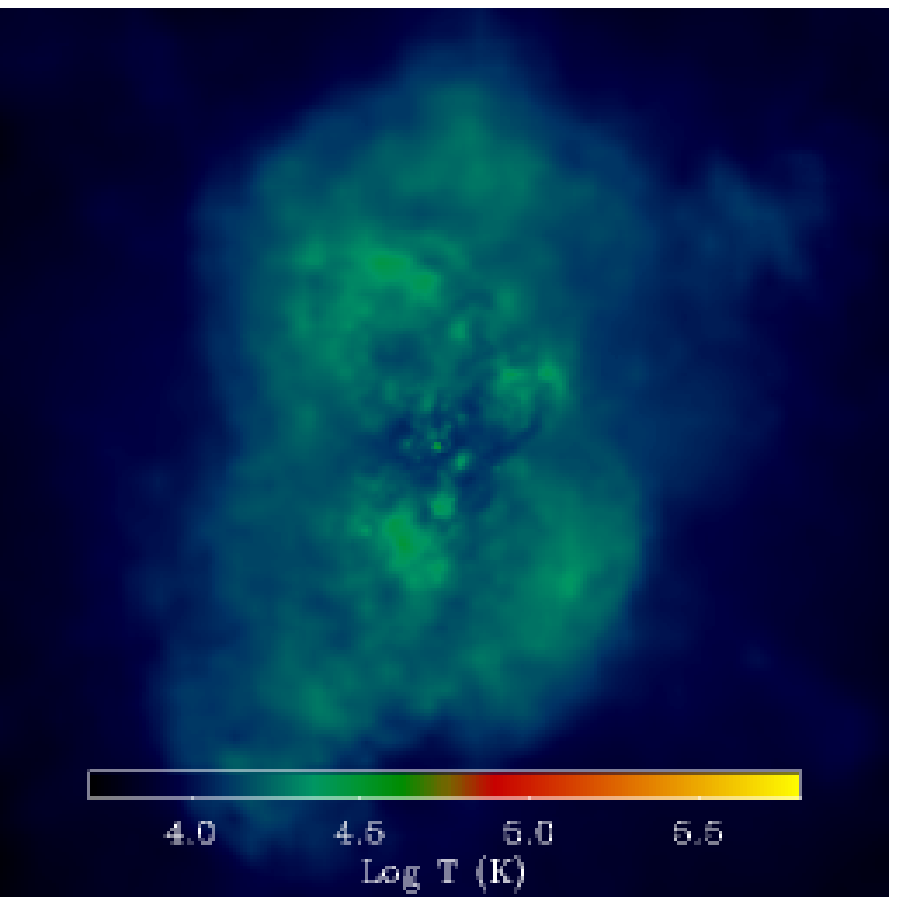}%
\includegraphics[width=0.3\textwidth]{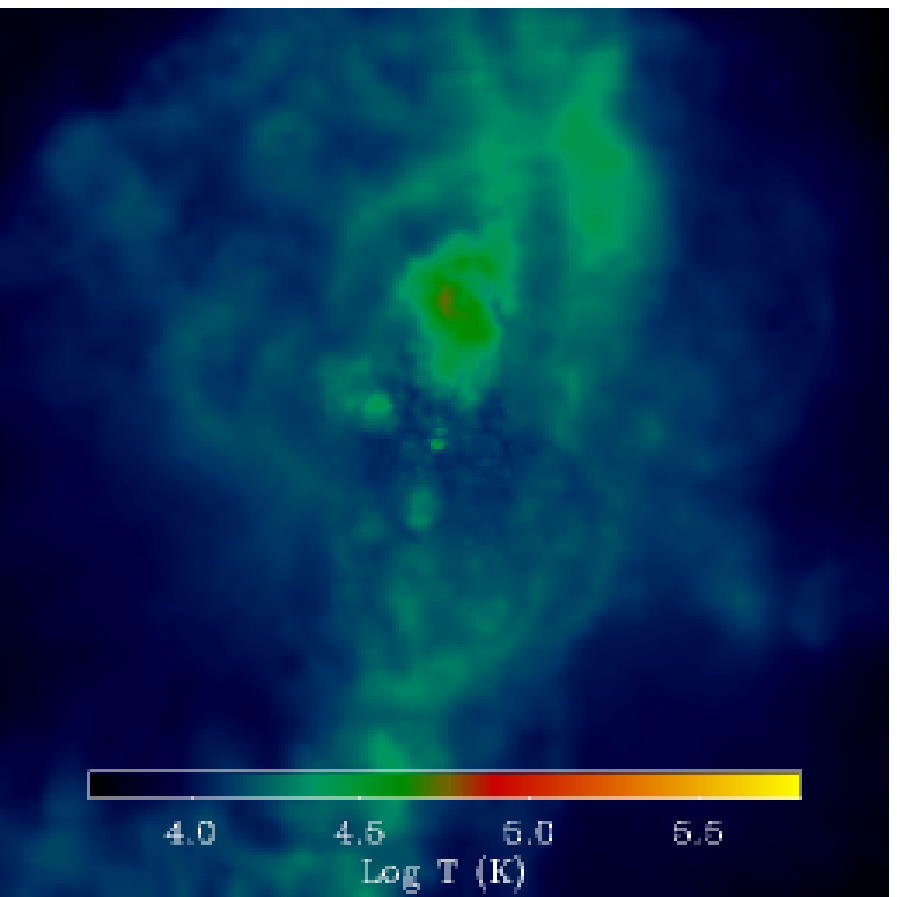}%
\includegraphics[width=0.3\textwidth]{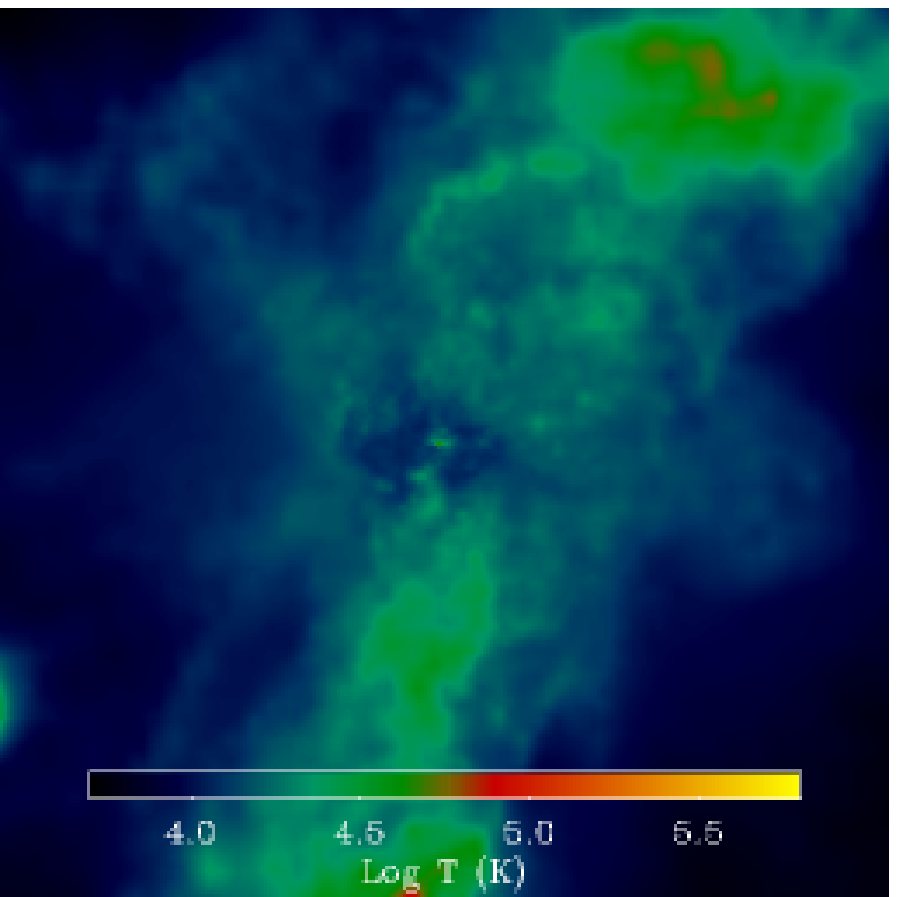}
\caption{Projections of the gas density (top row: edge-on; middle row:
face-on) and temperature (bottom-row: edge-on) for models
\textit{G10-100-65} (left column), the fiducial model
\textit{G10-100-75} (middle column) and \textit{G10-100-85}
(right column) at time $t=250~\Myr$. The white arrows in the top
row show the velocity field.  Images are $17.5\kpch$ on a side. The
colour coding is logarithmic in density ($-4.3<\log_{10}n_{\rm
H}/\cmmt<-0.3$) and temperature ($3.7<\log_{10}T/\K< 5.8$).  The
colour scale is indicated by the colour bars in each column.}
\label{fig:g10dens_comp}
\end{figure*}

The Kennicutt-Schmidt SF law is analytically converted and
implemented as a pressure law. As we demonstrated in
\citet{Schaye2008}, our method allows us to reproduce arbitrary input
SF laws for any equation of state without tuning any
parameters. We use the observed \citet{Kennicutt1998} law
\begin{equation}
\dot{\Sigma}_\ast = 1.5 \times 10^{-4}~\Msolyrkpcsq \left
    ({\Sigma_g \over 1~\Msolpcsq}\right )^{1.4},
\label{eq:KS}
\end{equation}
where the different normalisation accounts for the fact that we are
using a Chabrier IMF.

\begin{figure*}
\setlength{\tabcolsep}{0pt}
\begin{tabular*}{0.9\textwidth}{p{0.3\textwidth}p{0.3\textwidth}p{0.3\textwidth}}
\centering$\Delta T=10^{6.5}~\K$ & \centering$\Delta T=10^{7.5}~\K$ & \centering$\Delta T=10^{8.5}~\K$
\end{tabular*}
\includegraphics[width=0.3\textwidth]{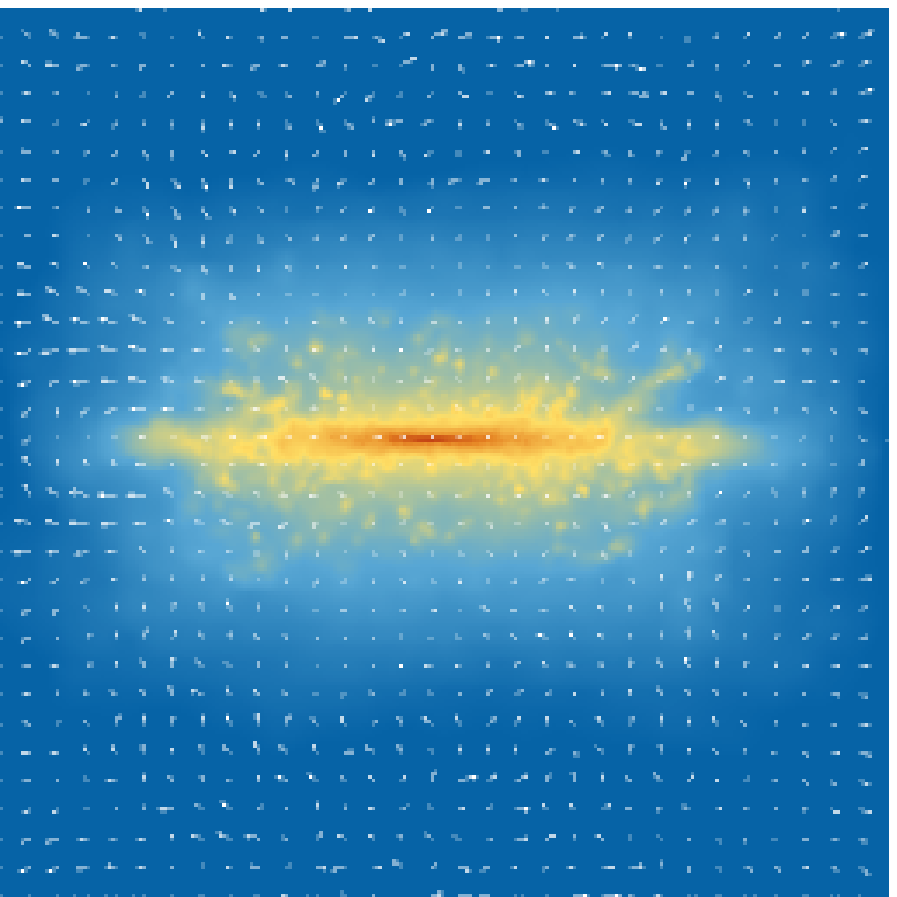}%
\includegraphics[width=0.3\textwidth]{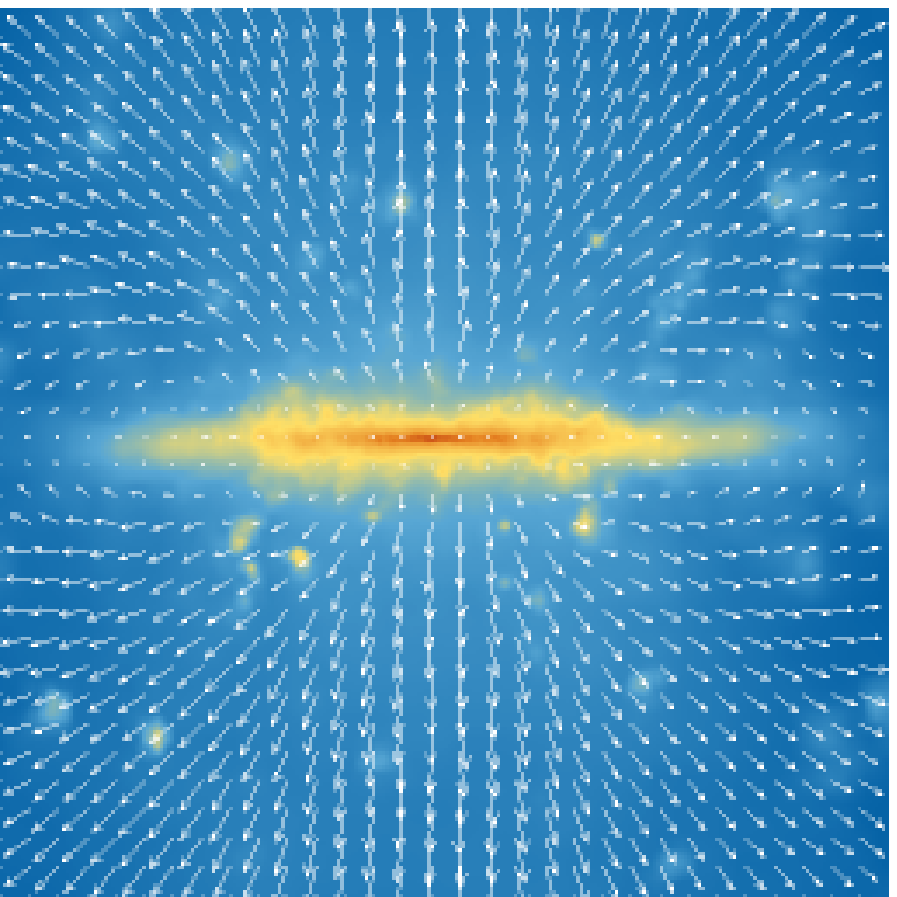}%
\includegraphics[width=0.3\textwidth]{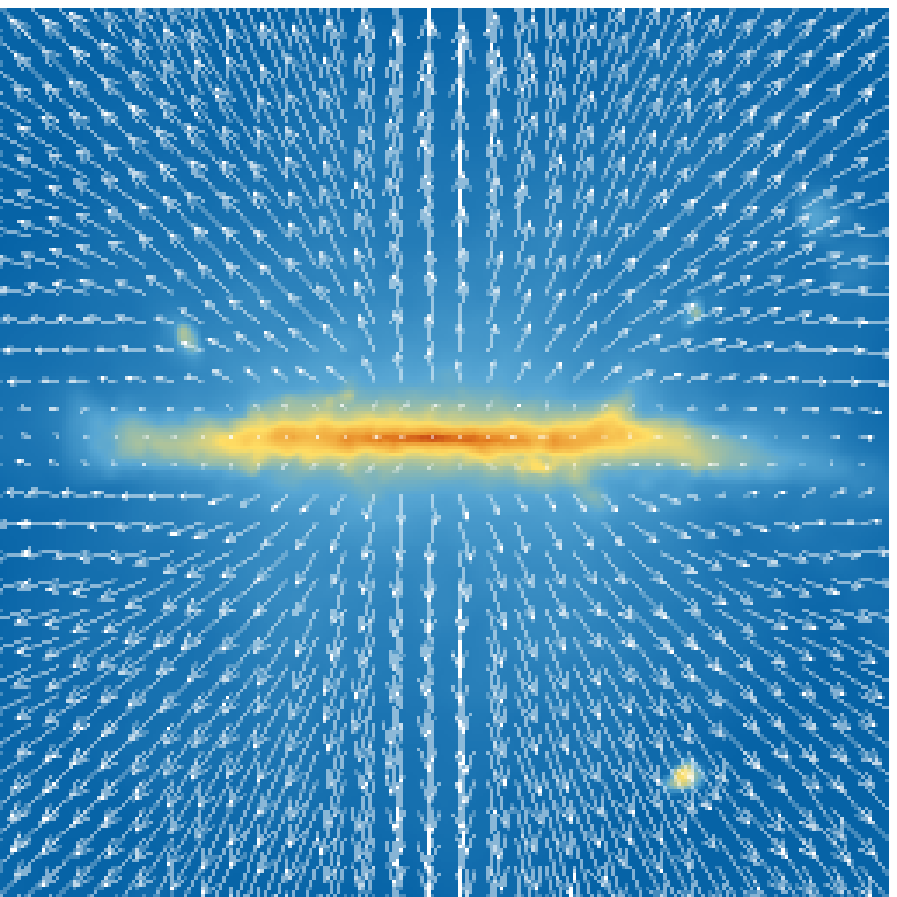}\\
\includegraphics[width=0.3\textwidth]{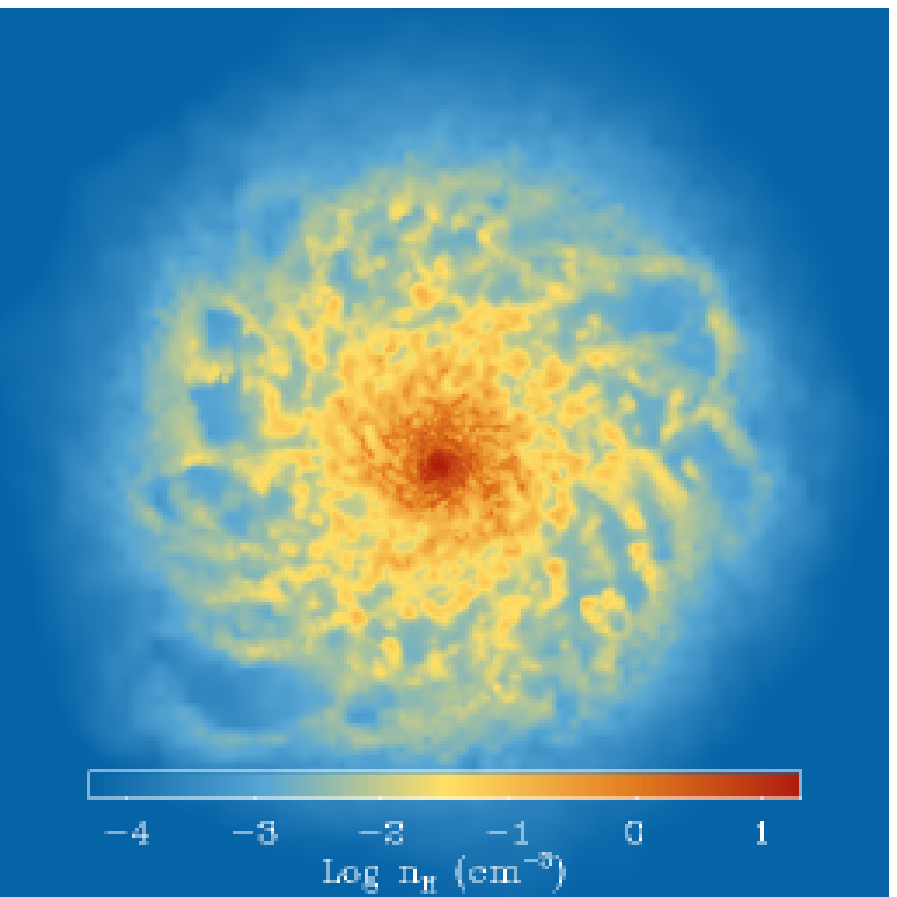}%
\includegraphics[width=0.3\textwidth]{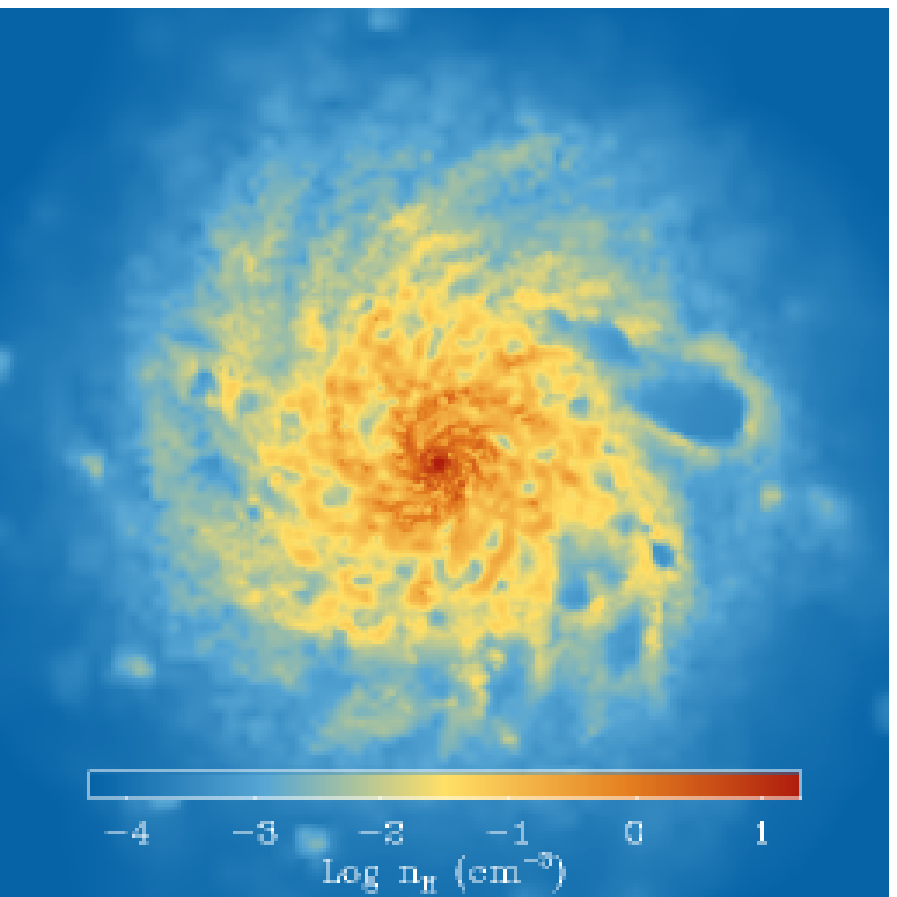}%
\includegraphics[width=0.3\textwidth]{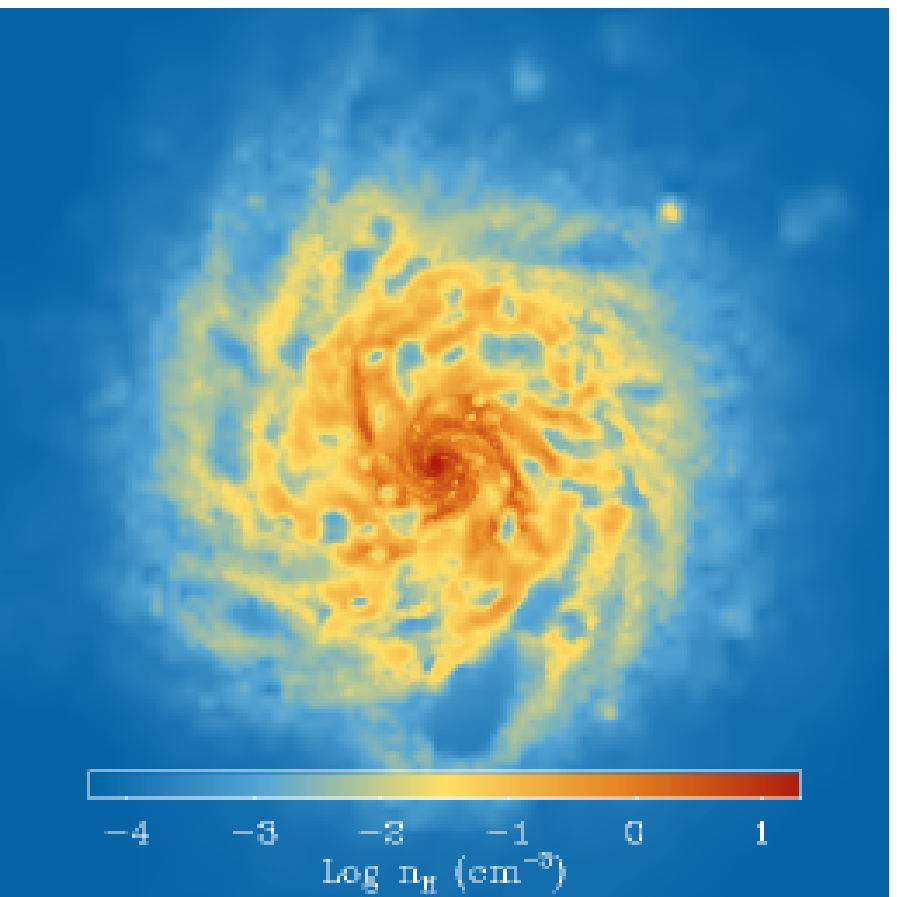}\\
\includegraphics[width=0.3\textwidth]{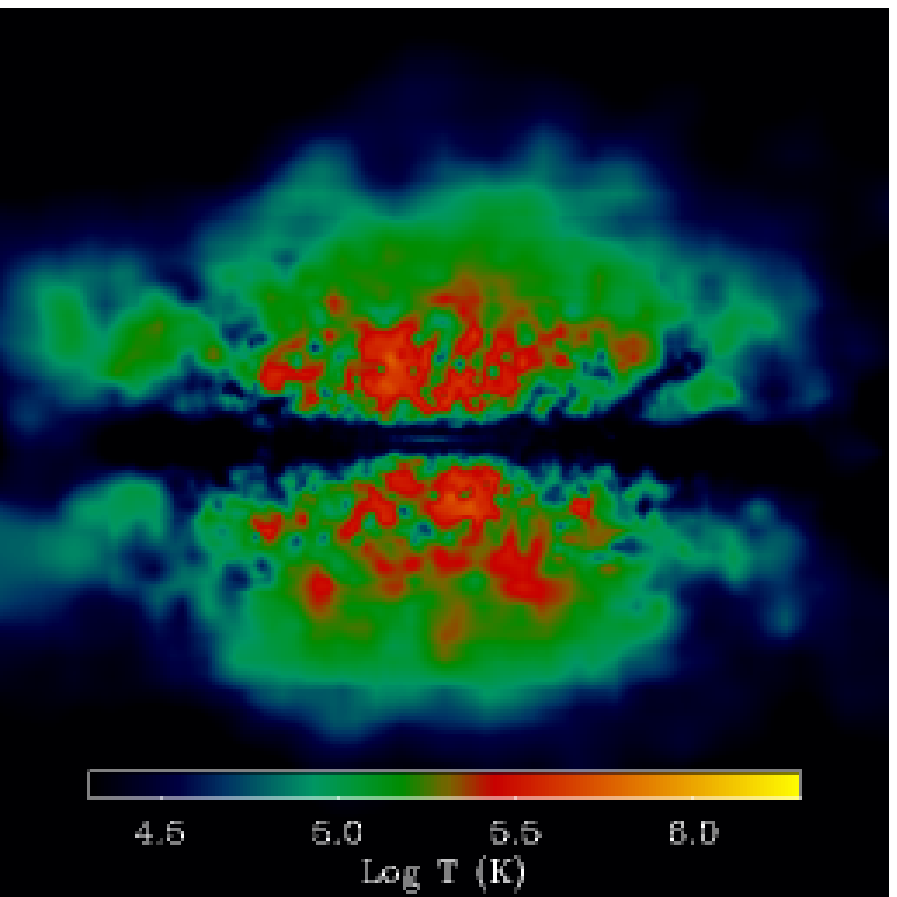}%
\includegraphics[width=0.3\textwidth]{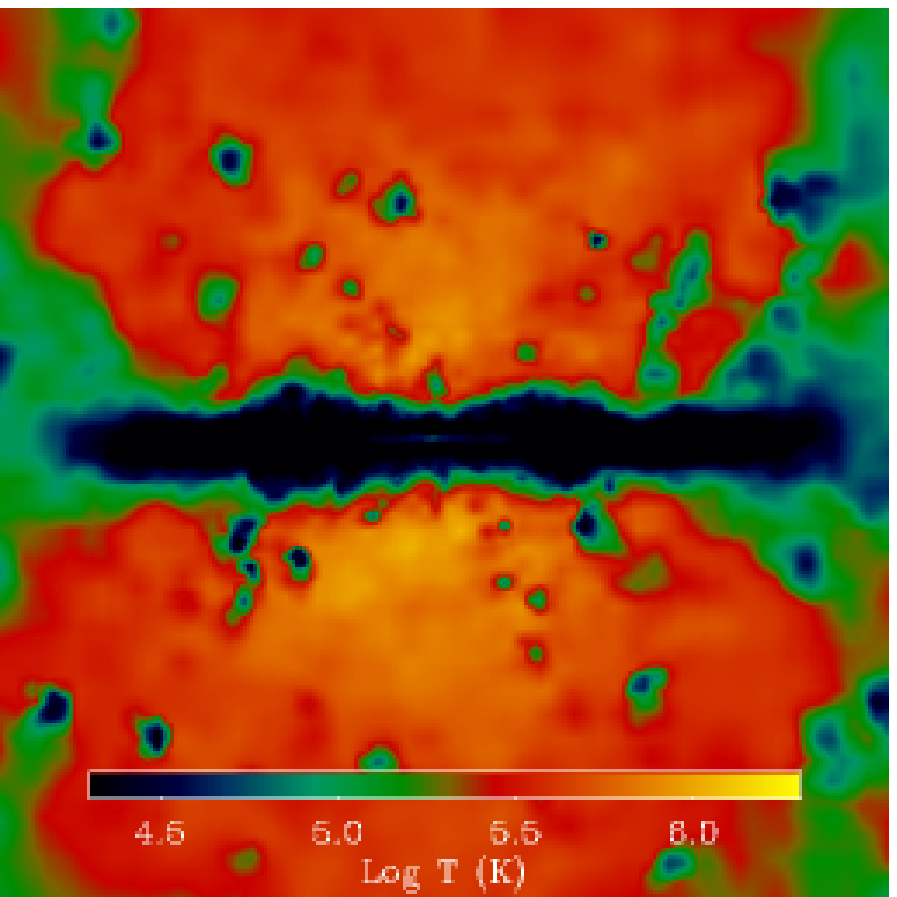}%
\includegraphics[width=0.3\textwidth]{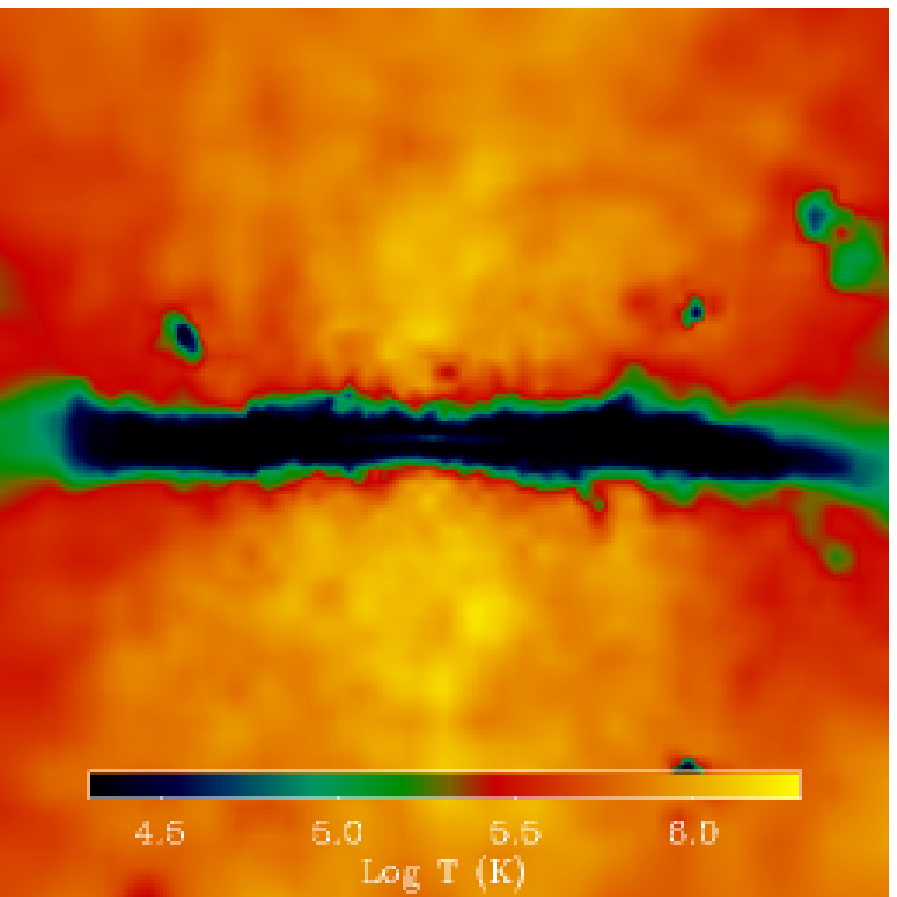}
\caption{Projections of the gas density (top row: edge-on; middle row:
face-on) and temperature (bottom-row: edge-on) for models
\textit{G12-100-65} (left column), the fiducial model
\textit{G12-100-75} (middle column) and \textit{G12-100-85}
(right column) at time $t=250~\Myr$. The white arrows in the top
row show the projected velocity field.  Images are $45\kpch$ on a
side. The colour coding is logarithmic in density
($-4.3<\log_{10}n_{\rm H}/\cmmt<1.3$) and temperature
($4.3<\log_{10}T/\K< 6.3$). The colour scale is indicated by the
colour bars in each column.}
\label{fig:g12dens_comp}
\end{figure*}

Radiative cooling and heating were included using tables for hydrogen
and helium from \citet{Wiersma2009a}. The cooling tables were generated
using the publicly available package \textsc{cloudy} \citep[version
07.02;][]{Ferland2000}, assuming ionisation equilibrium in the
presence of the \citet{Haardt&Madau2001} model for the $z=0$ UV
background radiation from quasars and galaxies, and the cosmic
microwave background.

We implemented a version of the time-stepping algorithm described by
\citet{Durier2011}. Inactive particles that receive feedback energy are
immediately activated so that they can respond promptly to their new
energetic state. Their and their active neighbours' signal velocities
\citep[see e.g.][]{Monaghan1997} are also updated in order to
calculate the size of the next time-step consistently. The new
time-step is then propagated to the inactive neighbours following the
scheme of \citet{Saitoh2009}. We refer the reader to \citet{Durier2011}
for a detailed discussion of the benefits introduced by the scheme.

However, we do not expect the integration accuracy of our simulations
to improve significantly compared with e.g. DS08. As argued by
\citet{Durier2011}, imposing a limit to the maximum allowed time-step
(e.g., by significantly reducing the gravitational softening, as we
have done here and in DS08), may also maintain good energy
conservation in the case of strong energy perturbations. Indeed, we
found only small deviations in the global properties of the outflows
when running the same fiducial models without the time-step limiter.

The initial conditions are based on the model of
\citet{Springeletal2005} and are described in DS08. The model consists
of a dark matter halo, a stellar bulge, and an exponential disc of
stars and gas. The circular velocities at the virial radii are $35.1$
and $163~\kms$ for the $10^{10}$ and $10^{12}\Msolh$ haloes,
respectively. The virial radii are $35.1$ and $163~\kpch$. The halo is rotating and has a dimensionless spin
parameter $\lambda=0.33$. The disc contains 4 percent of both the
total mass and the total angular momentum. The bulge contains 1.4
percent of the total mass and has a scale length one tenth of that of
the disc. The bulge has no net rotation. The initial gas fraction of
the disc is 30 percent, the remaining 70 percent of the disc mass is
made up of stars. The vertical distribution of the stellar disc has a
constant scale height of 10~percent of the radial disc scale length.

\begin{figure*}
\includegraphics[height=0.43\textwidth,trim=00mm 0 0 0mm,clip]{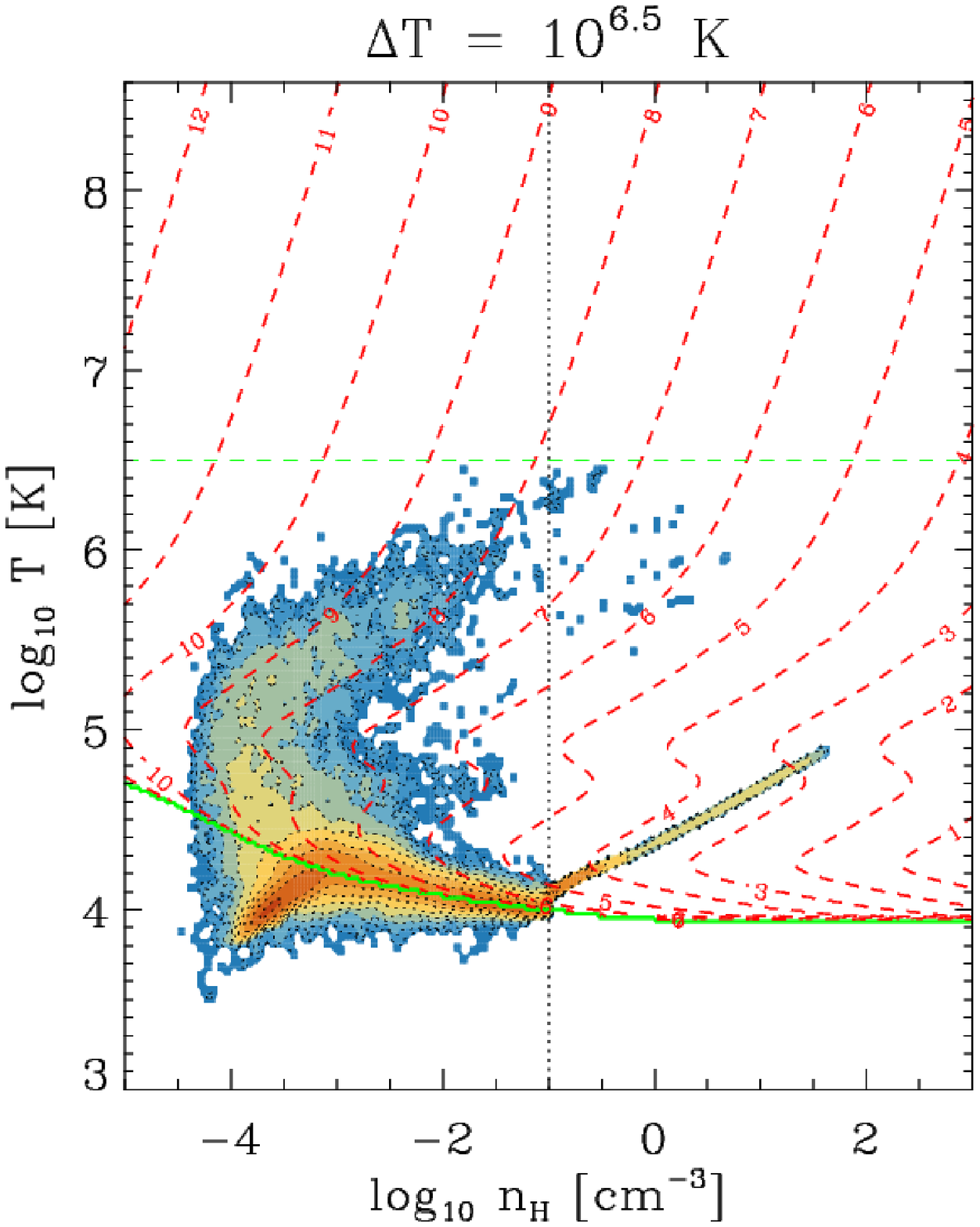}%
\includegraphics[height=0.43\textwidth,trim=25mm 0 0 0mm,clip]{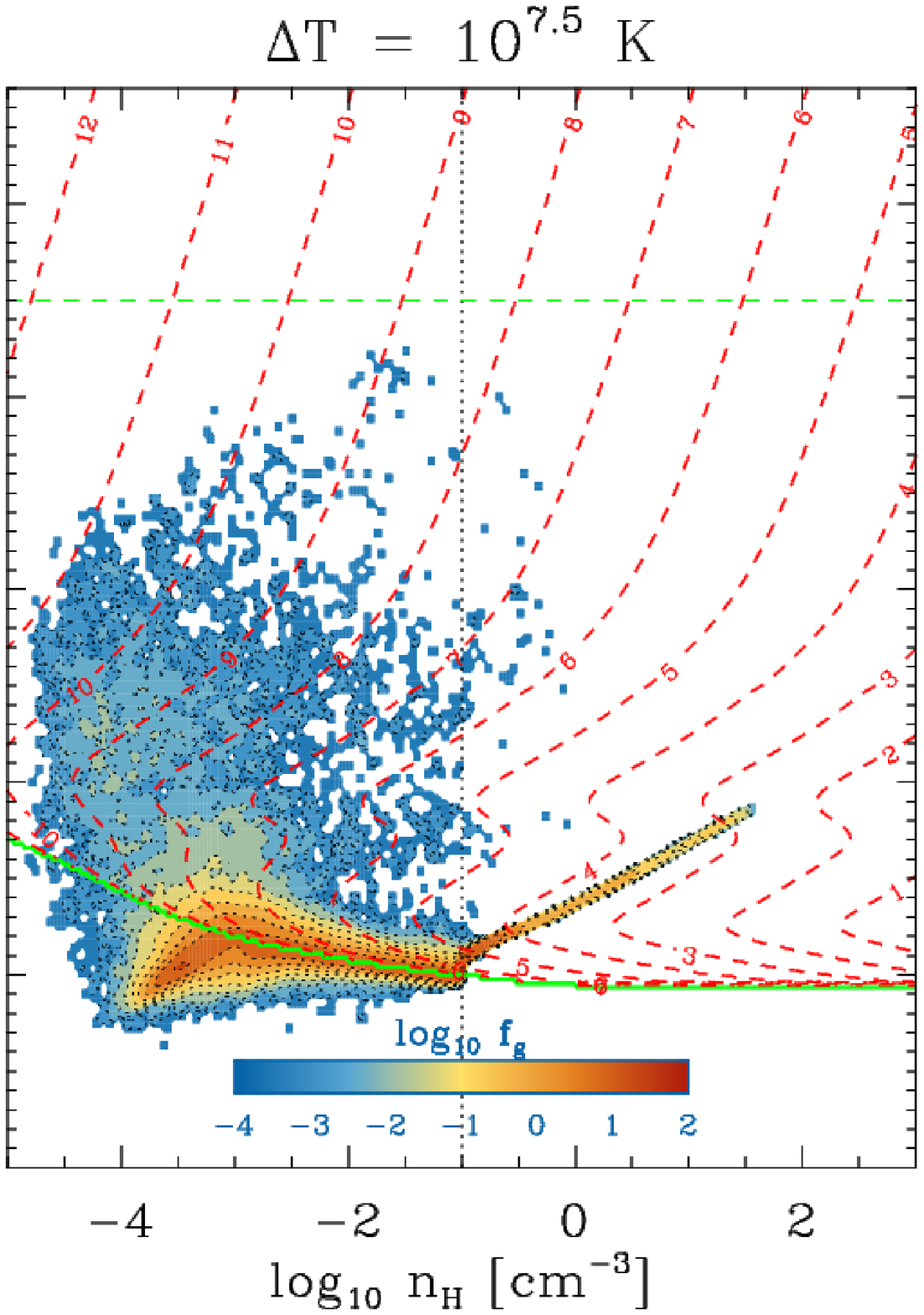}%
\includegraphics[height=0.43\textwidth,trim=25mm 0 0 0mm,clip]{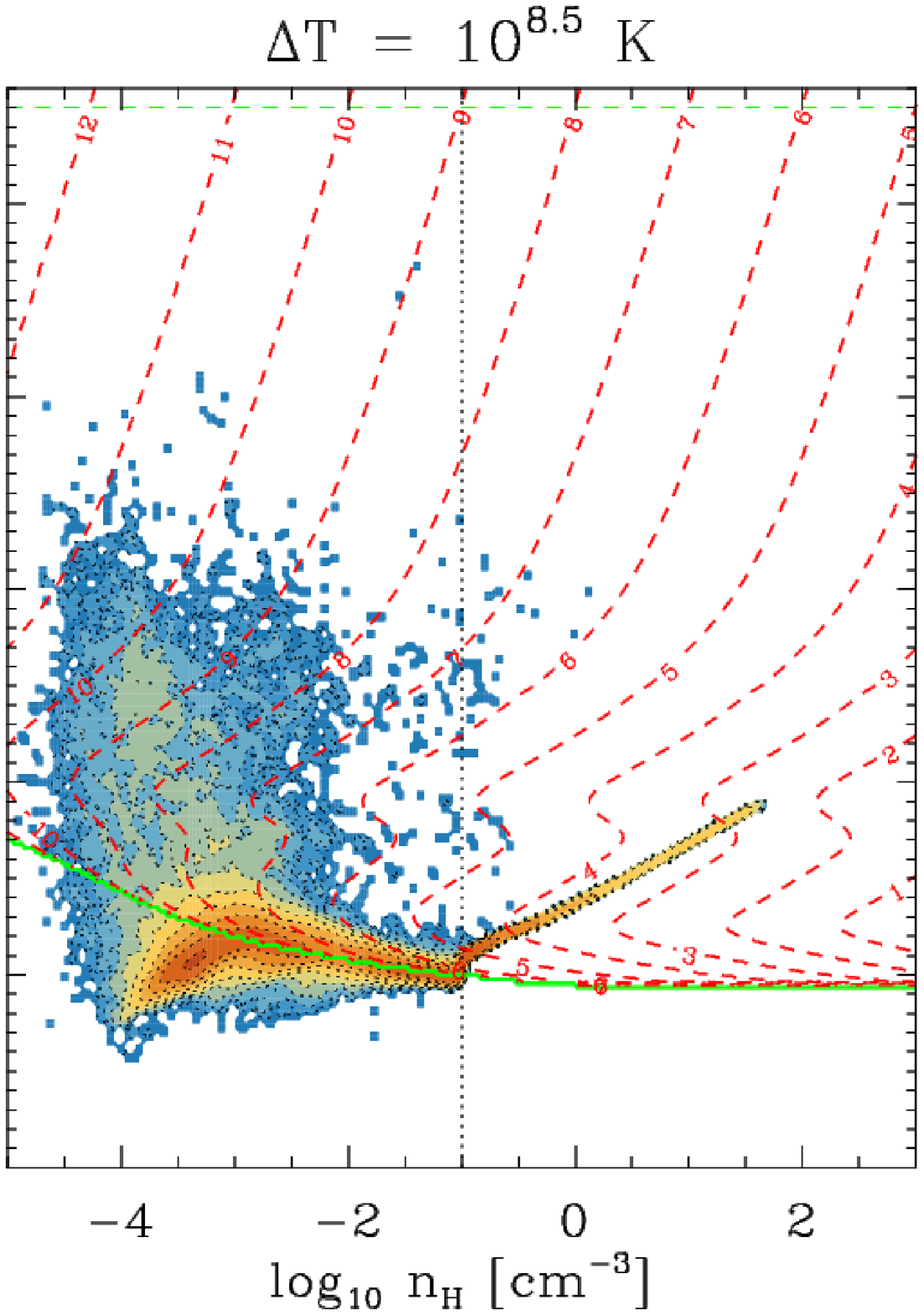}
\caption{Two-dimensional mass-weighted probability density distribution of the gas within $0.2 r_{\rm vir}$ of the dwarf galaxy in temperature-density space at time $t=250$~Myr. The colour coding indicates $f_{\rm g}=({\rm d}M/M)/{\rm d}\log_{10}n_{\rm H}/{\rm d}\log_{10}T$. The three panels correspond to the same three models as were shown in Fig.~\protect\ref{fig:g10dens_comp}. The twelve contours of $f_{\rm g}$ (black, dotted) are equally spaced in the range showed by the colour bar. The red, dashed curves indicate radiative cooling time contours and their labels indicate $\log_{10}(t_{\rm c}/\yr)$. Below the solid, green line photo-heating by the UV background dominates over radiative cooling. The horizontal, dashed line indicates the heating temperature $\Delta T$. The vertical, dotted line marks the threshold for star formation. The imposed, effective equation of state is visible for densities larger than the threshold density.}
\label{fig:g10_phase}
\end{figure*}

Except for our low-resolution runs, the total number of particles in
each simulation is 5,000,494, of which 235,294 are gas particles in
the disc. The baryonic particle mass for the $10^{10}\Msolh$
($10^{12}\Msolh$) halo is $m_{\rm b}=5.1\times 10^2\Msolh$ ($m_{\rm
  b}=5.1\times 10^4\Msolh$).  The gravitational softening length was
set to $\epsilon_b=10\pch$ for the baryons and to $(m_{\rm DM}/m_{\rm
  b})^{1/3}\epsilon_b\approx 17\pch$ for the dark matter in the
$10^{10}\Msolh$ halo. The softening for the massive galaxy is scaled
up in proportion to $m_{\rm DM}^{1/3}$.

\subsection{Simulation parameters}

Table~\ref{tbl:params} lists the simulations we have performed. Each
simulation was evolved for $500~\Myr$. The simulations are labelled
with the prefix \textit{G10} and \textit{G12} for the $10^{10}$ and
the $10^{12}\Msolh$ haloes, respectively, followed by the percentage of
the SN energy that is injected and by the logarithm of the temperature
increase. For example, \textit{G10-040-75} refers to the
$10^{10}\Msolh$ halo, with the SN feedback injecting $40\%$ of
the total available energy and increasing the gas particle temperature
by $\Delta T=10^{7.5}~\K$. We have run several variations of the
fiducial models, \textit{G10-100-75} and \textit{G12-100-75}. The list
of all the models follows:
\begin{itemize} 
\item One run without SN feedback (\textit{G[10,12]-NOFB}).
\item One run injecting $40$ percent of the available SN energy,
  \textit{G[10,12]-040-75}. This is used for comparison with the
  kinetic feedback simulations of DS08, which also used 40 percent of the energy.
\item One set of runs varying the temperature increase $\log_{10} \Delta
  T=[6.5,7.0,7.5,8.0,8.5]$. These runs are labelled
  \textit{G[10,12]-100-[65,70,75,80,85]}, respectively.
\item Two runs in which the number of particles was decreased by
  factors of 8 and 64, respectively (\textit{G[10,12]-100-75-LR08},
  \textit{G[10,12]-100-75-LR64}).
\end{itemize}

To eliminate differences other than the feedback recipe from the comparison with DS08, we repeated the simulations of models
\textit{m12} and \textit{m12nowind} of DS08 with the new version of
the code. We changed the softening to the one used in this work, and
used the time-step limiter for the feedback run. 


\begin{figure*}
\includegraphics[height=0.43\textwidth,trim=00mm 0 0 0mm,clip]{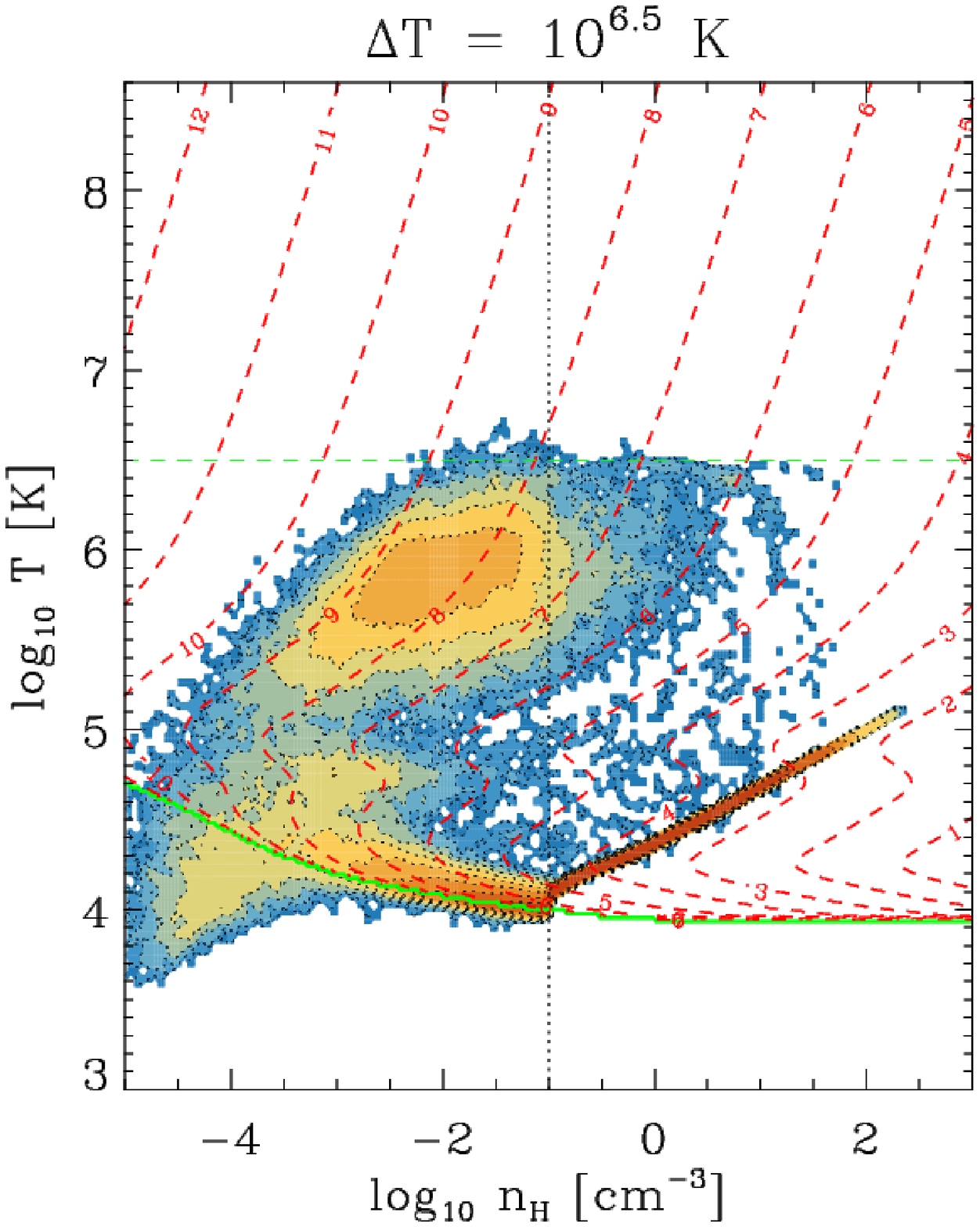}%
\includegraphics[height=0.43\textwidth,trim=25mm 0 0 0mm,clip]{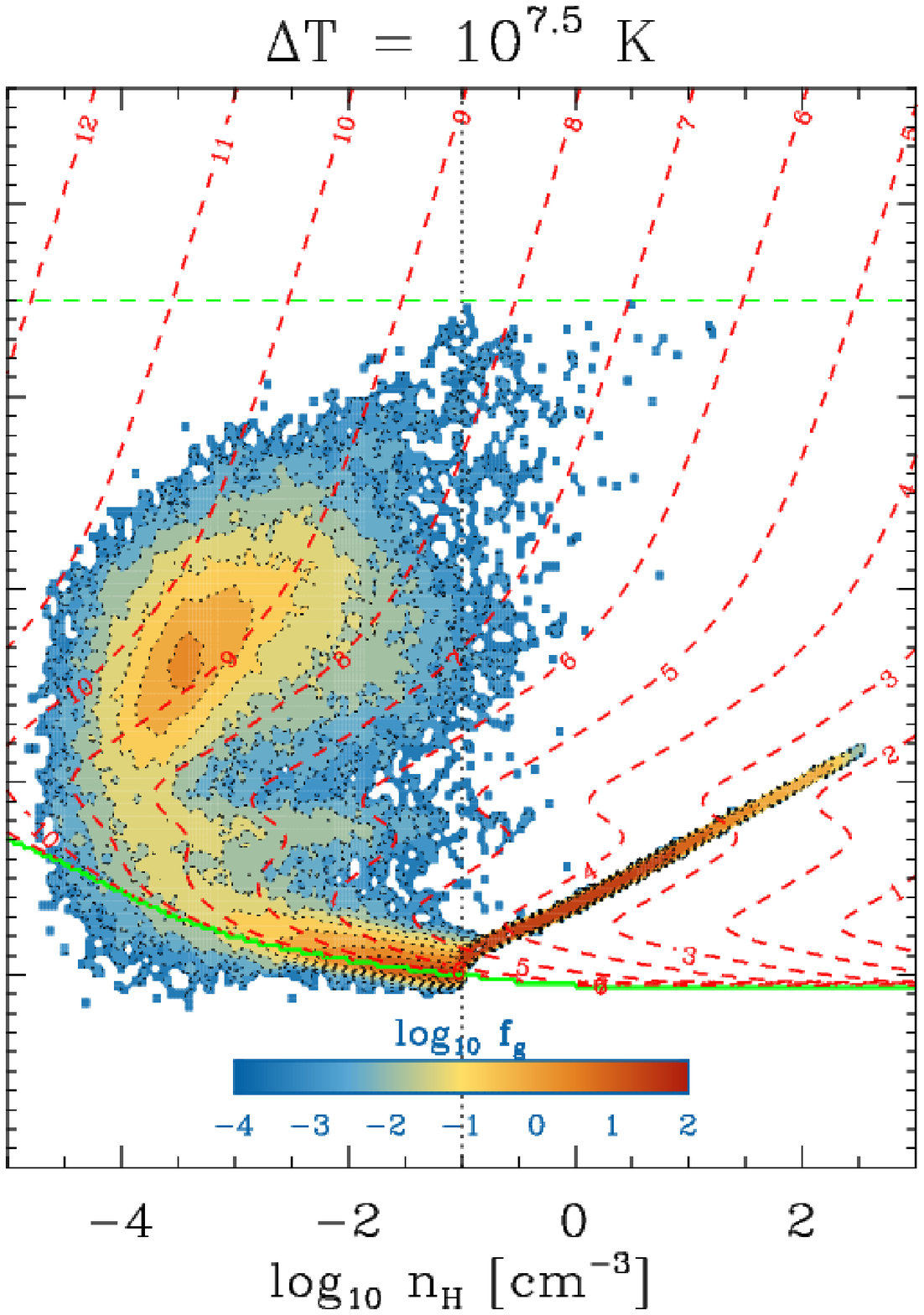}%
\includegraphics[height=0.43\textwidth,trim=25mm 0 0 0mm,clip]{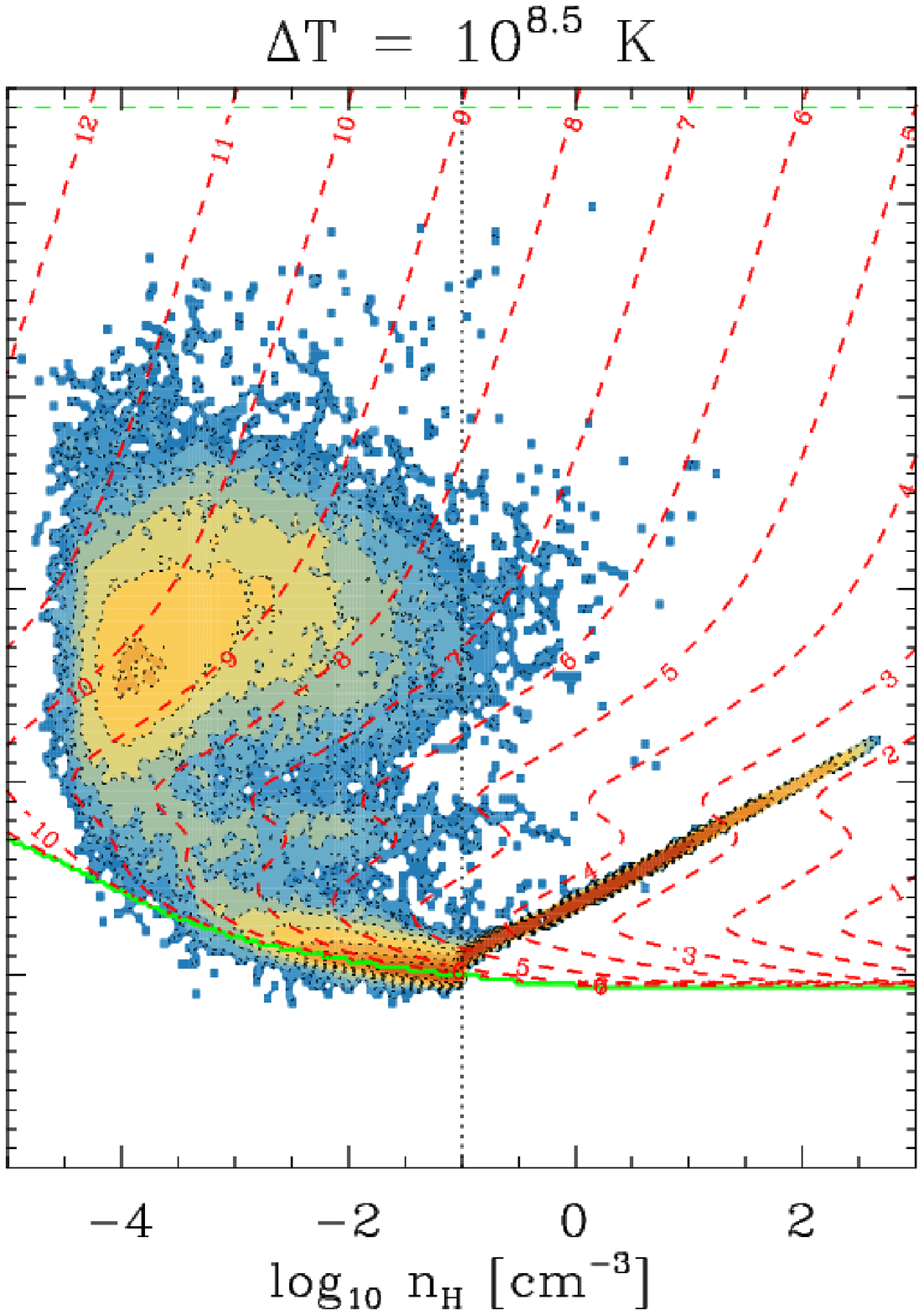}
\caption{As Fig.~\protect\ref{fig:g10_phase}, but for the models of the massive galaxy shown in Fig.~\protect\ref{fig:g12dens_comp}.}
\label{fig:g12_phase}
\end{figure*}

\begin{figure}
\includegraphics[width=0.46\textwidth]{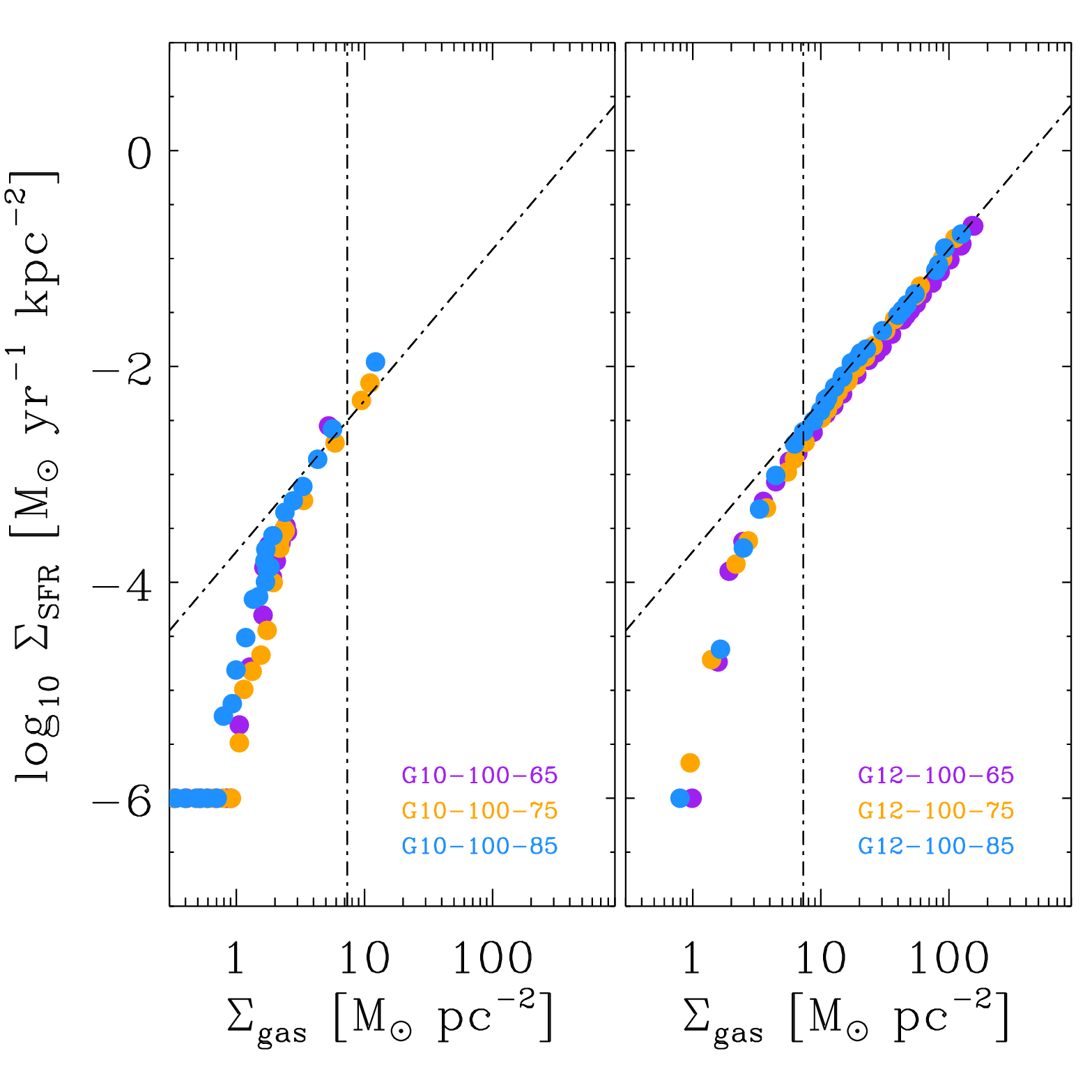}
\caption{The Kennicutt-Schmidt SF relation for a selection of \textit{G10} (left panel) and \textit{G12} (right panel) models at $t=250~\Myr$. Surface densities are computed in cylindrical annuli containing a constant number of particles and including all particles with vertical coordinate $|z|<2~\kpc$. The tilted line shows the observed Kennicutt SF law, equation (\ref{eq:KS}), whereas the vertical line shows the gas surface density below which SF is observed to become inefficient \citep[for more details see][]{Schaye2008}. All models are in excellent agreement with the observations.}
\label{fig:kh}
\end{figure}

\section{Simulation results}
\label{sec:res}

In this section we describe tests of our implementation of thermal
feedback. We first show the effect of varying the temperature
increase. We proceed with a comparison to the kinetic feedback model of DS08. We conclude the section by reporting the results of
resolutions tests.

\subsection{Dependence on the temperature increase}

We ran a set of simulations with $\FTH=1$ and varying $\log_{10}
\Delta T=[6.5,7.0,7.5,8.0,8.5]$ to study the dependence on the
temperature increase.
We first describe the morphology of the galaxies and their outflows,
and proceed with discussing the SF histories and the outflow
properties quantitatively.

\subsubsection{Morphology}
\label{sec:morphology}

For each halo, we compare here the fiducial model, which has
$\log_{10} \Delta T=7.5$, to our most extreme models, which use $\log_{10} \Delta
T=[6.5,8.5]$. Fig.~\ref{fig:g10dens_comp} shows projections of
the density (top and middle rows) and temperature (bottom row) for the
dwarf galaxy at time $t=250~\Myr$ for three different values of the
temperature increase (from left to right, $\log_{10} \Delta
T=[6.5,7.5,8.5]$). Recall that all models inject the same amount of
energy per unit stellar mass. 

The morphology of the galaxy is
irregular in all cases. As the heating temperature is increased, the low-density bubbles in the ISM and in the circumgalactic medium increase in size and open up vertical channels through which the outflows can escape. Consequently, the outflow becomes more collimated along the vertical axis as $\Delta T$ increases, enhancing its bipolarity. The velocity field overplotted in the top row shows that the flow is faster for larger $\Delta T$, while the bottom row shows that the outflowing gas is also hotter.

Fig.~\ref{fig:g12dens_comp} shows edge-on projections of the density
(top row) and temperature (bottom row) for the massive galaxy for the
same three different values of the temperature increase. The
dependence on $\Delta T$ is more evident for this more massive galaxy.
For $\Delta T=10^{6.5}~\K$ (left column),
most of the outflowing gas is confined to a region
around the disc. The disc looks puffed up as the expelled gas is
deposited just outside it, and the gas breaks up in little blobs. The velocities are small and the velocity field does not show a clear preferential
orientation. There is a galactic fountain, but no large-scale galactic
wind.

In contrast, the fiducial model ($\Delta T=10^{7.5}~\K$; middle column) shows a
clear bipolar outflow, which is sustained until the end of the run.
The outflow is mostly driven from the inner part of the disk where a large fraction of the SF is taking place, and it is collimated by the disc which impedes motion within its plane.
The temperature map shows several cold blobs above and
below the galactic disc. The blobs come from the disc and are moving
outward, thus the outflow is ejecting parcels of cold gas. Cold
blobs are seen falling back onto the disc at large radii.

The highest $\Delta T$ run ($\Delta T=10^{8.5}~\K$; right column) looks similar to the fiducial model, but the wind is hotter and moving faster and the circumgalactic medium contains fewer cold
blobs.

Videos illustrating the time evolution of the model galaxies are
available at this web address: \texttt{http://www.strw.leidenuniv.nl/DS12/}

\subsubsection{Gas phase distribution}
\label{sec:gasdistr}

The mass-weighted probability distribution function (PDF) of the gas at $t=250$~Myr in the $(n_{\rm H},T)$ plane is shown in Figs.~\ref{fig:g10_phase} and \ref{fig:g12_phase} for the models shown in Figs.~\ref{fig:g10dens_comp} and \ref{fig:g12dens_comp}, respectively.
To limit the effects of the vacuum boundary conditions (mainly adiabatic cooling to extremely low temperatures and densities), we include only the gas inside a spherical volume of radius
$0.2 r_{\rm vir}$, but we normalise the PDFs to the total gas mass. 
The vertical, dotted line marks our threshold density
for SF, and at higher densities the imposed, effective equation of
state is clearly visible. Contours of constant radiative cooling time (dashed, red lines) are over-plotted.\footnote{Note that the $t_{\rm c}$ contours have been calculated assuming a constant mean molecular weight, $\mu=0.6$.} The contour labels indicate integer values of $\log_{10}(t_{\rm c}/\yr)$. The equilibrium between radiative cooling and photo-heating by the UV background is shown by the solid, green curve. Below this curve radiative heating dominates over radiative cooling. Note that gas can reach temperatures lower than the equilibrium value if it cools adiabatically. In the absence of feedback, none of the gas below the SF threshold would have temperatures significantly above the equilibrium value. Finally, the horizontal, dashed (green) lines indicate the feedback heating temperatures $\Delta T$. Gas in which feedback energy has been injected initially resides near these lines and, as long as radiative cooling is unimportant, will expands adiabatically, exiting the star-forming region at different temperatures. Indeed, the phase diagrams show different distributions of shock-heated gas for different $\Delta T$, with the low-density, high-temperature regions being more populated for increasing $\Delta T$.

For the dwarf galaxy (Fig.~\ref{fig:g10_phase}) the temperature-density distributions are similar for all values of $\Delta T$, although there are some differences in the high temperature regime. We will show later that the SF histories and the outflows are also very similar for all values of $\Delta T$ and that, given the resolution, this is in accord with the results of section~\ref{sec:coolexp}. There is very little hot gas. The surface density of the disc, and hence the pressure, is too low to confine the heated gas, which therefore immediately blows out of the disc and cools adiabatically.

For the massive galaxy changing $\Delta T$ has a much more dramatic impact on the distribution of shock-heated gas ($T>10^5~\K$). For the model with the lowest value of $\Delta T$ (left panel of Fig.~\ref{fig:g12_phase}), the peak of the PDF lies within the density range $10^{-3}$ to $10^{-2}~\cmmt$. This confirms the qualitative result shown in Fig.~\ref{fig:g12dens_comp}: most of the outflowing gas accumulates in a region around the disc. If we increase $\Delta T$ (middle and right panels), the peak in the PDF moves to lower densities ($\sim10^{-4}~\cmmt$) thanks to the development of a large-scale outflow that moves gas away from the disc. The total fraction of shock-heated gas decreases with the $\Delta T$ because less gas resides within $0.2 r_{\rm vir}$ if the wind velocity is higher.

\subsubsection{Star formation history}
\label{sec:sfr}

Before discussing the SF histories, we show the predicted Kennicutt-Schmidt SF relations in Fig.~\ref{fig:kh}. The left (right) panel shows the same three different models as were shown in Fig.~\ref{fig:g10dens_comp} (Fig.~\ref{fig:g12dens_comp}) for the dwarf (massive) galaxy. Gas mass and SF surface densities were computed in annuli containing a constant number of gas particles and including all particles with vertical coordinate $|z|<2~\kpc$. The observed Kennicutt-Schmidt law (eq.~(\ref{eq:KS}); tilted line) and the steepening at the SF threshold density (vertical line) are well matched. This success is not unexpected, as we already showed in \citet{Schaye2008} that the observed SF law can be implemented directly in the form of a pressure law and that this enables the simulations to reproduce the observations without the need to tune any parameters and irrespective of whether strong feedback is present. While feedback determines the surface density of the gas, it does not affect the efficiency of star formation at a fixed surface density  in our models, since pressure and surface density are closely related in self-gravitating systems. 

\begin{figure*}
\includegraphics[width=0.45\textwidth]{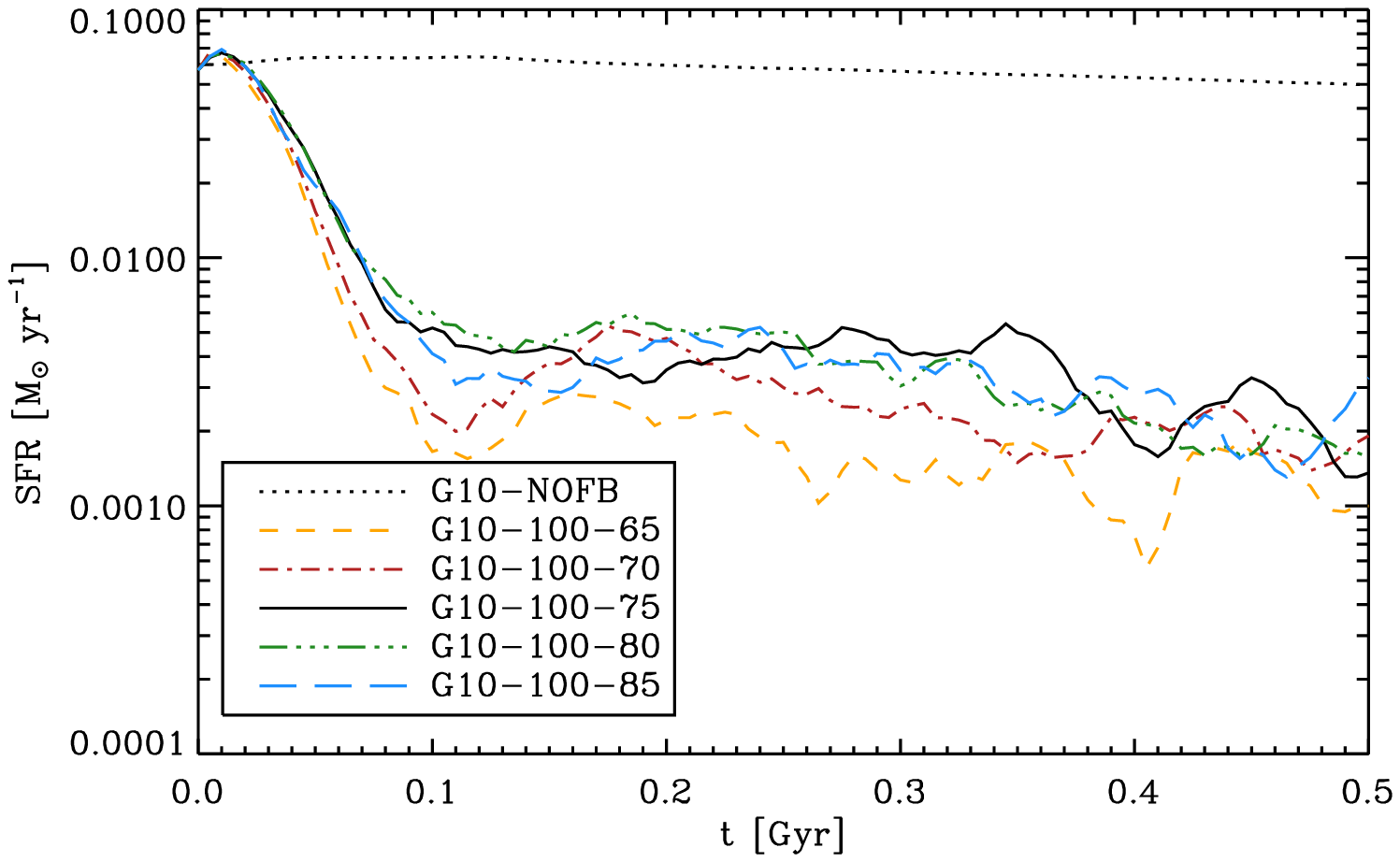}%
\includegraphics[width=0.45\textwidth]{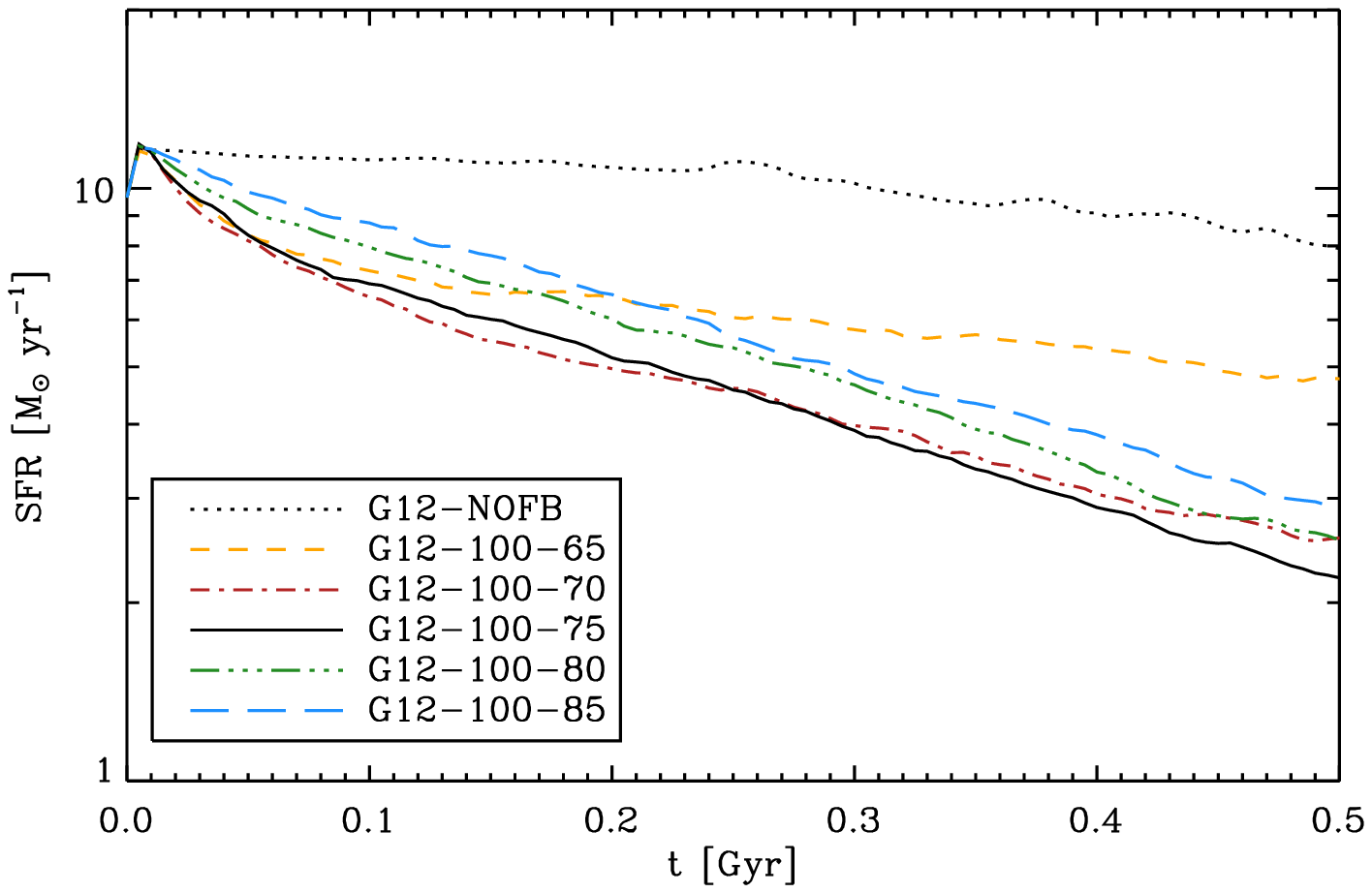}
\caption{SFR as a function of time for the galaxies in the $10^{10}$
  and the $10^{12}\Msolh$ haloes (left- and right panels,
  respectively). We vary the temperature increase due to feedback events in steps of
  $0.5~\dex$ over the range $\log_{10} \Delta T/\K=[6.5,8.5]$. The dwarf
  galaxy's SF history is insensitive to the value of $\Delta T$, while for the massive
  galaxy the feedback is less efficient for the lowest heating temperature.}
\label{fig:sfr}
\end{figure*}

The dependence of the SF histories on $\Delta T$ is shown in
Fig.~\ref{fig:sfr}. For the dwarf galaxy (left panel), the star formation rate (SFR)
drops sharply within the first $100~\Myr$ due to the strong feedback
produced by the initial burst of SF, and remains nearly constant thereafter. The fact that the sharp drop is due to feedback can be seen by comparing to the nearly horizontal black, dotted curve, which shows the SF history for a run without feedback. 
The factor by which the feedback reduces the SFR is insensitive to $\Delta T$. This is in accord with the calculations presented in Section~\ref{sec:coolexp}. The dwarf galaxy has a low surface density and forms most of its stars at densities close to the SF threshold of $n_{\rm H} = 10^{-1}~{\rm cm}^{-3}$ (see Fig.~\ref{fig:g10_phase}). For such densities and primordial abundances we expect cooling losses to be small even for heating temperatures as low at $10^{6.5}$~K. Indeed, for our particle mass of $7\times 10^2~$M$_\odot$ and a heating temperature of $T=10^{6.5}$~K, equation~(\ref{eq:nhcrit}) tells us that cooling losses should be unimportant for densities $n_{\rm H} < 10~{\rm cm}^{-3}$. 

However, as Fig.~\ref{fig:tscale} shows, the situation could be different for solar abundances. While the cooling time is insensitive to the metallicity for $T\ga 10^{7.5}$~K, for $T = 10^{6.5}$~K it is about an order of magnitude smaller for solar metallicity than it is for primordial abundances. A factor ten increase in the cooling rate would reduce the maximum density for which the feedback is expected to be effective to $n_{\rm H} \approx 0.3~{\rm cm}^{-3}$ (eqs.~[\ref{eq:tratio}] and [\ref{eq:nhcrit}]). 
We would therefore expect a significant fraction of the feedback energy to be radiated away before it can be converted into kinetic form, if we were to run the dwarf galaxy simulation with $\Delta T = 10^{6.5}$~K and solar metallicity. Indeed, we have performed such a run (not shown) and find the SFR to be much higher. After 400~Myr it is about 0.04 M$_\odot \, {\rm yr}^{-1}$ which is closer to the run without feedback than to the run with the same $\Delta T$ but primordial abundances.

For the massive galaxy the decline in the SFR is more gradual (right panel). After a few hundred Myr, all runs predict roughly the same SFR except for the one adopting $\Delta T = 10^{6.5}~$K, which has a substantially higher SFR. The morphological comparison (Fig.~\ref{fig:g12dens_comp}) shows that in this run the gas is
unable to escape to large radii. Instead, it accumulates around the disc, and eventually falls back onto it. This is expected, because equation~(\ref{eq:nhcrit}) shows that radiative losses will become important at densities that are 10 times lower than for the dwarf galaxy, because the particle mass is 100 times higher. Moreover, the densities in the ISM are higher in the massive galaxy, with many star particles forming in gas with densities $n_{\rm H} \sim 1-10~{\rm cm}^{-3}$ (Fig.~\ref{fig:g12_phase}). Comparing these numbers to the maximum density for which cooling losses are small at this resolution, $n_{\rm H} \lesssim 31 ~{\rm cm}^{-3} ~ (T/10^{7.5})^{3/2}$ (eq.~[\ref{eq:nhcrit}]), we expect small changes for $\Delta T = 10^7~$K and a substantial reduction of the feedback efficiency for $\Delta T = 10^{6.5}~$K. 

Models \textit{G12-100-70} and
\textit{G12-100-75} have similar SF histories, suggesting convergence of the results. However, models \textit{G12-100-80} and
\textit{G12-100-85} have SFRs that are larger than that
of the fiducial model by factors of $\simeq 25$ and 40 percent,
respectively. The trend for the largest $\Delta T$'s likely arises from poor sampling of the distribution of SN energy in the disc. Indeed, the expectation value for the
number of heated neighbours decreases with $\Delta T$, and is (from
eq.~[\ref{eqn:nheat}]) 0.42 and 0.13 for \textit{G12-100-80} and
\textit{G12-100-85}, respectively. This shows the importance of
locally linking the feedback events with the formation of star
particles by depositing the star particle SN energy into at least
one of its neighbours. The effect is less severe for the dwarf galaxy
because star formation is restricted to smaller scales, these scales are resolved with many more particles, and it is easier to eject gas in this case due to the lower ISM pressure and the shallower gravitational potential well. 

Finally, we note that the SFRs are lower than for the kinetic feedback runs of
DS08. This difference is, however, due to the different amount of energy injected rather than to the manner in which this is done (i.e.\ thermal vs.\ kinetic). We inject more energy here ($\FTH=1$
whereas DS08 used $\FTH=0.4$), thus a stronger quenching is expected.
In Section~\ref{sec:kin} we will show that the two methods are in fact
in good agreement with each other.

\begin{figure*}
\includegraphics[width=0.45\textwidth]{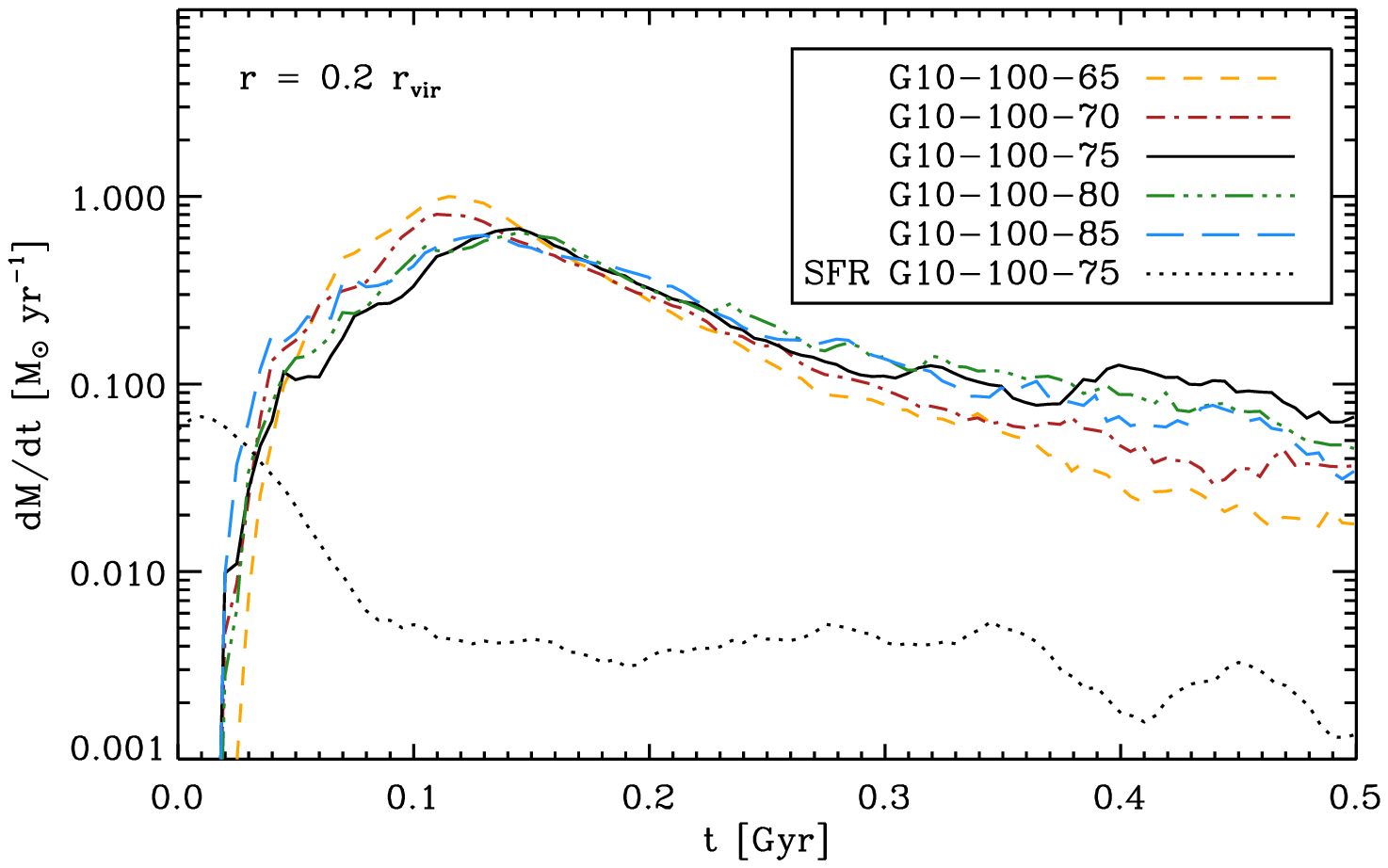}%
\includegraphics[width=0.45\textwidth]{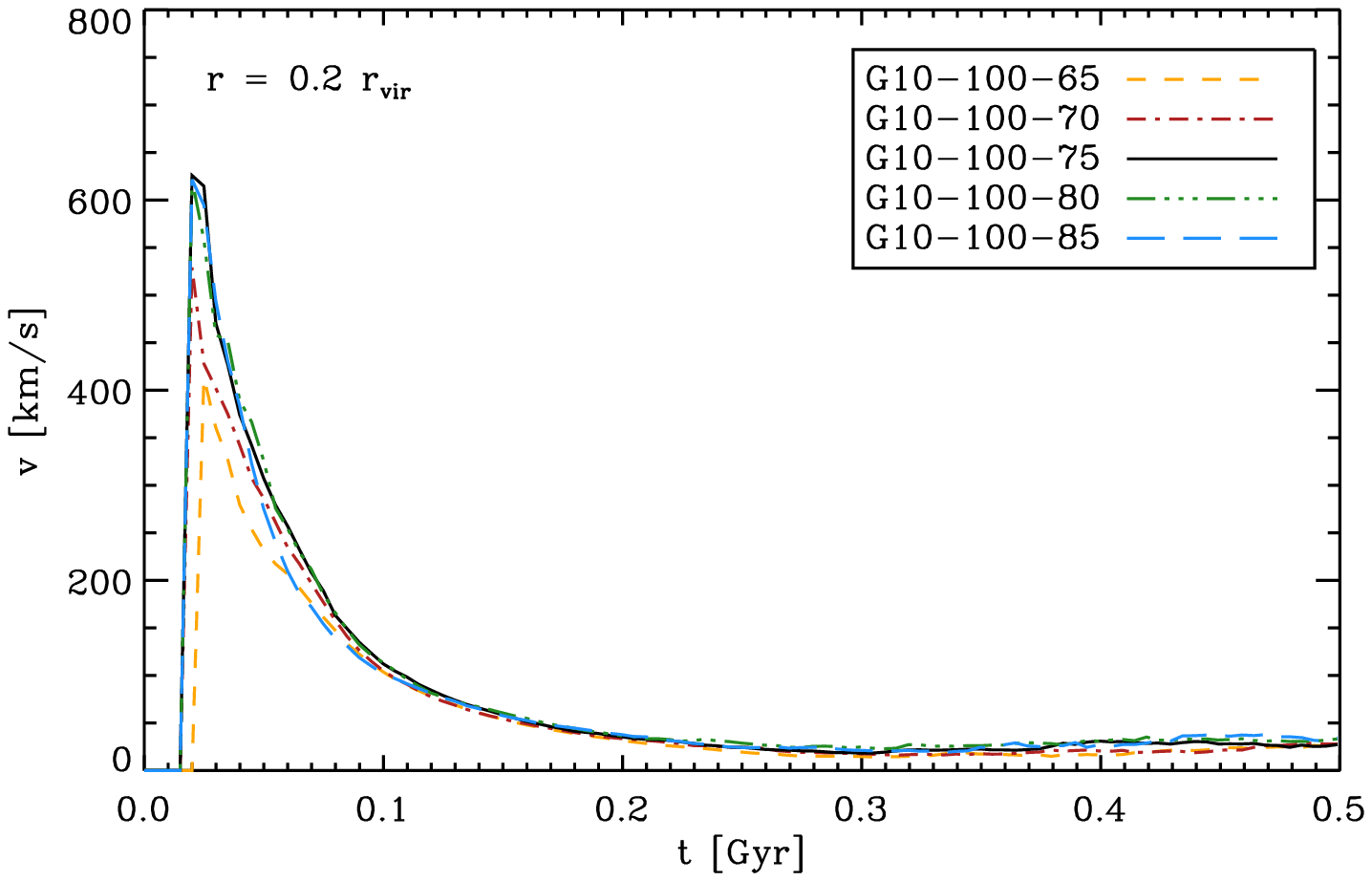}\\
\includegraphics[width=0.45\textwidth]{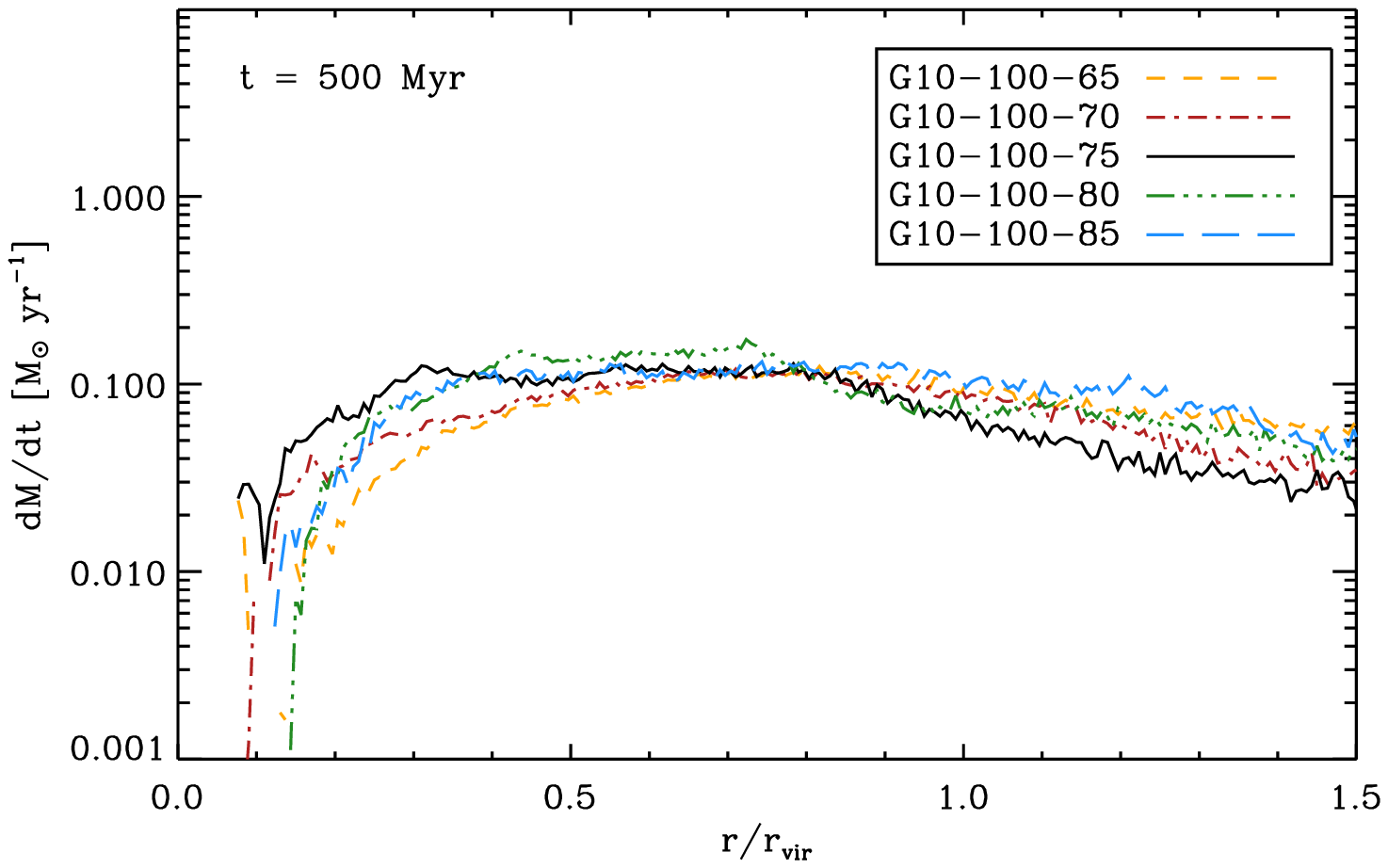}%
\includegraphics[width=0.45\textwidth]{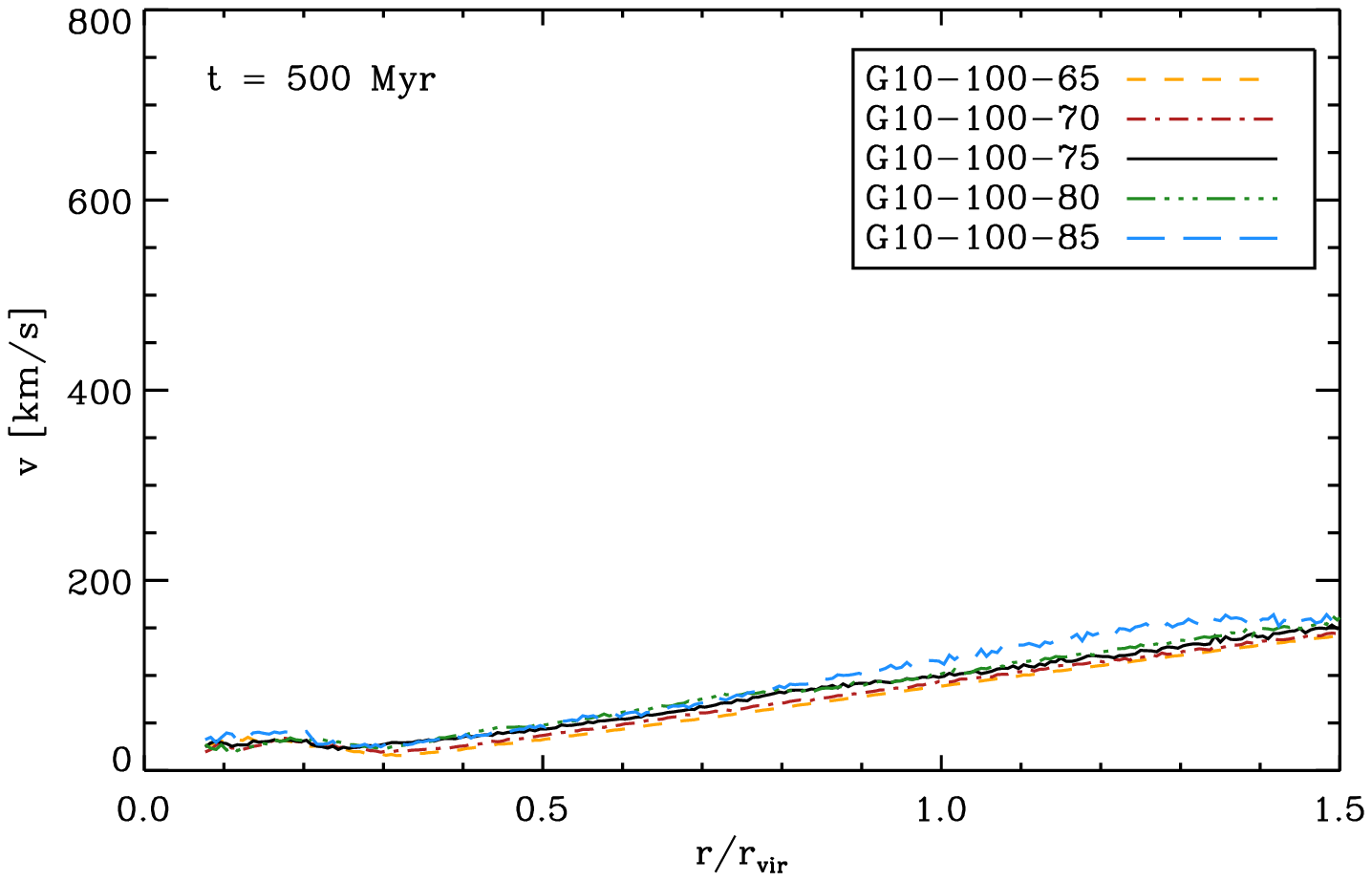}
\caption{Mass outflow rate (left column) and average outflow
velocity (right column) measured through a spherical shell at
radius $r=0.2r_{\rm vir}$ as a function of time (top row) and at
$t=500~\Myr$ as a function of radius (bottom row) for models
\textit{G10-100-[65,70,75,80,85]}. The dotted line in the top-left
panel indicates the SFR of model \textit{G10-100-75}. All other curves
are labelled in the legends. }
\label{fig:g10outflow}
\end{figure*}

\subsubsection{Mass outflow rate and wind velocity}
\label{sec:winds}

In this section we measure the wind velocity and mass loading as a function of time and radius. We first briefly describe the method, which is identical to the one employed in DS08.

\begin{figure*}
\includegraphics[width=0.45\textwidth]{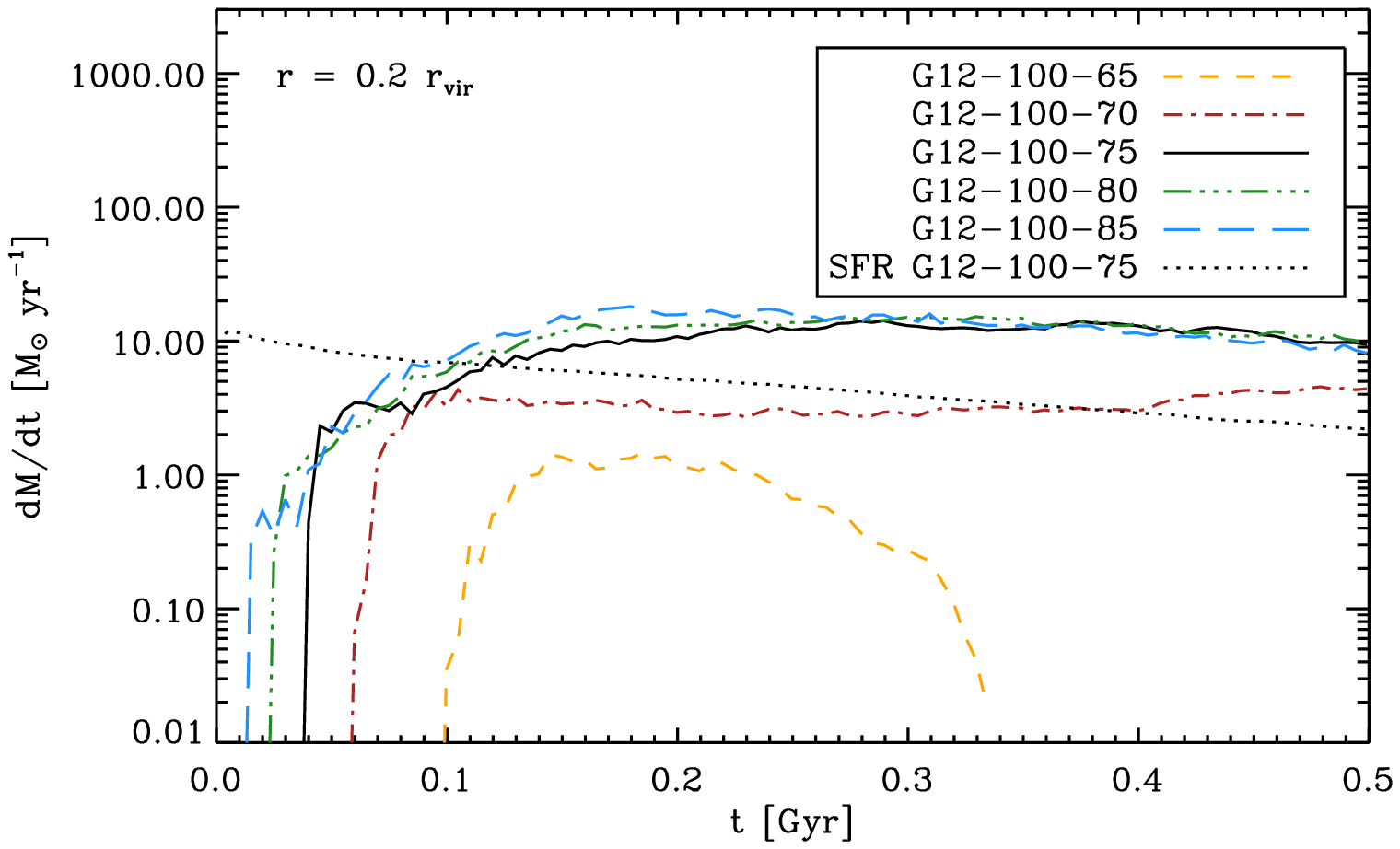}%
\includegraphics[width=0.45\textwidth]{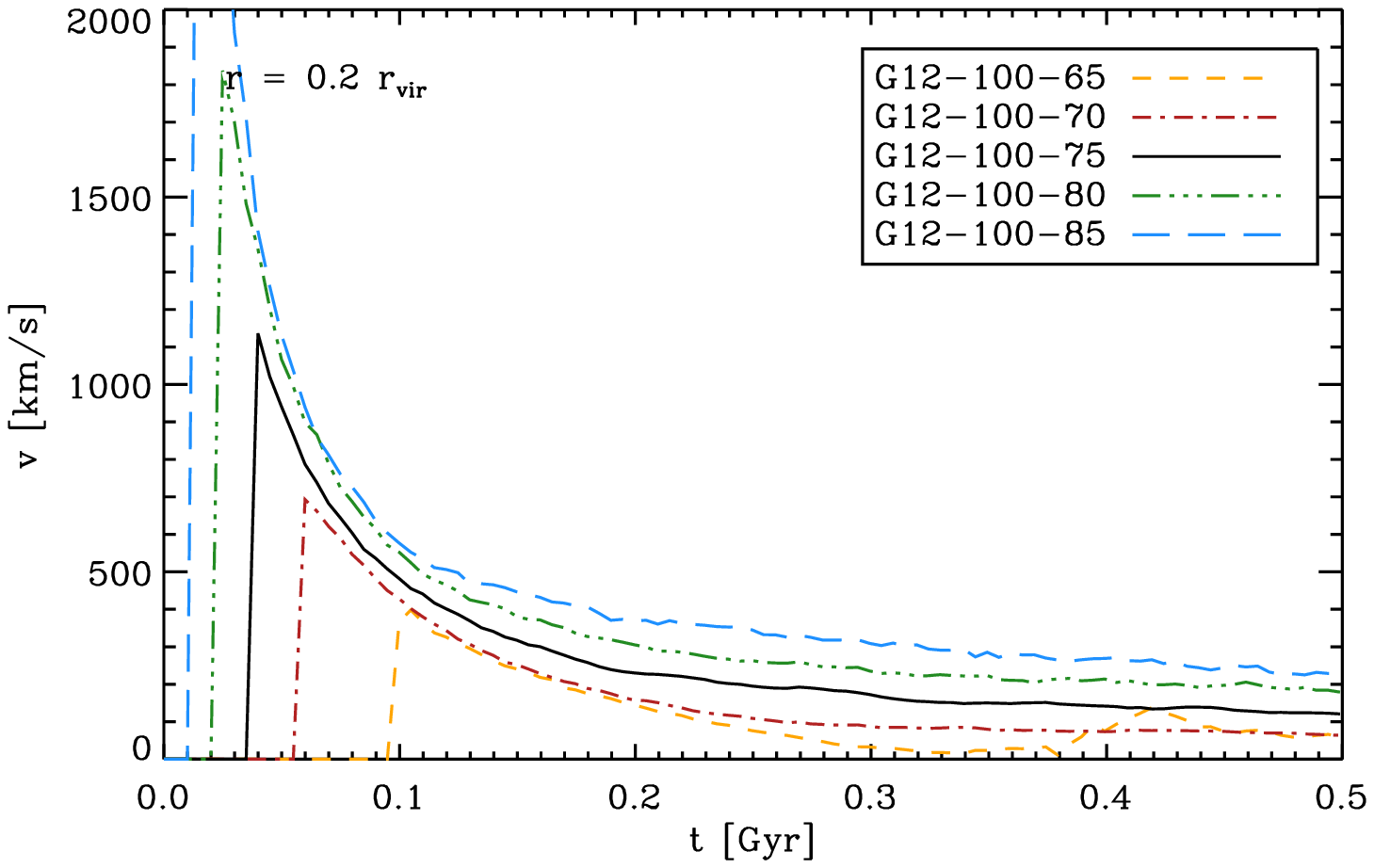}\\
\includegraphics[width=0.45\textwidth]{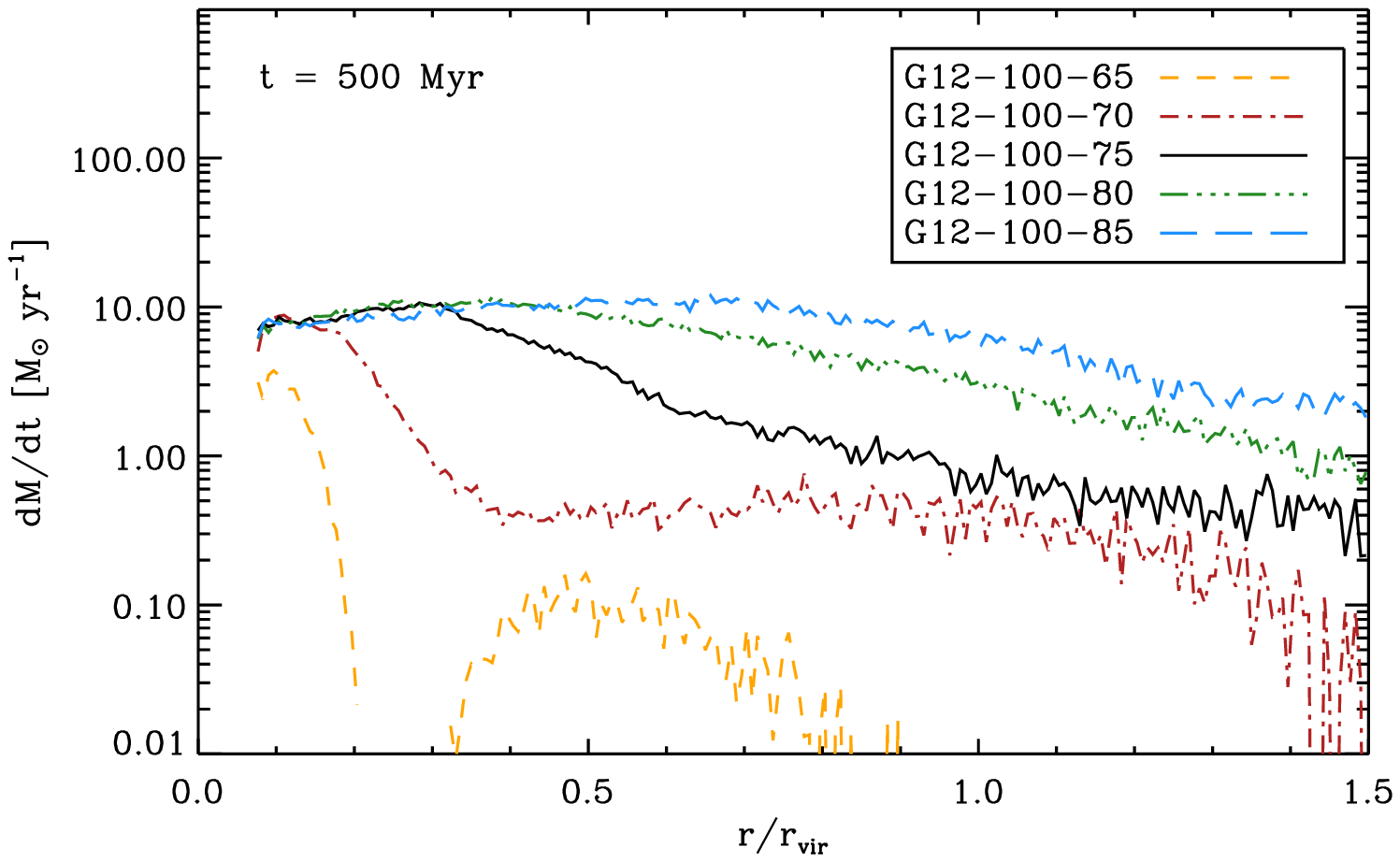}%
\includegraphics[width=0.45\textwidth]{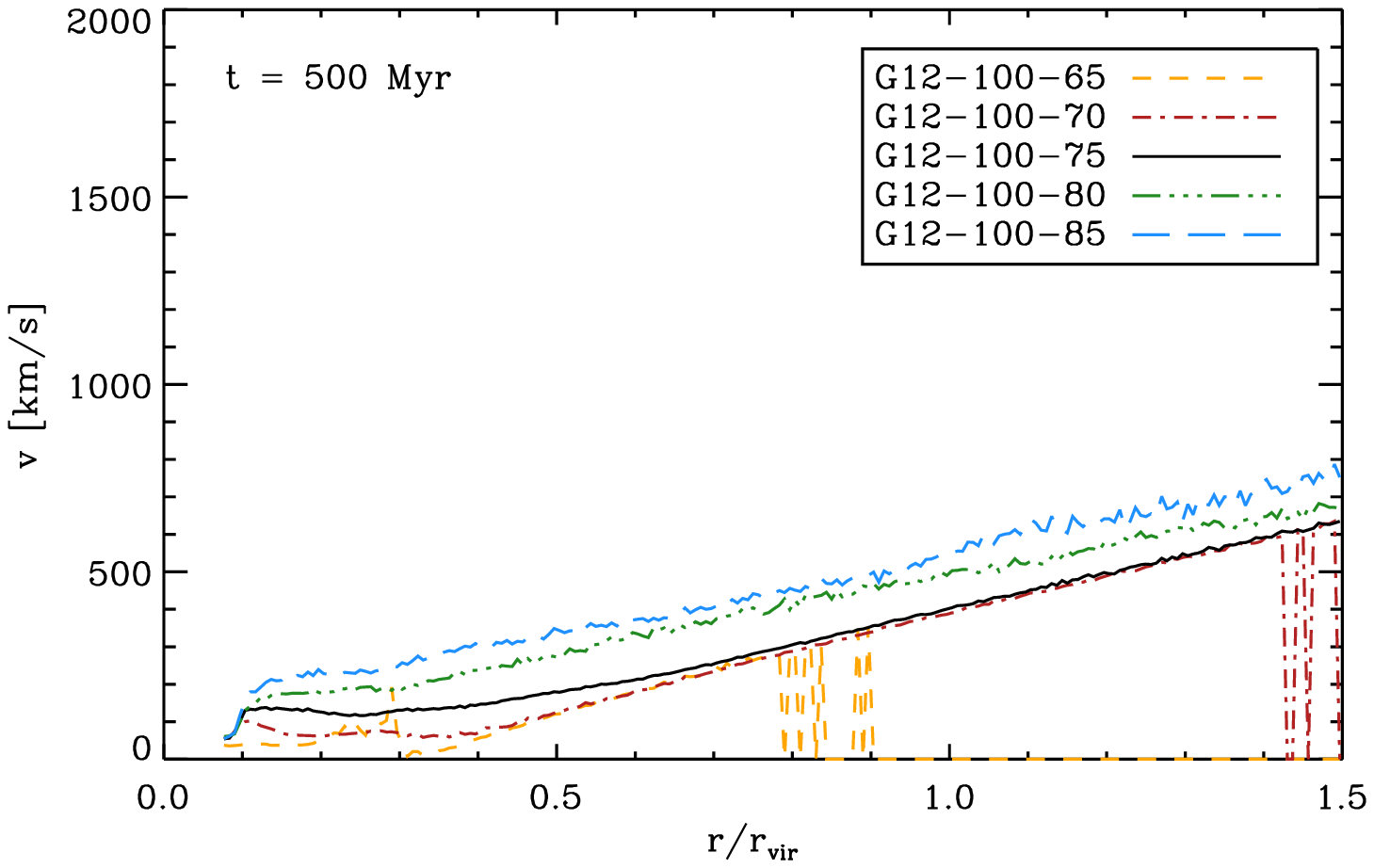}
\caption{Mass outflow rate (left column) and average outflow
velocity (right column) measured through a spherical shell at
radius $r=0.2r_{\rm vir}$ as a function of time (top row) and at
$t=500~\Myr$ as a function of radius (bottom row) for models
\textit{G12-100-[65,70,75,80,85]}. The dotted line in the top-left
panel indicates the SFR of model \textit{G12-100-75}. All other curves
are labelled in the legends.}
\label{fig:g12outflow}
\end{figure*}

We discretise the following integral equation of the net mass outflow rate
through a surface $S$:
\begin{equation}
\dot{M} = \int_S \rho \mathbf{v}\cdot{\rm d}\mathbf{S},
\end{equation}
where $\rho$ is the gas density and $\mathbf{v}$ is the gas velocity
at any position on the surface. Given a spherical shell of radius $r$
and thickness $\Delta r$ centred on the origin, the above equation
becomes
\begin{equation}
\dot{M}(r,\Delta r)=\frac{1}{\Delta r}\sum_{i=1}^{N_{\rm shell}}m_i
\mathbf{v}_i\cdot\frac{\mathbf{r}_i}{r_i},
\end{equation}
where $N_{\rm shell}$ is the total number of particles within the
shell, and $m_i$ and $\mathbf{r}_i$ are their mass and position,
respectively. We consider all particles and use $\Delta r=r_{\rm vir}/150$.

The average outflow velocity is also given in discrete form, and is
the mass-weighted, average radial velocity:
\begin{equation}
\left <v\right > (r,\Delta r) =\frac{\sum_{i=1}^{N_{\rm shell}}m_i(\mathbf{v}_i\cdot\frac{\mathbf{r}
_i}{r_i})_+}{\sum_{i=1}^{N_{\rm shell}}m_i},
\end{equation}
where $\mathbf{v}_i$ is the particle velocity. We consider only
particles moving away from the origin, as indicated by the
subscript ``$+$'' in the above equation.

To investigate the dependence of the outflow properties on the
temperature increase $\Delta T$, we show in Figs.~\ref{fig:g10outflow}
(dwarf galaxy) and \ref{fig:g12outflow} (massive galaxy) the mass
outflow rate (left column) and the average outflow velocity
(right column). The top row shows the evolution of the wind at radius $r=0.2r_{\rm vir}$ and the bottom row shows the dependence on radius at time $t=500~\Myr$.

We first consider the dwarf galaxy (Fig.~\ref{fig:g10outflow}). Apart from the peak velocity, which is reached after only a few tens of Myr, all models look very similar. Apparently, the properties of the wind are nearly independent of the temperature increase $\Delta T$. As discussed in the previous section, this is expected for such a high resolution and for a primordial composition. However, for solar abundances the equations derived in Section~\ref{sec:coolexp} predict that radiative losses do become important for $\Delta T = 10^{6.5}$~K. Indeed, we find that for this heating temperature the mass outflow rate is reduced by more than an order of magnitude if we assume the metallicity to be solar (not shown). 

The wind is highly mass-loaded, with mass outflow rates at $0.2 r_{\rm vir}$ that are a factor of $10-100$ higher than the SFR (the SFR of the fiducial model is shown as the black, dotted curve in the top-left panel). The mass flux builds up quickly in the first 100~Myr and declines gradually thereafter. 

The top-right panel shows that the wind velocity peaks at different values at the beginning of the simulations, but converges to similar values after $\sim 50~\Myr$. The peak outflow velocity depends strongly on
the temperature increase, with the smallest (largest) value of $\sim
400~\kms$ ($\sim 600~\kms$) corresponding to the smallest (largest)
$\Delta T$. The peak velocity is a measure of the
average velocity of gas particles that are able to move freely to large
radii. Because our simulations start without a gaseous halo, it is not clear how meaningful the early evolution is, given that the high wind velocities would have resulted in strong shocks if a gaseous halo had been present. However, because the winds fill the haloes with gas, the results quickly become insensitive to the artificial initial conditions. After about 100~Myr the wind velocities of all models converge at about $100~\kms$ after which they decline gently to several tens of kilometres per seconds.

The convergence in the SFR, mass outflow rate and outflow velocity for different values of $\Delta T$
indicates that similar amounts of star-forming gas has been extracted from the
disc over time. The plots in the bottom row of
Fig.~\ref{fig:g10outflow} confirm that. At each radius the outflow rates and
velocities are similar for all models, and the gas can
efficiently escape into the halo, and eventually beyond the virial
radius. 

As was already noted by DS08, the fact that the wind velocity becomes proportional to the radius as we move away from the galaxy (bottom-right panel) is due to travel time effects. Because the wind has only been blowing for a finite amount of time $t$, only gas with a velocity greater than
\begin{equation}
v = 98~ {\rm km}\,{\rm s}^{-1}~ \left(\frac{r}{10~{\rm kpc}}\right) \left (\frac{t}{10^8~{\rm yr}}\right )^{-1},
\end{equation}
would be present at radius $r$ if the wind velocity is constant with radius. Although the wind may in reality accelerate or decelerate, our results suggest that travel time effects may well be important too. Hence, wind velocities that are observed to increase with distance do not necessarily imply that the wind is accelerating. 

Fig.~\ref{fig:g12outflow} shows that the picture is different for the massive galaxy. The initial wind velocity is sensitive to the heating temperature and determines the time at which the outflow first passes through the shell at $r=0.2r_{\rm vir}$ (corresponding to the sharp rise in the top panels). For $\Delta T \ge 10^{7.5}$~K the mass outflow rates converge at about five times the SFR (the SFR of model \textit{G12-100-75} is indicated by the dotted curve in the top-left panel), so the winds are much less mass-loaded than for the dwarf galaxy. For lower heating temperatures the mass flux is substantially lower. The outflow rate of model \textit{G12-100-65} even becomes
negative after about 320~Myr, indicating net infall (dotted curve in
the top-left panel). The wind cannot escape the inner region of the
halo, and is confined within a fraction of the virial radius. This model therefore also predicts a much higher SFR (Fig.~\ref{fig:sfr}). The fact that the mass flux drops strongly for $\Delta T < 10^{7.5}$~K can be understood by noting that the gas in the central regions and spiral arms has densities $n_{\rm H} \sim 1-10~{\rm cm}^{-3}$ (Fig.~\ref{fig:g12dens_comp}) and that equation~(\ref{eq:nhcrit}) shows that cooling losses should make the feedback inefficient for densities $n_{\rm H} < 31 ~{\rm cm}^{-3} ~ (T/10^{7.5})^{3/2}$.

The wind velocities are a factor of a few higher than for the low-mass galaxy and, at least for $\Delta T \ge 10^{7.5}$~K, depend strongly on the temperature increase. Figure~\ref{fig:g12dens_comp} showed that for these high heating temperatures, the wind blows channels through which it can freely escape to very large distances. The effect of drag by halo gas is therefore smaller than for the dwarf galaxy and for the low-$\Delta T$ versions of the massive galaxy, which all predict much puffier gas disks. Hence, in the high-$\Delta T$ simulations of the massive galaxy the outflow speed is predominantly regulated by gravity (as opposed to gas drag) and by the initial velocity. While the potential is always the same, the initial wind velocity is set by the heating temperature. 

The radial dependence of the outflow is shown in the bottom
row of Fig.~\ref{fig:g12outflow}. Interestingly, the models that predict similar SF histories, also predict similar outflow rates for $r<0.2r_{\rm vir}$. However, at larger radii the outflow rates diverge, with higher heating temperatures giving larger mass fluxes. Except for model \textit{G12-100-65}, the velocities increase linearly with radius beyond about $0.5~r_{\rm vir}$, suggesting that they are determined by travel time constraints, as was the case for the dwarf galaxy. 

\begin{figure*}
\includegraphics[width=0.45\textwidth]{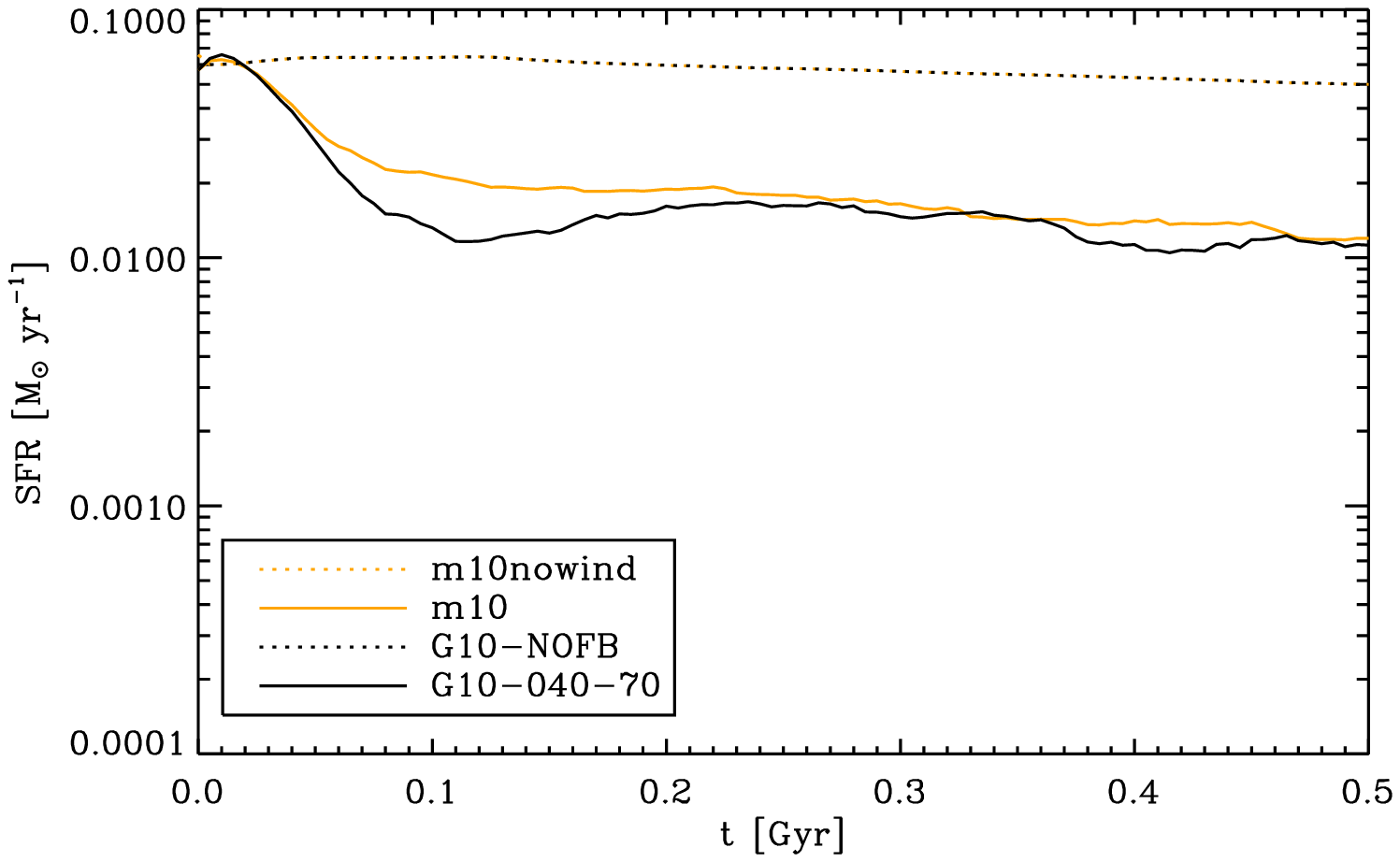}%
\includegraphics[width=0.45\textwidth]{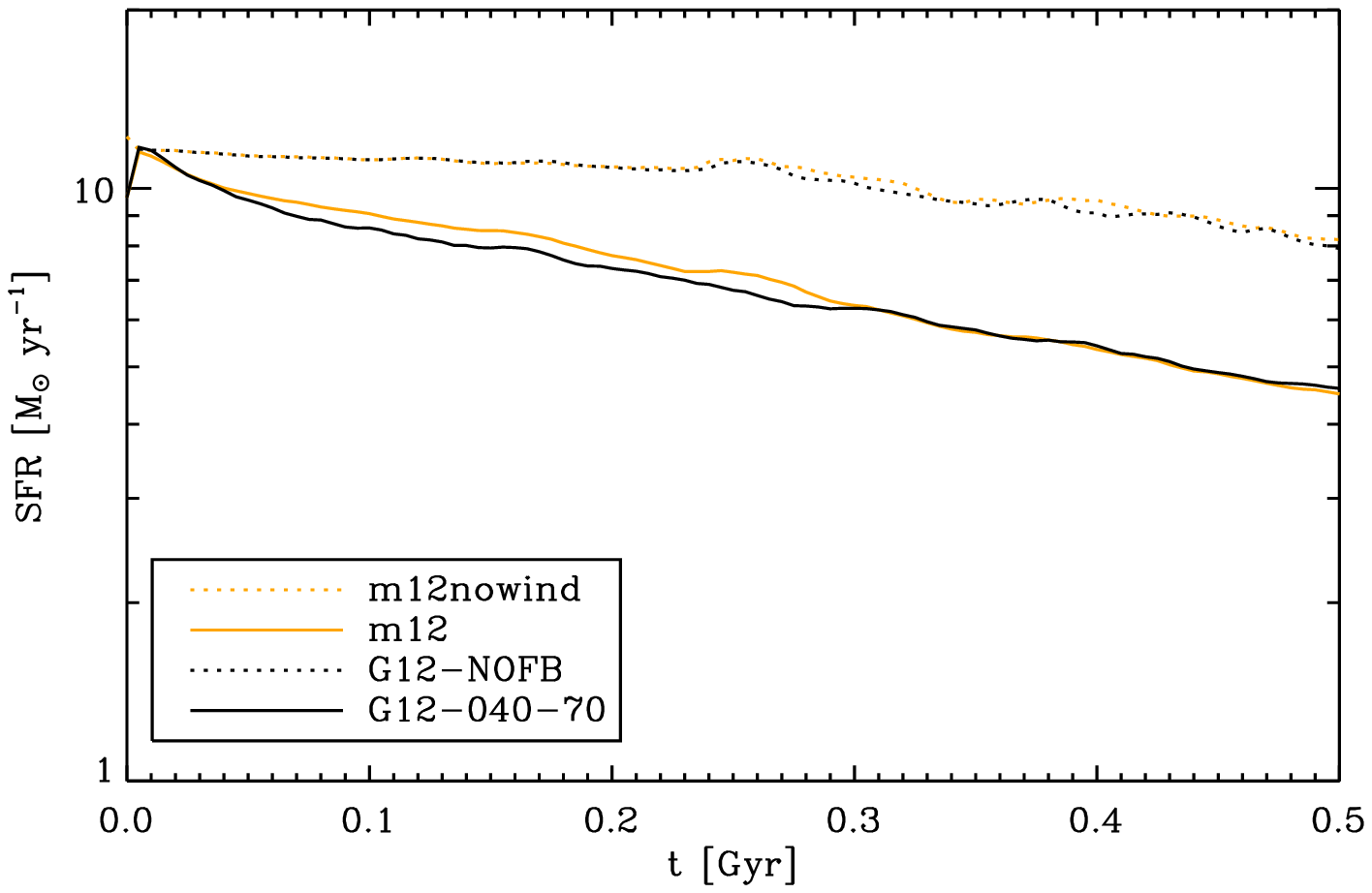}
\caption{Comparison of the SF histories in simulations employing thermal (\textit{G10-040-70} and \textit{G12-040-70}) and kinetic (\textit{m10} and \textit{m12}) feedback. All models inject 40 percent of the SNII energy. The kinetic feedback assumes an initial wind velocity of $600~\kms$ and the thermal feedback a temperature increase of $10^7$~K, which is close to the post-shock temperature for a shock velocity of $600~\kms$. The left and right panels show the SFR as a function of time for the $10^{10}$ and $10^{12}\Msolh$ haloes, respectively. The very small difference between the no-feedback models (that are only noticeable for the massive galaxy) are due to the different treatment of star-forming gas (see section~\ref{sec:code}). The two feedback implementations result in very similar SF histories.}
\label{fig:sfr_comp}
\end{figure*}

\begin{figure*}
\includegraphics[width=0.45\textwidth]{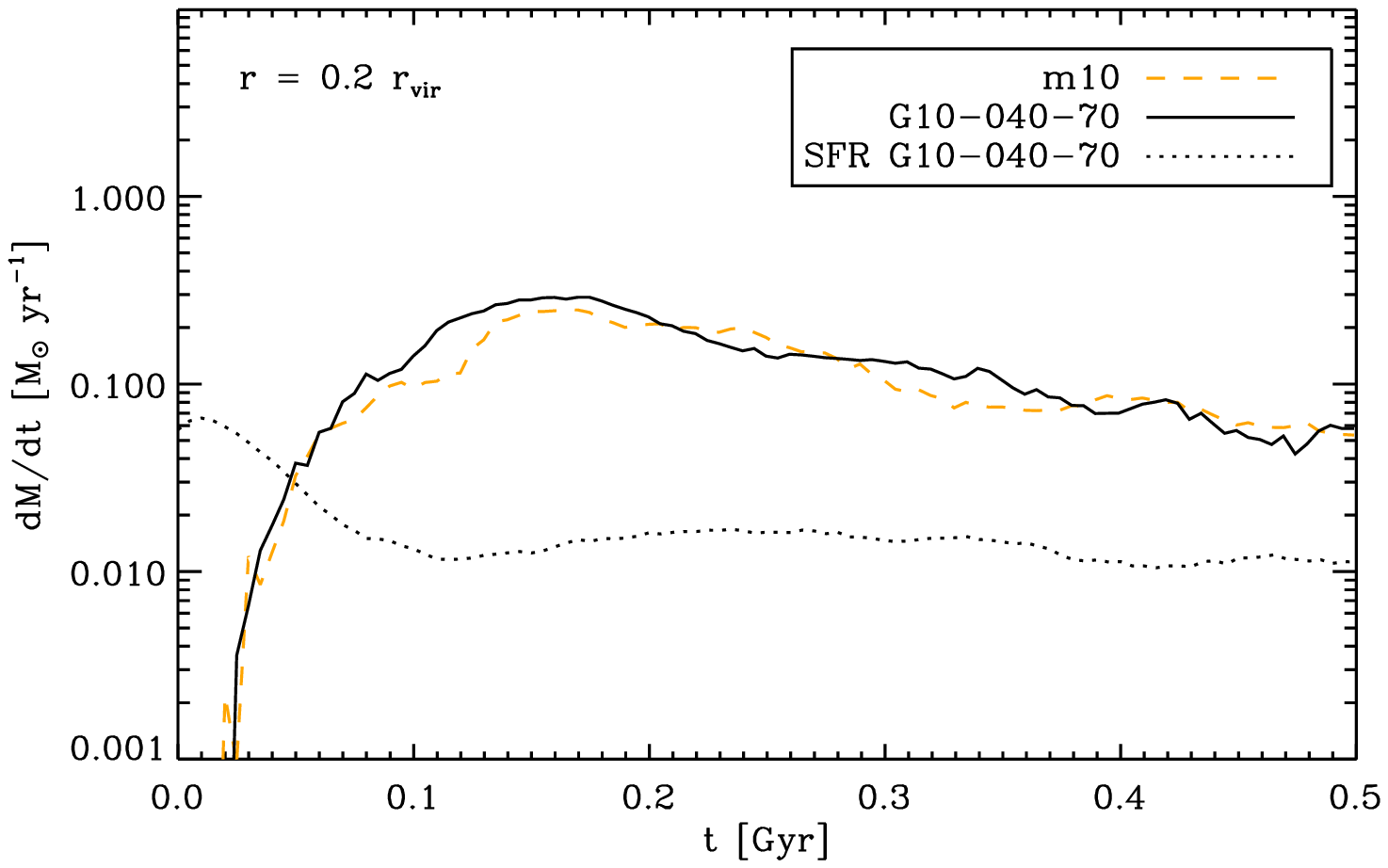}%
\includegraphics[width=0.45\textwidth]{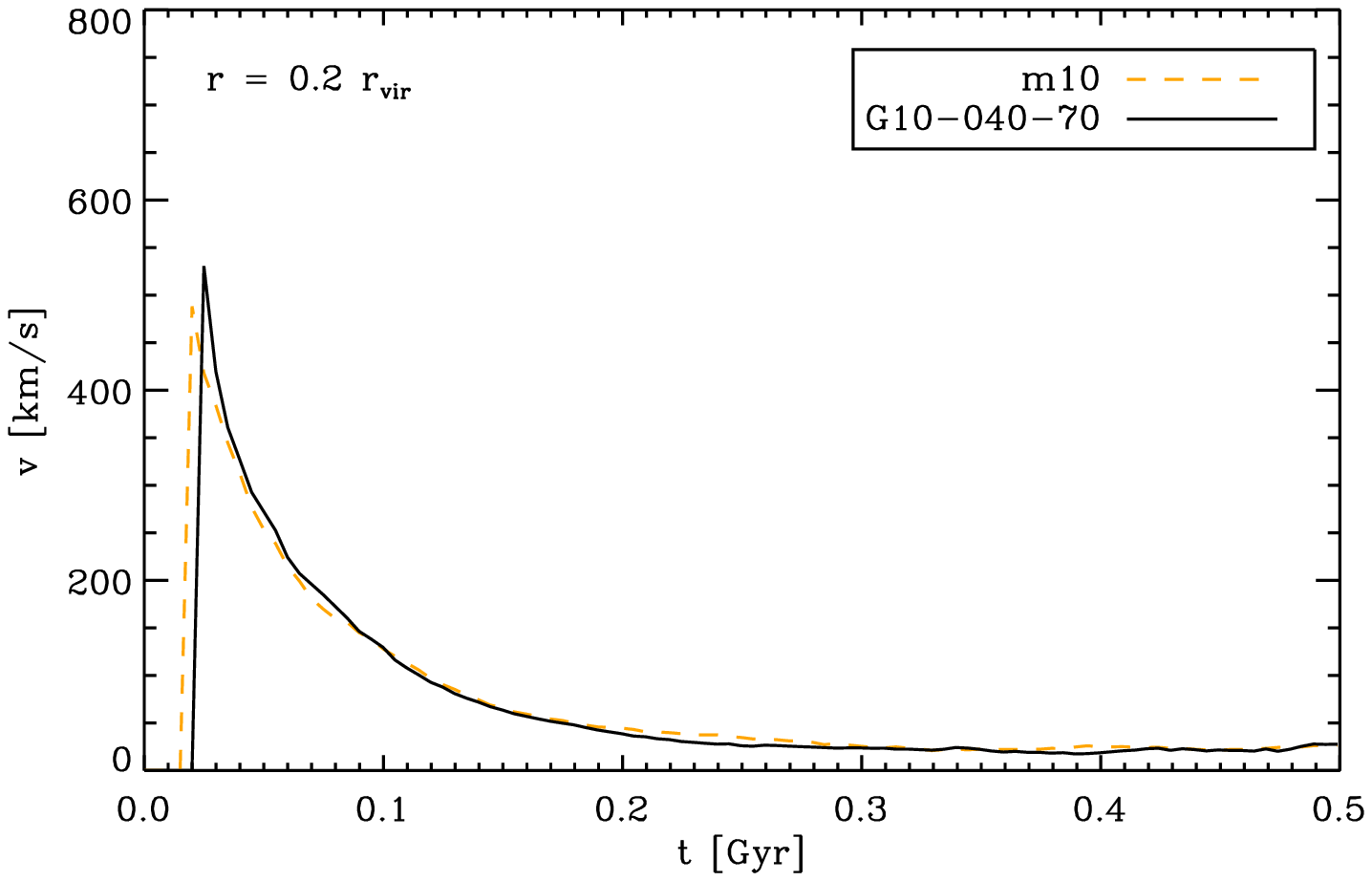}\\
\includegraphics[width=0.45\textwidth]{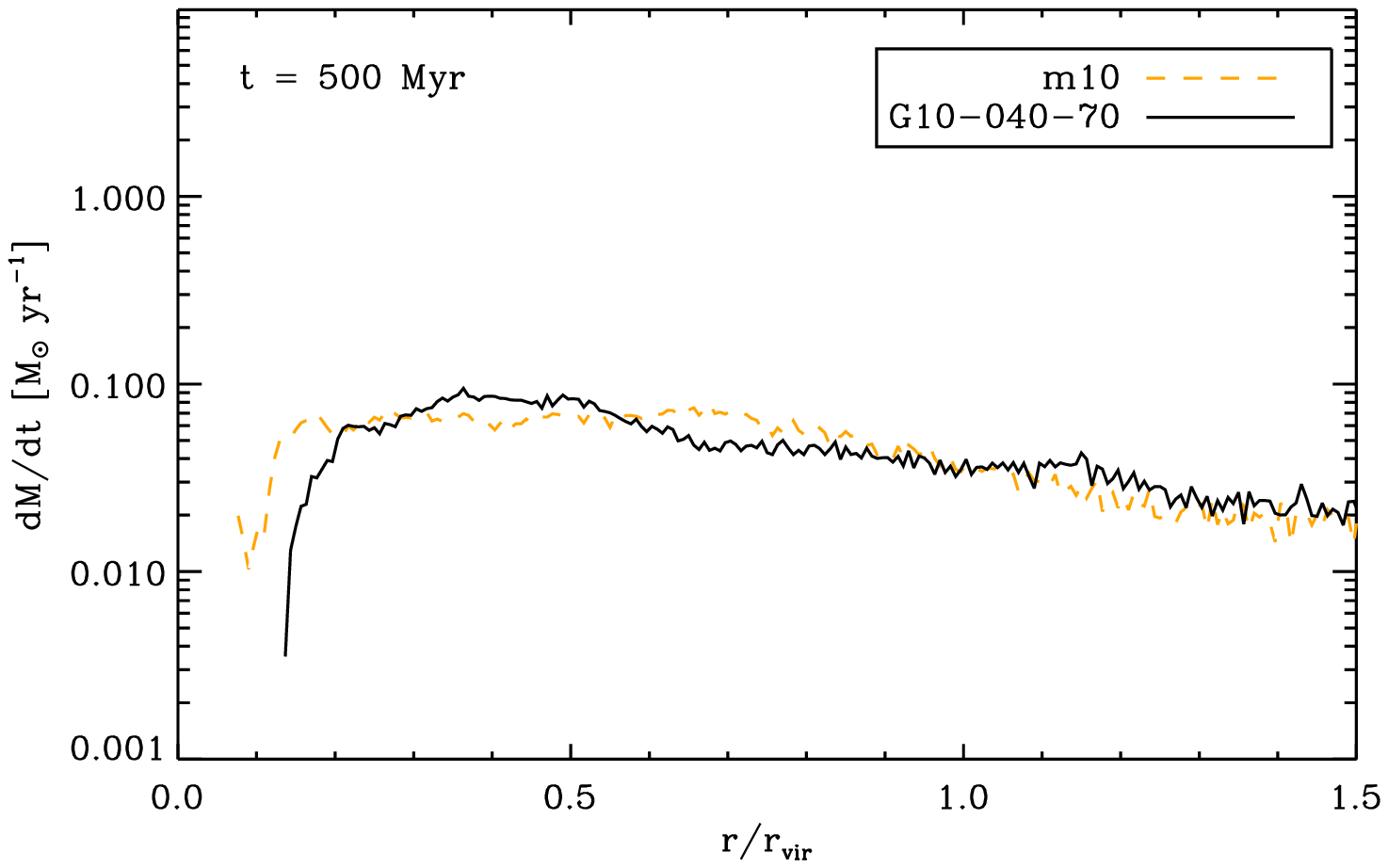}%
\includegraphics[width=0.45\textwidth]{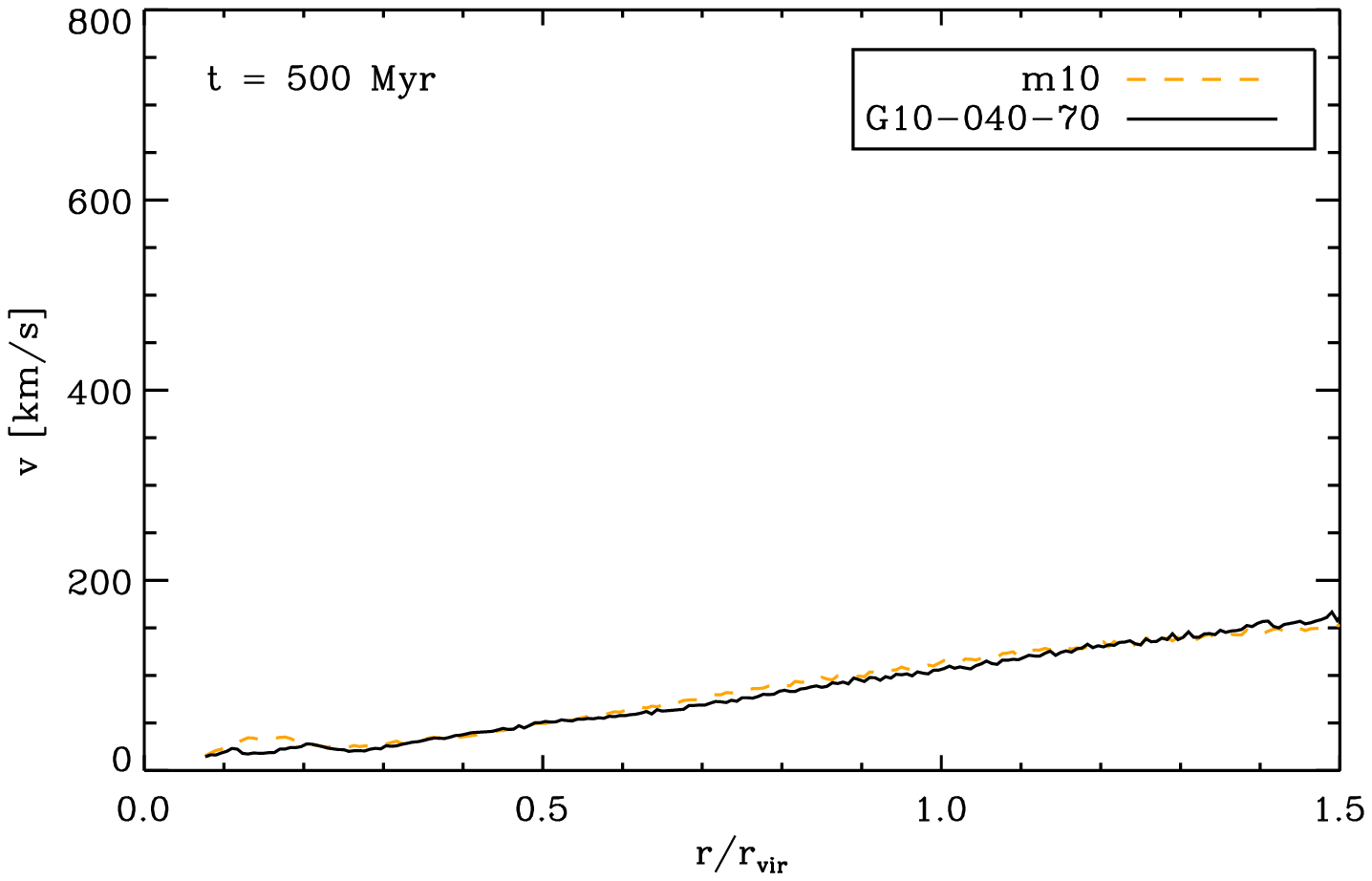}
\caption{Comparison of the mass outflow rate (left column) and average outflow
  velocity (right column) measured through a spherical shell at
  radius $r=0.2r_{\rm vir}$ as a function of time (top row) and at
  $t=500~\Myr$ as a function of radius (bottom row) between 
  model \textit{G10-040-70}, which employs thermal feedback, and model \textit{m10}, which uses kinetic feedback. The dotted curve in the top-left panel indicates the
  SFR of model \textit{G10-040-70}. All other curves are
  labelled in the legends. The agreement between the outflows predicted by simulations that inject the SNII energy in thermal and kinetic form is generally excellent.}
\label{fig:g10outflow_comp}
\end{figure*}

\begin{figure*}
\includegraphics[width=0.45\textwidth]{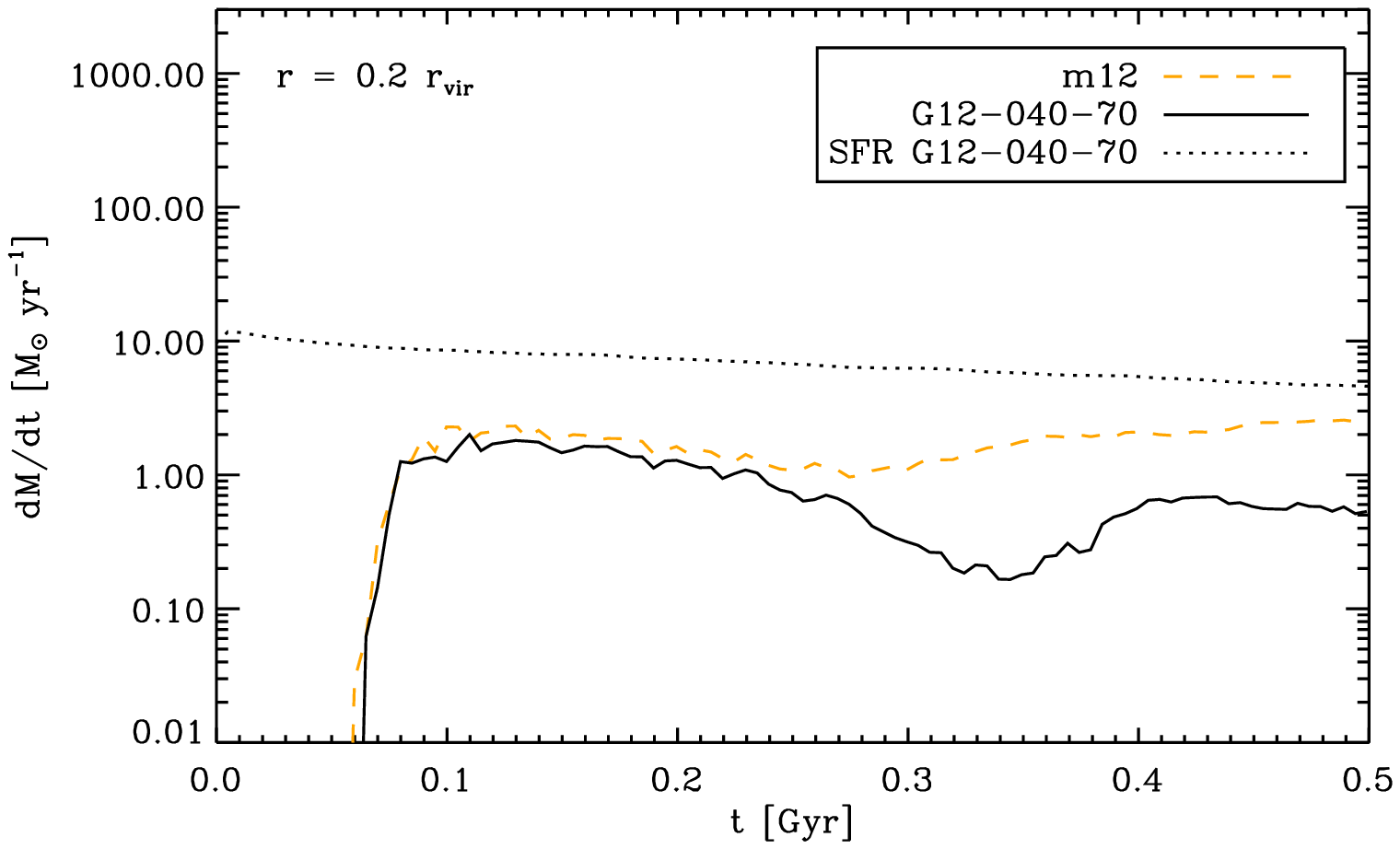}%
\includegraphics[width=0.45\textwidth]{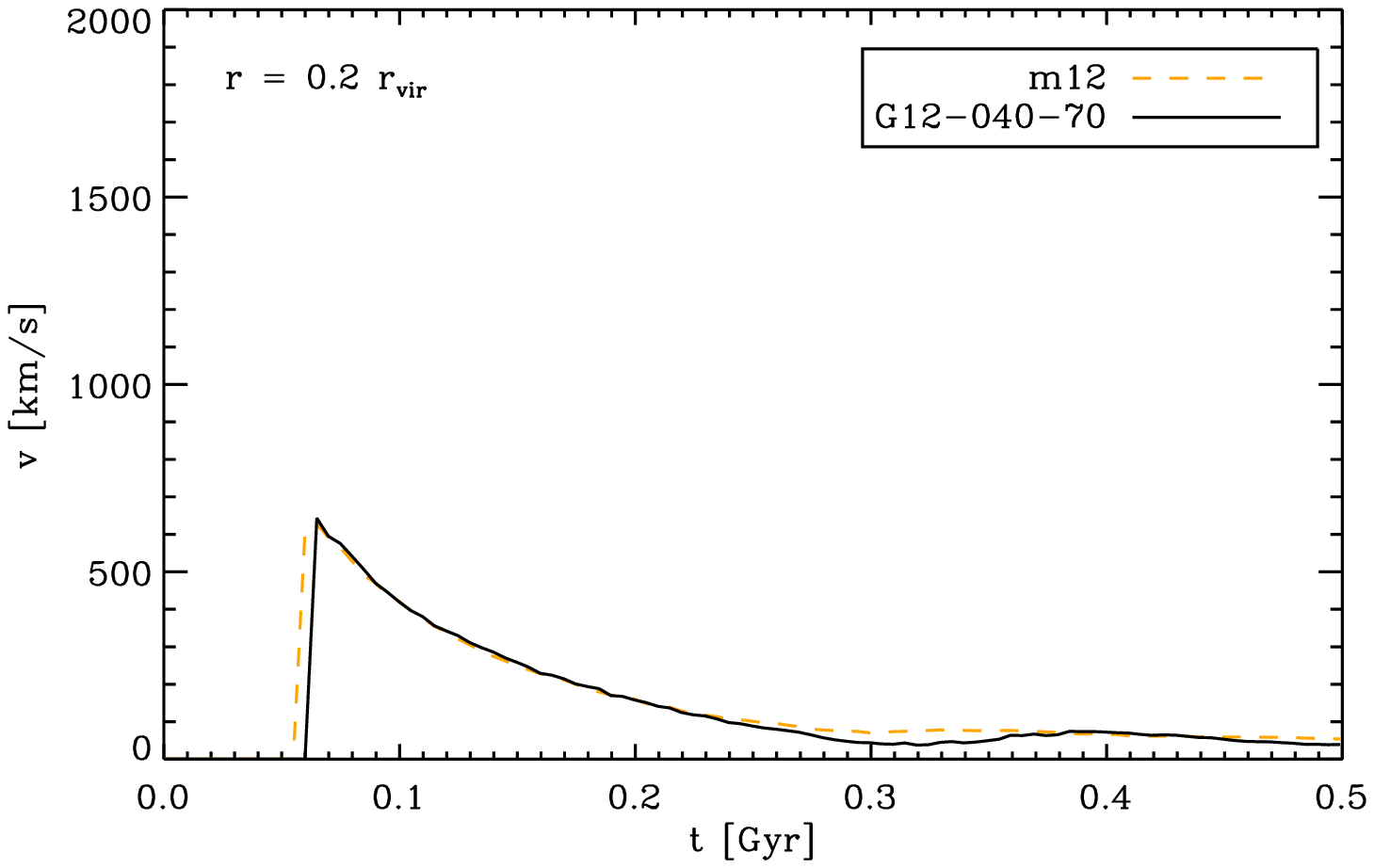}\\
\includegraphics[width=0.45\textwidth]{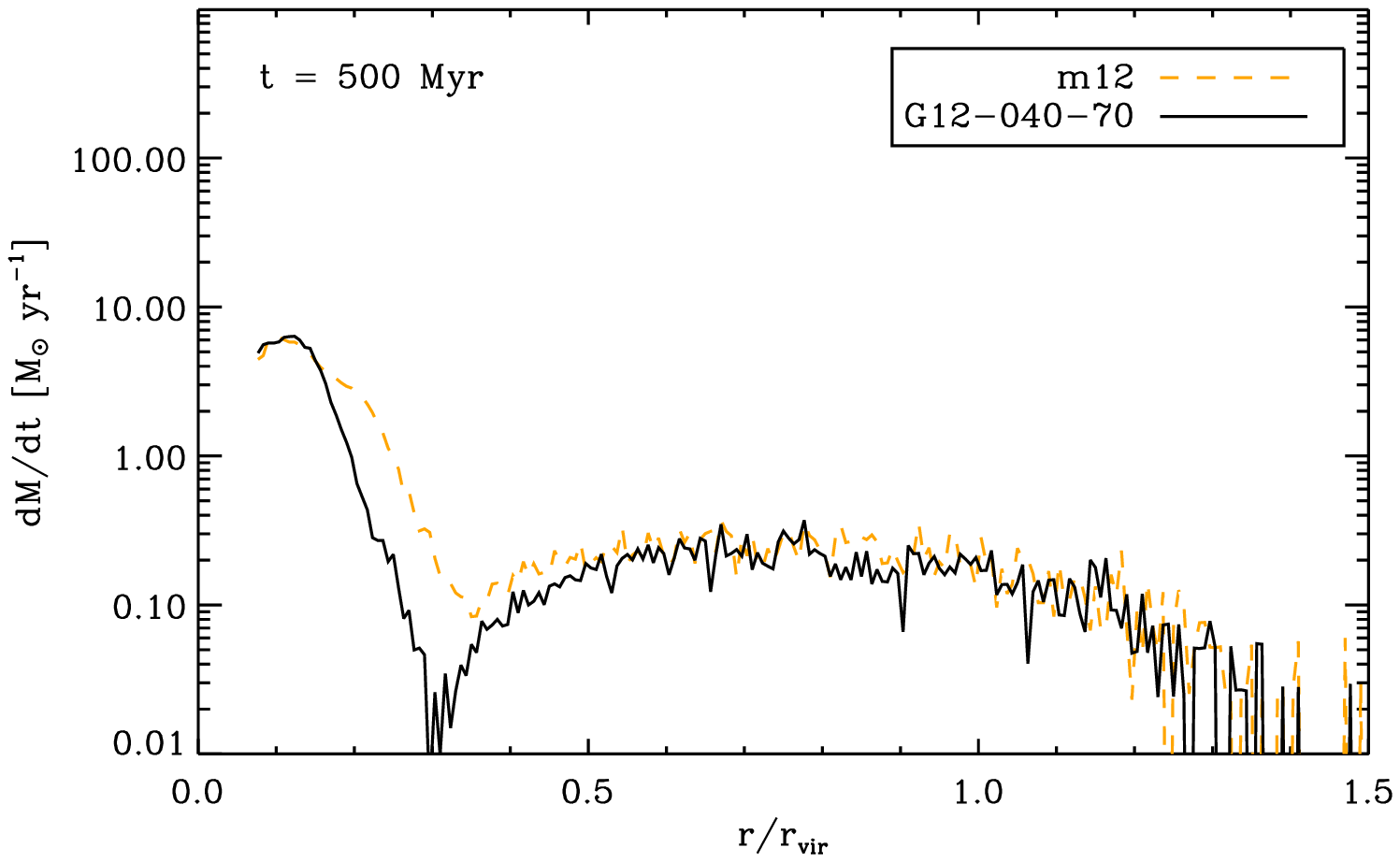}%
\includegraphics[width=0.45\textwidth]{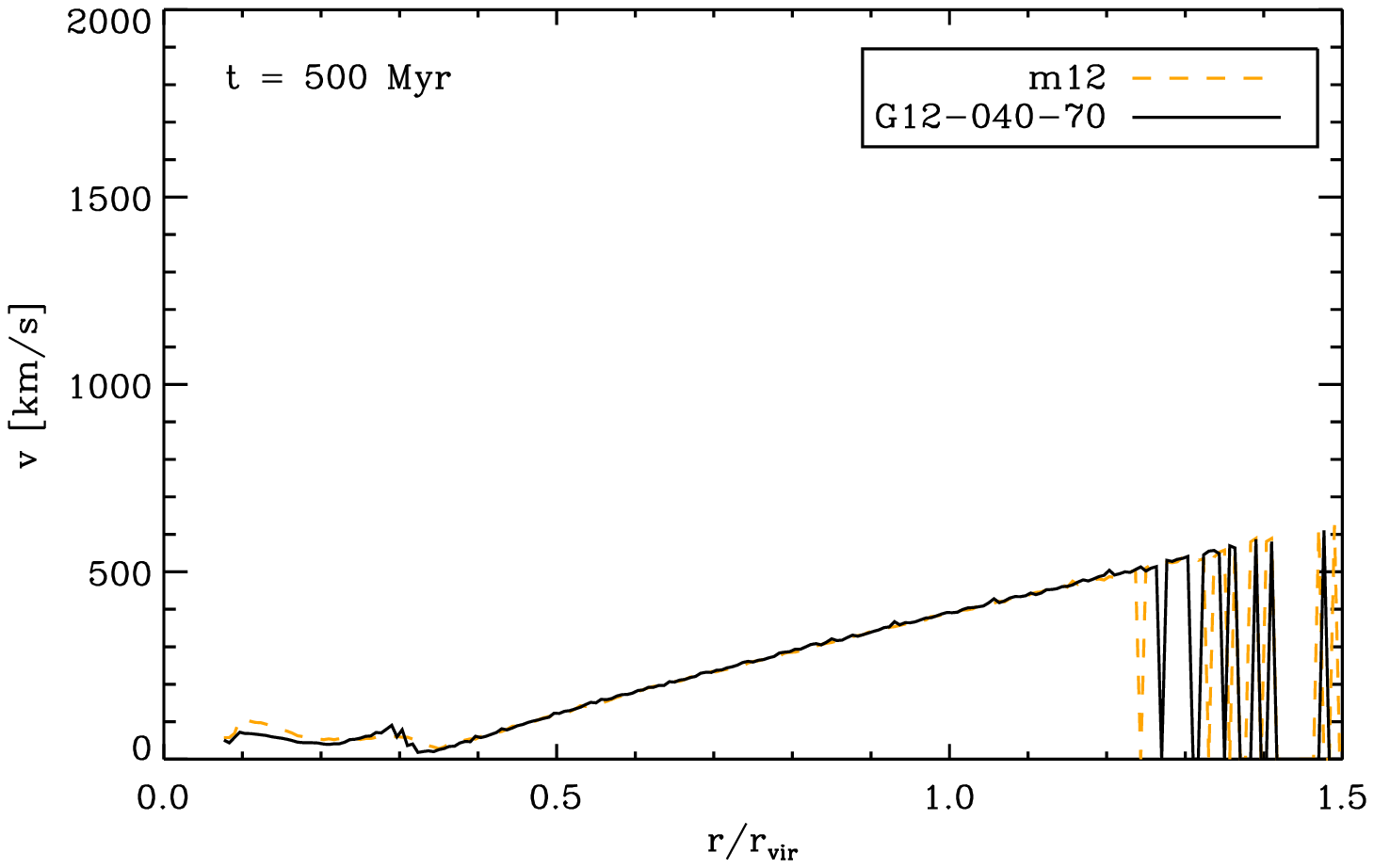}
\caption{As Fig.~\ref{fig:g10outflow_comp}, but for the $10^{12}\Msolh$ halo. The agreement between the thermal and kinetic implementations of SNII feedback is very good for the wind velocities, but the mass outflow rates differ substantially at late times and intermediate radii. This probably reflects the fact that these wind parameter values yield results that are intermediate between the galactic fountain and efficient, large-scale winds regimes, making the outcome very sensitive to the exact values of the wind parameters.}
\label{fig:g12outflow_comp}
\end{figure*}

\subsection{Comparison with the kinetic feedback model}
\label{sec:kin}

In this section we will compare the results of simulations using our new thermal feedback prescription with runs employing the kinetic feedback of DS08. As discussed in detail in DS08, the kinetic feedback recipe works as follows. Once a star particle reaches an age of $3\times 10^7$~yr, its neighbouring gas particles each have a probability of $\eta m_\ast/\Sigma_{i=1}^{N_{\rm ngb}}m_i$ of receiving a randomly oriented kick of velocity $v_{\rm w}$. For the case of equal mass particles, the mass loading factor $\eta$ equals the average number of particles kicked per star particle. We use DS08's fiducial values of $v_{\rm w} = 600~\kms$ and $\eta = 2$, which correspond to 40 percent of the available energy. 

We compare the kinetic feedback runs to thermal feedback simulations that use the same fraction of the available SN energy (i.e.\ $f_{\rm th} = 0.4$ as opposed to 1.0 for our fiducial model). We use a temperature increase of $\Delta T = 10^7$~K (as opposed to $10^{7.5}$~K for our fiducial model) as this is close to the post-shock temperature for a shock velocity of $600~\kms$. 

To eliminate potential differences other than the feedback recipe, we re-ran models \textit{m12} and \textit{m12nowind} of DS08 with the
new code and employing the same softening lengths as used here. 
 
\subsubsection{Star formation history}

Fig.~\ref{fig:sfr_comp} compares the SF histories of the thermal feedback runs \textit{G10-040-70} and
\textit{G12-040-70} with the equivalent kinetic feedback runs \textit{m10} and \textit{m12} of DS08. 

We first verify that the new criterion for identifying star-forming gas (see Section~\ref{sec:code}) gives results that are consistent with the implementation used in DS08. In order to eliminate differences due to the feedback implementation, we compare runs without feedback. Comparison of the black and orange dotted curves in Fig.~\ref{fig:sfr_comp}, which show the SF histories predicted with the new and old prescriptions, respectively, shows that differences due the slight change in the recipe for SF are negligible.

Comparing the orange and black solid curves, which show the SF histories in the runs with kinetic and thermal feedback, respectively, we see that the two methods for injecting the energy from SNII are generally in good agreement.

\subsubsection{Mass outflow rate and wind velocity}

Fig.~\ref{fig:g10outflow_comp} shows the mass outflow rate (left
column) and the mean outflow velocity (right column) for models
\textit{G10-040-70} and \textit{m10}. The top row shows the evolution measured at $r=0.2 r_{\rm vir}$, while the bottom row
illustrates the dependence on radius at time $t=500~\Myr$.
The agreement is generally excellent. This implies that, at the resolution used for the dwarf galaxy, the average quantities of the outflow are insensitive to the form in which the energy is injected. Note that this is not a consequence of a fortunate choice of parameters, as we showed in
Section~\ref{sec:winds} that the outflow is insensitive to the temperature increase $\Delta T$. The agreement between simulations injecting kinetic and thermal energy is consistent with the results of \citet{Durier2011}. 

Fig.~\ref{fig:g12outflow_comp} shows the same plots for the
$10^{12}\Msolh$ halo. While the agreement for the velocities is again very good, there are in this case some noticeable
differences between the mass outflow rates. Compared with the thermal feedback simulation, the kinetic model predicts a higher outflow rate at radii $0.2 \lesssim r/r_{\rm vir} \lesssim 0.4$ at late times. The differences may imply that the resolution is too low to achieve convergence between kinetic and thermal feedback prescription (recall that the mass resolution is two orders of magnitude lower than for the dwarf galaxy). Another reason, is however, that, unlike for the dwarf galaxy, for the massive galaxy the outflow does depend on the wind parameters. Moreover, for this massive halo $\Delta T = 10^7$~K (or $v_{\rm w} = 600~\kms$) marks the transition between the regimes of galactic fountains (lower $\Delta T$ or $v_{\rm w}$) and efficient large-scale winds (higher $\Delta T$ or $v_{\rm w}$), as can be seen from Figs.~\ref{fig:sfr} and \ref{fig:g12outflow}. Hence, the results are very sensitive to small differences in the input parameters and we could have obtained better agreement by fine-tuning the value of $\Delta T$ (or $v_{\rm w}$). 

\subsection{Resolution tests}
\label{sec:resol}

\begin{figure*}
\includegraphics[width=0.45\textwidth]{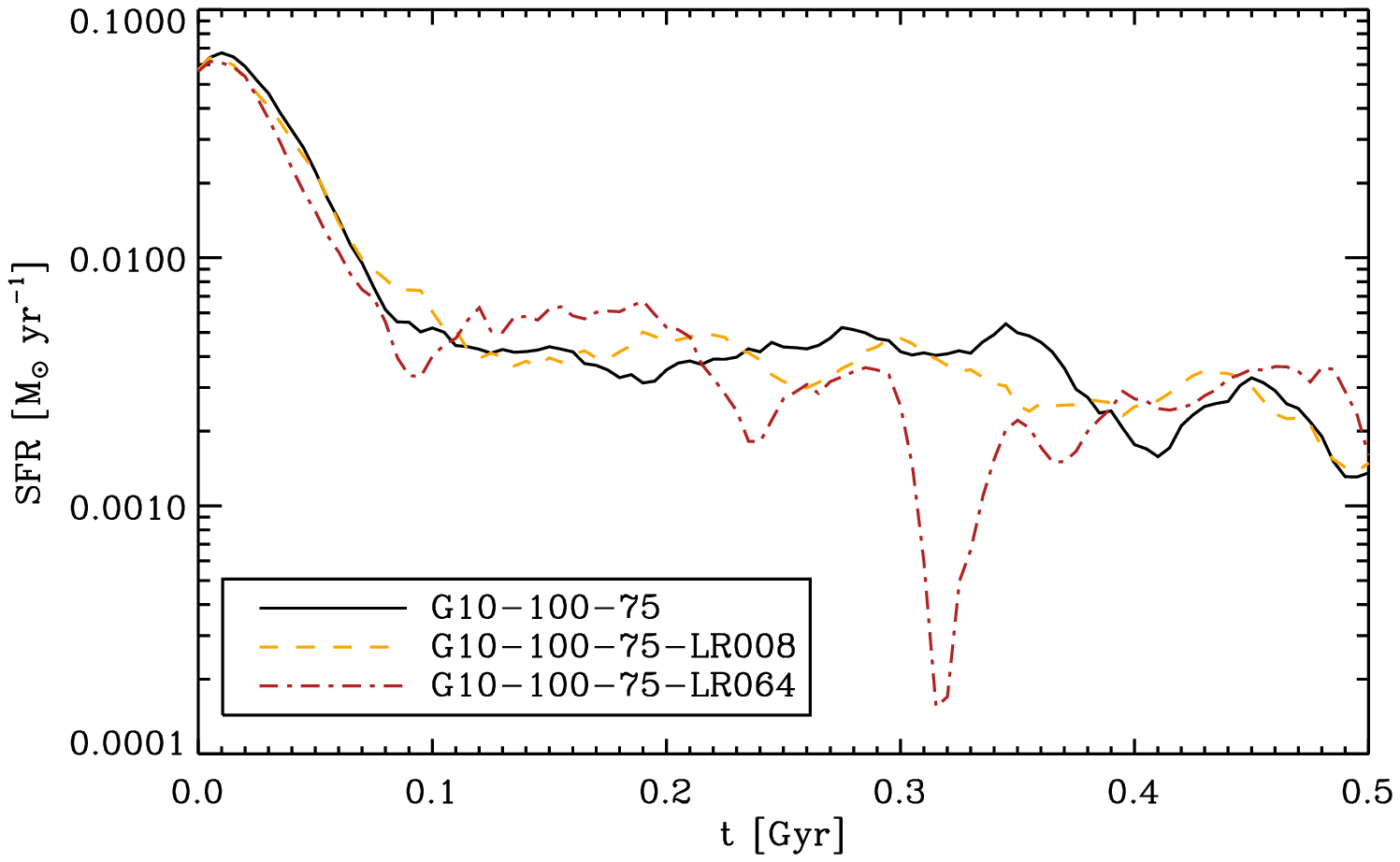}%
\includegraphics[width=0.45\textwidth]{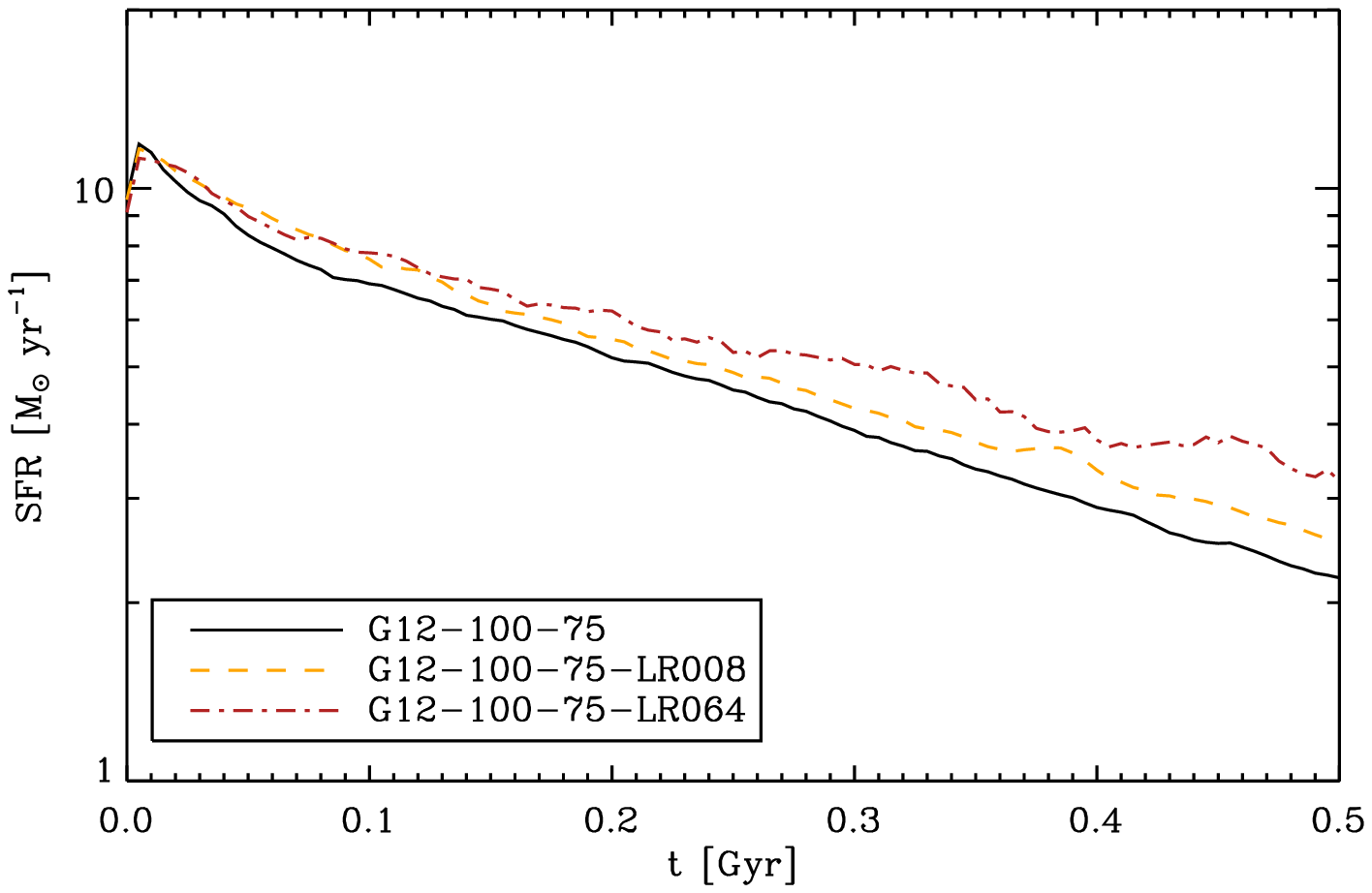}
\caption{Numerical convergence of the SF histories of the $10^{10}\Msolh$ and $10^{12}\Msolh$ haloes (left- and right
panels, respectively). The mass resolution is decreased by factors of
8 (dashed curves) and 64 (dash-dotted curves). The convergence is
good, although there is a small, systematic increase of the SFR with decreasing resolution for the massive galaxy.}
\label{fig:resolution-sfr}
\end{figure*}

\begin{figure*}
\includegraphics[width=0.45\textwidth]{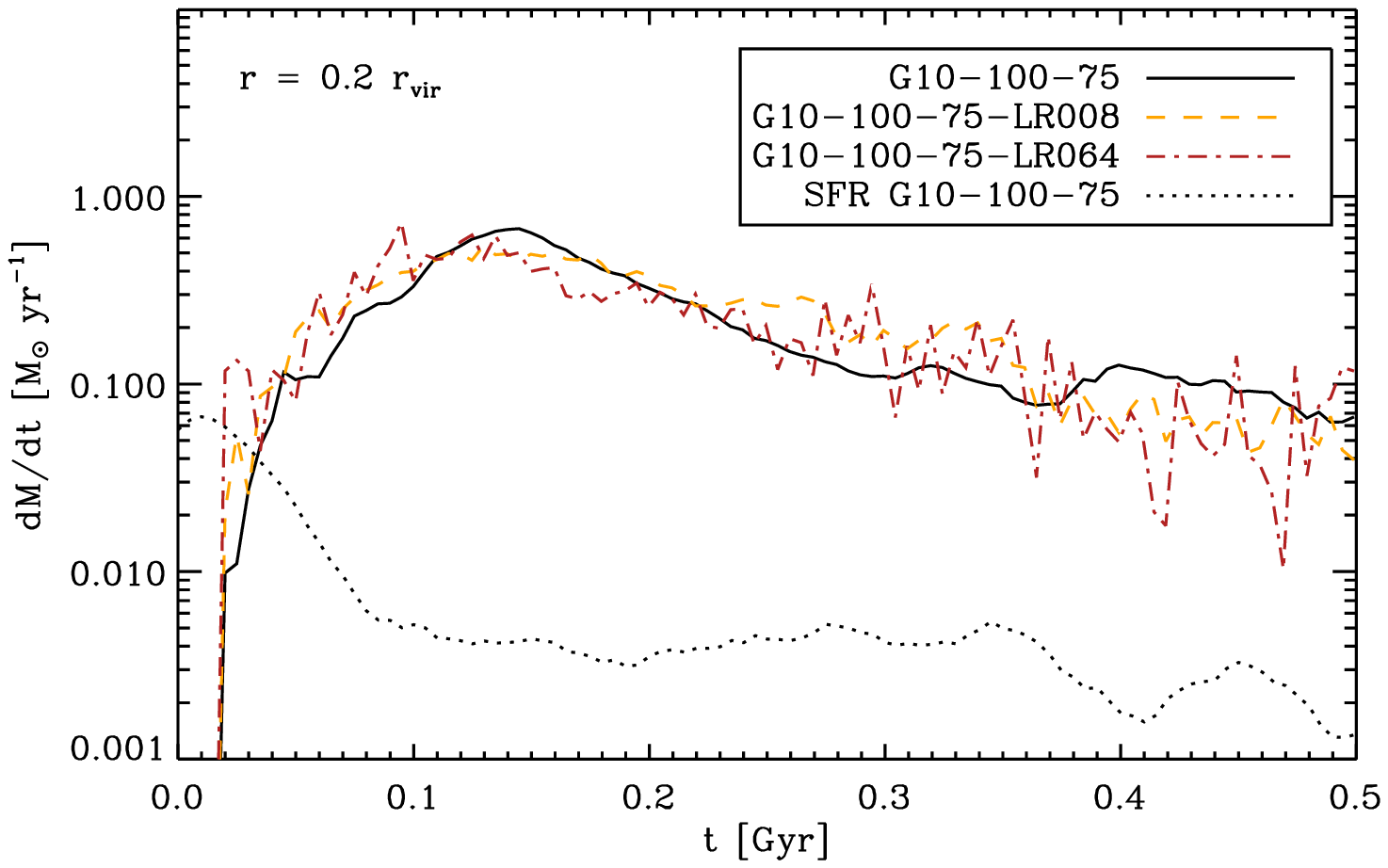}%
\includegraphics[width=0.45\textwidth]{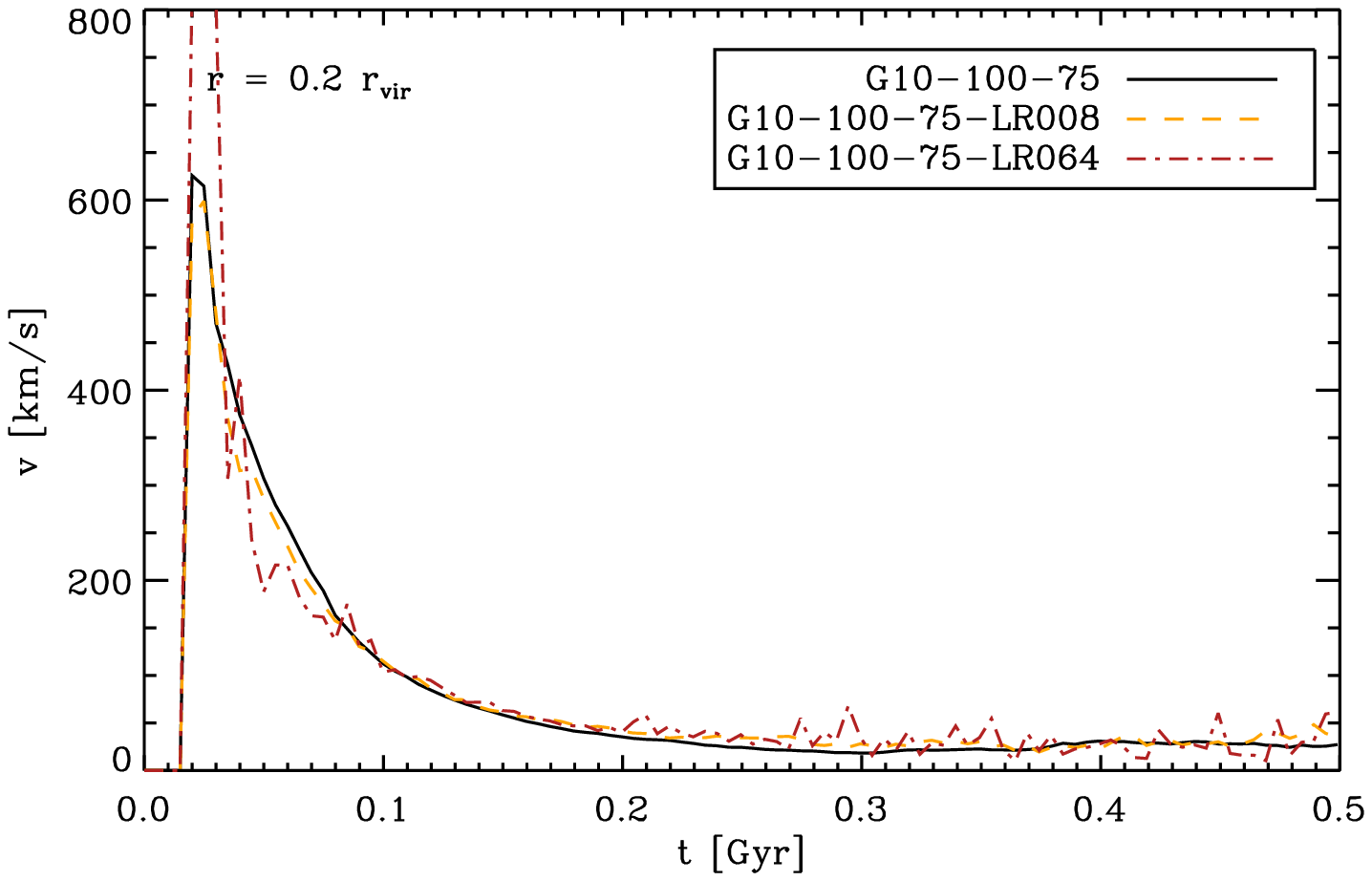}\\
\includegraphics[width=0.45\textwidth]{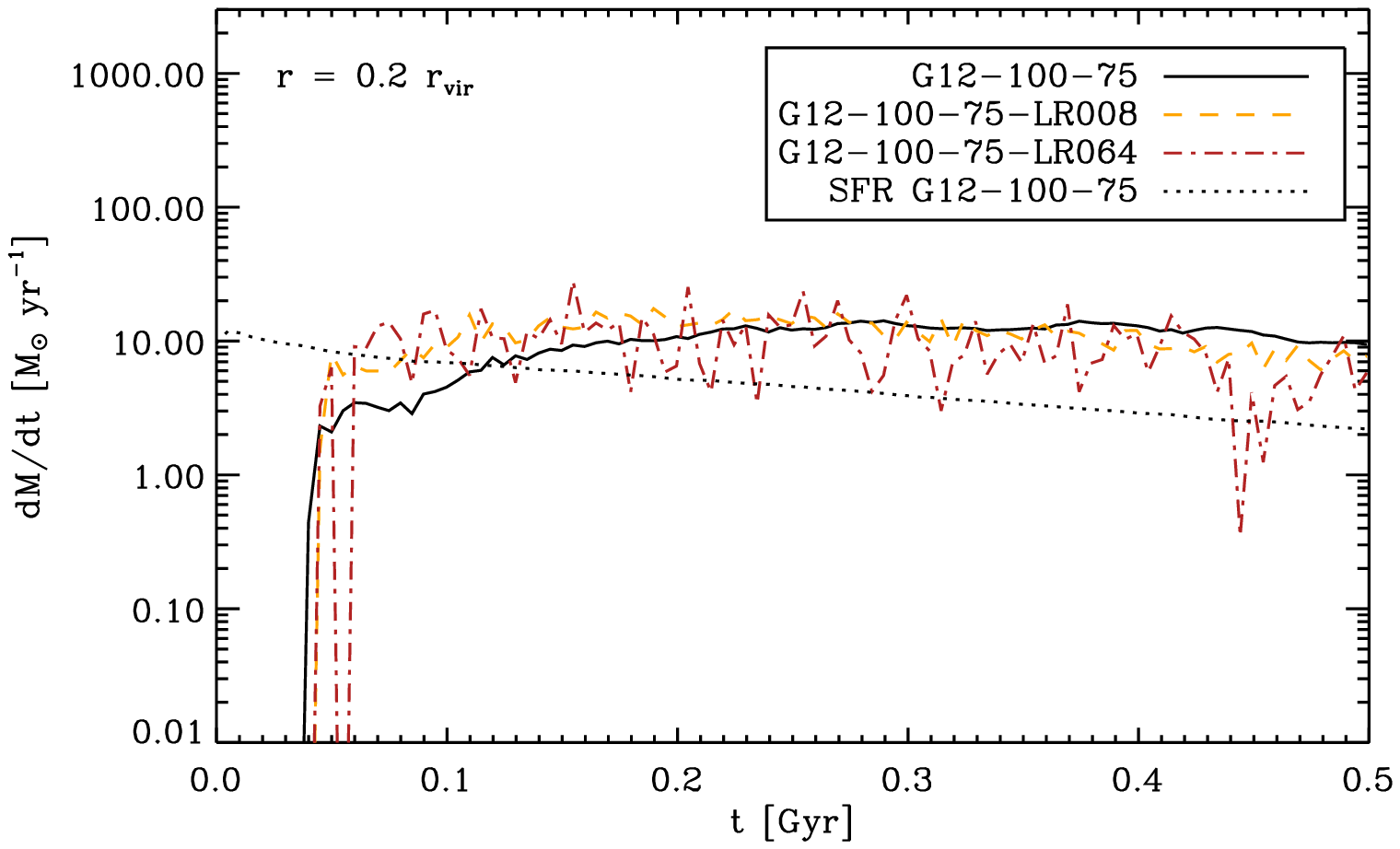}%
\includegraphics[width=0.45\textwidth]{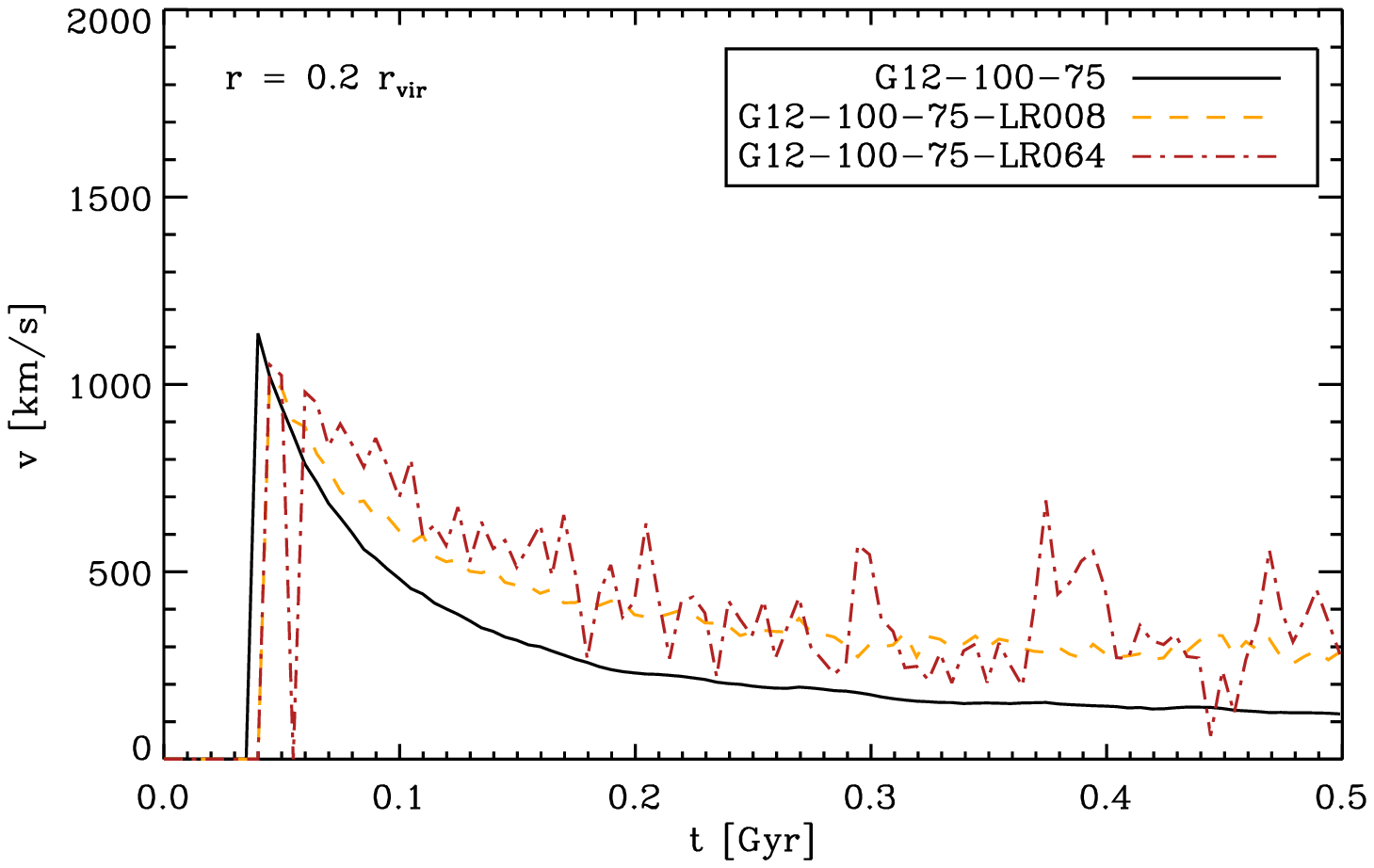}
\caption{Resolution dependence of the mass outflow rate (left
column) and mean outflow velocity (right column) measured through
a spherical shell at radius $r=0.2r_{\rm vir}$ as a function of time
for the $10^{10}$ (top row) and the $10^{12}\Msolh$ (bottom row)
haloes. Note that
the particle mass in \textit{G10-100-75-LR064} is still lower than
that in \textit{G12-100-75}. While the predictions for the low-mass galaxy are converged, decreasing the resolution yields mostly higher outflow
velocities for the high-mass galaxy, but has little effect on the mass outflow rates.}
\label{fig:outflow_res}
\end{figure*}

We tested the numerical convergence of our implementation of thermal feedback by decreasing the particle numbers by factors of 8 and 64. We will denote the corresponding runs by appending `LR008' resp.\ `LR064' to the simulation names. Hence, the particle masses are increased by factors of 8 and 64, while the gravitational softening lengths and, for a fixed density, the SPH smoothing kernels are increased by factors of 2 and 4, respectively. Before showing the results, it is useful to consider what we may expect. To do so, we have to check whether the simulations resolve the Jeans scales and whether we expect radiative cooling losses to be significant.

As discussed in \citet{Schaye2008}, for star-forming gas in our fiducial simulation of the massive halo the ratio of the SPH kernel mass to the Jeans mass is smaller than $1/6$ and the ratio of the SPH kernel size to the Jeans length is at most $1/(48)^{1/3}\approx 0.28$. For the low-mass halo the mass and length ratios are lower by factors of 100 and $100^{1/3}\approx 4.64$, respectively. Note that the maximum possible values of these ratios are independent of the density because star-forming particles cannot have temperatures below a power-law effective equation of state with polytropic index $\gamma_{\rm eff} = 4/3$. Thus, while even our lowest-resolution simulation of the dwarf galaxy resolves the Jeans scales, the same is only true for our highest-resolution simulation of the massive galaxy.

In Section~\ref{sec:coolexp} we demonstrated that, for $\Delta T = 10^{7.5}$~K and our fiducial resolution for the massive galaxy, radiative losses should have little impact on the efficiency of the feedback if the energy is injected in gas with density $n_{\rm H} < 31~{\rm cm}^{-3}$ and that this critical density is inversely proportional to the squareroot of the particle mass (eq.~[\ref{eq:nhcrit}]). Hence, for the intermediate- and low-resolution models the densities above which radiative losses may prevent efficient feedback are about 11 and $4~{\rm cm}^{-3}$, respectively, and for the low-mass galaxy these densities are 10 times higher. Comparing this to the actual densities of the star-forming gas in the high-resolution simulations (Figs.~\ref{fig:g10_phase} and \ref{fig:g12_phase}), we see that radiative losses
may not be negligible for the lower-resolution simulations of the
high-mass galaxy, but that the feedback should remain efficient in the low-resolution models of the dwarf galaxy. 

Thus, both Jeans and radiative cooling arguments suggest that even the lowest-resolution simulation of the dwarf galaxy should give converged results. On the other hand, we do expect the lower-resolution versions of the massive galaxy to show some difference. Furthermore, given that the lowest-resolution models have fewer than 3700 gas particles in the disk, we expect the results to become noisy due to the poor sampling. 

Figs.~\ref{fig:resolution-sfr} shows how the SF history depends on the numerical resolution for the fiducial models \textit{G10-100-75} (left) and \textit{G12-100-75} (right). Similarly, Fig.~\ref{fig:outflow_res} illustrates the resolution dependence of the predicted evolution of the mass outflow rate (left column) and wind velocity (right column) for the dwarf (top row) and massive (bottom row) galaxies. Clearly, our expectations are borne out.

Whereas the dwarf galaxy runs are very well converged, for the massive galaxy the predicted SFR increases with resolution, although the effect is small. The first 250~Myr the mass outflow rate is somewhat higher in the lower resolution simulations of the massive galaxy, but the situation reverses at later times. Except for the first 10~Myr, the differences are, however, small. The wind velocity is more sensitive to the resolution and is generally higher in the lower-resolution models. The predictions for the outflow properties become noisy for the lowest resolution simulations. 

It is interesting that the convergence with numerical resolution is better than found by DS08 for the kinetic feedback versions of the same simulations (c.f.\ Figs.~10 and 11 of DS08). For example, with kinetic feedback the outflow rate for the massive galaxy increased by a factor of about 3 as the mass resolution was decreased by a factor of 8 and the SFR of the dwarf galaxy increased substantially when the particle mass was decreased by a factor of 64. It should be kept in mind, however, that this could in part be due to the fact that the fiducial wind velocity used by DS08 ($600~\kms$) corresponds to post-shock temperature jumps that are a factor of a few lower than our fiducial value of $\Delta T = 10^{7.5}$~K. 


\section{Discussion}
\label{sec:disc}

Models of galaxy formation and evolution require feedback from star
formation to reproduce the observed properties of galaxies. Cosmological simulations do not have sufficient resolution to resolve individual SN explosions and must therefore resort to sub-grid recipes. The simplest method, injecting the SN energy released by a star particle (i.e.\ a simple stellar
population; SSP) during each time step into its surroundings, does not lead to efficient feedback because the injected thermal energy is quickly radiated away. Successful recipes generally either inject the energy in kinetic form, turn off radiative cooling temporarily, or inject the energy in a hot sub-grid phase that is decoupled from the colder phases by hand.

We demonstrated that the catastrophic radiative losses suffered by simple thermal feedback recipes are due to a mismatch between the gas mass in which the energy is injected and the mass of the SSP that produced the energy. In the real Universe, one SNII is produced for every $\sim 10^2~\Msol$ of stars, and the energy released in the explosion is initially carried by $\ll 10^2~\Msol$ of ejecta. Because the ratio between the mass of the ejecta and the mass of the stellar population that released the energy is small ($\ll 1$), the SN energy per unit mass of ejecta is high. Hence, the ejecta move at very high velocity ($\gg 10^3~\kms$) and the post-shock temperatures are sufficiently large for the radiative cooling time to be long. Indeed, observations do provide evidence for hot gas associated with fast galactic outflows (e.g.~\citealt{Heckman1987,Strickland2000}; see also the review by \citealt{Veilleux2005}).

In simulations, however, the ratio between the gas mass receiving the SN energy and the stellar mass that produced it is much larger. Even if all the SNII energy of an SSP is injected at once, the ratio between the mass of the `ejecta' and that of the SSP will typically be large. For example, if, for the case of SPH, the energy is shared by all $N_{\rm SPH}$ neighbours, then the ratio will be $N_{\rm SPH}\gg 10$ and even larger if multiple star particles are spawned per gas particle or if the feedback energy is released over multiple time steps. Consequently, the temperature of the heated particles will be relatively low and most of the thermal energy will be radiated away before it is able to do $PdV$ work on its surroundings. It is important to note that the ratio between the gas mass receiving the energy and the mass of the SSP that produced it is independent of the resolution.

The realisation that the cause of the inefficient thermal feedback is a mismatch between the simulated and observed ratios of heated mass to stellar mass, rather than a straightforward lack of numerical resolution, immediately indicates the solution: we need to decrease this ratio in the simulations. By decreasing the mass ratio, we increase the temperature of the heated gas and hence its cooling time. If the cooling time is long compared with the sound crossing time scale across a resolution element, the heated gas will begin to expand adiabatically and the injected thermal energy will be efficiently converted into kinetic energy. We showed that in the temperature regime for which Brehmsstrahlung dominates the radiative cooling rate ($\ga 10^7$~K for solar abundances), the ratio of the cooling time and the sound crossing time remains constant if the gas expands adiabatically, so that adiabatic cooling does not invalidate the argument.

We showed analytically (see eq.~[\ref{eq:nhcrit}]) that, for the case of SPH, the maximum density for which radiative losses are small is $n_{\rm H} \sim 26~{\rm cm}^{-3}(T/10^{7.5}~{\rm K})^{3/2}(m/10^5~\Msol)^{-1/2}$, where $T$ is the temperature of the heated gas after receiving the feedback energy and $m$ is the mass of a gas resolution element (and we assumed that $\Delta T$ is sufficiently high for Brehmsstrahlung to dominate the radiative cooling). The maximum density is nearly the same for AMR simulations with cell sizes that are at least a factor of 4 smaller than the local Jeans length (evaluated in gas with this density and a temperature of $10^4$~K, see eq.~[\ref{eq:nhcritamr}]). Hence, by specifying the desired temperature jump $\Delta T$, we can guarantee that the feedback is efficient up to some gas density that we can estimate analytically. 

We implemented this idea in the SPH code \textsc{gadget} in the form of stochastic thermal feedback. A fixed $\Delta T$ then translates into a fixed probability of receiving energy, which we evaluate for each SPH neighbour of a star particle that has just crossed the critical age $t_{\rm SN} = 3\times 10^7~{\rm yr}$, corresponding to the maximum lifetime of stars that end their lives as SNII. For a Chabrier IMF and assuming equal mass particles, the expectation value of the number of heated particles is $1.34 f_{\rm th} (\Delta T/10^{7.5}~{\rm K})^{-1}$, where $f_{\rm th}$ is the fraction of the SNII energy that is injected (see eq.~[\ref{eq:nheat}]). 

Note that the parameter $\Delta T$ plays a similar role as the initial
wind velocity $v_{\rm w}$ for the case of kinetic feedback. For a fixed $v_{\rm w}$, the mass loading factor used in kinetic feedback
implementations sets $f_{\rm th}$. The fraction of the injected SN energy can also be used as a second free
parameter for the case of thermal feedback. For a fixed temperature
jump, the ``initial mass loading'' is then the ratio of the heated mass
per unit stellar mass formed and this ratio is proportional to the
parameter $f_{\rm th}$. 

The combination of stochastic feedback and a fixed increase in the energy per unit mass of the gas receiving the feedback energy also works for kinetic feedback: we can specify $v_{\rm w}$ and give each neighbouring resolution element of a star particle a probability of being kicked in the wind that is proportional to the mass loading factor or, equivalently, to $f_{\rm th}$. This is in fact exactly the implementation of kinetic feedback that we used in DS08. 

The equivalence of the roles of $\Delta T$ and $v_{\rm w}$ suggests that the efficiency of the kinetic feedback does not depend directly on the ratio of the initial wind velocity and the escape velocity, contrary to what is often assumed. Indeed, in DS08 we showed that if $v_{\rm w}$ is too low, the wind stalls in the ISM, before it has even begun to climb out of the gravitational potential. As we have shown, the efficiency of the wind depends on the radiative losses and hence, for a fixed value of $v_{\rm w}$, on the numerical resolution and the gas density. Because the typical gas densities increase with the gas pressure and thus with the depth of the potential well, the escape velocity does matter indirectly (also because drag forces increase with the pressure). If the wind manages to blow out of the ISM, then the ultimate efficiency of the feedback does depend on the ratio of the velocity of the wind leaving the ISM and the escape velocity, because the ejected gas will rain back onto the galaxy if it cannot escape the galaxy's potential well.

We presented analytic derivations of the resolution criteria, both for SPH and AMR simulations. We tested our recipe for thermal feedback on SPH simulations of isolated disc galaxies in dark matter haloes of total mass $10\Msolh$ and $10^{12}\Msolh$ using the same set-up as we used to study kinetic SN feedback in DS08. We explored the effect of the feedback on the gas distribution, the star formation history, the mass outflow rate, and the wind velocity. The results were in accord with our analytic predictions. 

For sufficiently high $\Delta T$ and for sufficiently high resolution, the thermal feedback strongly reduces the star formation rate and results in a strong, large-scale, bi-polar outflow. Reassuringly, the results converge with both $\Delta T$ and resolution and the converged results also agree well with simulations employing kinetic feedback\footnote{The thermal feedback models only agree with kinetic feedback simulations if the kicked wind particles are \emph{not} temporarily decoupled from the hydrodynamics, see DS08 for a critical discussion of this common practice.} (with sufficiently high $v_{\rm w}$). 

However, if $\Delta T$ and/or the resolution are too low, then the results become sensitive to both. For a fixed resolution, higher values of $\Delta T$ will then result in more efficient feedback. Hence, in this regime the ability to choose $\Delta T$ or, for the case of kinetic feedback $v_{\rm w}$, implies a considerable freedom. This freedom associated with the implementation of feedback from star formation is currently the limiting factor for the predictive power of cosmological simulations \citep[e.g.][]{Schaye2010,Scannapieco2011}. 

Given that a higher $\Delta T$ yields smaller radiative losses, one may ask why we do not use ultra-high values. Indeed, even in low-resolution simulations the feedback could be made efficient locally by increasing $\Delta T$ (or $v_{\rm w}$ for the case of kinetic feedback). There are, however, several reasons why it is undesirable to increase this parameter to values $>10^8$~K. First, even for $f_{\rm th}=1$ such high heating temperatures imply that, on average, each star particle will heat less than one neighbouring gas particle. Such a situation breaks the locality of the feedback and may lead to sampling problems. That this can have grave
consequences is easy to see by considering the limiting case in which the number of heated gas particles per star particle (i.e.\ the mean initial mass loading) is $\ll 1$. Most heavy elements released by massive stars will then no longer be injected in a wind. Many generations of star particles can form in a given gas cloud before a single feedback event takes place. Conversely, if, despite the low probability, a star particle forming in a region with a low star formation density does heat a neighbour, the energy injected may be sufficiently large to do catastrophic damage. Clearly, we should avoid the regime in which the expectation value for the number of heated gas elements per star particle formed is much less than one, at least for galaxies resolved with relatively small numbers of particles. 

At present, large-volume cosmological simulations typically have particle masses $\ga 10^6~\Msol$. At this resolution even heating one gas particle per star particle (which corresponds to $\Delta T \sim 10^{7.5}$~K for a Chabrier IMF and $f_{\rm th}=1$) results in strong radiative losses for densities $n_{\rm H} > 10~{\rm cm}^{-3}$, which are routinely reached in such simulations.
Hence, the predictions are still sensitive to the values of the feedback prescription and are thus uncertain. This undesirable limitation can be turned into an advantage if one takes an approach similar in spirit to semi-analytic models: by varying $\Delta T$ (or $v_{\rm w}$) with halo mass or with the local physical conditions, the feedback can be tuned to reproduce the desired galaxy formation efficiency. However, the arguments given above demonstrate that the results may change with increasing resolution\footnote{Turning off the hydrodynamical forces in the high-density regime could make the results insensitive to resolution for kinetic feedback \citep{Springel2003}, but the predictions will in that case disagree with converged, self-consistent high-resolution simulations.}. If the resolution and the value of $\Delta T$ are sufficiently high for cooling losses to be small, then the feedback can still be tuned by varying $f_{\rm th}$ with the local conditions.

We have demonstrated that, contrary to common wisdom, thermal feedback can be efficient without turning off radiative cooling. For sensible parameter choices overcooling can already be avoided for densities typical of the warm ISM (i.e.\ $n_{\rm H}\sim 1~{\rm cm}^{-3}$) at the resolution achievable for large-scale cosmological simulations ($m\sim 10^7~\Msol$) and for simulations of individual (low- to intermediate-mass) galaxies we can already afford the resolution ($m\sim 10^2~\Msol$) required for the feedback to remain efficient up to densities typical of molecular clouds ($n_{\rm H}\sim 10^3~{\rm cm}^{-3}$). We have also shown that with sufficient resolution, the results become insensitive to the problematic parameter of the feedback implementation (i.e.\ $\Delta T$ for thermal feedback and $v_{\rm w}$ for kinetic feedback) and the form in which the energy is injected, thus removing some of the most important uncertainties in the ingredients of hydrodynamical simulations of galaxy formation. 


\section*{Acknowledgements}

We are very grateful to Volker Springel for allowing us to use
\textsc{gadget} and his initial conditions code for the simulations
presented here. We thank Rob Crain and Jarrett Johnson for a careful reading of the manuscript, and the anonymous referee for a helpful report.
The simulations presented here were run
on the Cosmology Machine at the Institute for Computational Cosmology
in Durham as part of the Virgo Consortium research programme and on the TMoX cluster 
at the Rechenzentrum Garching of the Max Planck Society. This work was supported by Marie Curie Reintegration Grant FP7-RG-256573 and by the Marie Curie Initial Training Network CosmoComp (PITN-GA-2009-238356).

\end{document}